\documentclass[12pt]{article}
\pdfoutput=1
\usepackage[utf8]{inputenc}
\usepackage{amsfonts,amsmath}
\usepackage{jheppub}

\usepackage{color}
\usepackage{tabularx}

\usepackage{bbm}

\usepackage{tikz}
\usetikzlibrary{positioning,arrows}
\usetikzlibrary{decorations.pathmorphing,calc}
\usetikzlibrary{decorations.markings}
\newif\ifmirrorsemicircle
\usepackage{float}
\usepackage[justification=centering]{caption}

\usepackage{ifthen}
\usepackage{enumerate}
\usepackage{amsmath}
\usepackage{amssymb}
\usepackage[mathscr]{eucal}
\usepackage[errorshow]{tracefnt}
\usepackage{amsmath}
\usepackage{slashed}
\usepackage{amssymb}
\usepackage{array}
\usepackage{amsthm}
\usepackage{eucal}
\usepackage{epsf}
\usepackage{epsfig}
\usepackage{psfrag}
\usepackage{multirow}
\usepackage{wasysym} 
\usepackage{tikz-cd} 
\usepgfmodule{shapes}
\usetikzlibrary{backgrounds, arrows,calc,shapes,decorations.pathreplacing, decorations.markings, automata,positioning}

\newcommand{\be}{\begin{equation}}
\newcommand{\ee}{\end{equation}}
\newcommand{\bea}{\begin{eqnarray}}
\newcommand{\eea}{\end{eqnarray}}
\newcommand{\ba}{\begin{array}}
\newcommand{\ea}{\end{array}}
\newcommand{\bpic}{\begin{tikzpicture}}
\newcommand{\epic}{\end{tikzpicture}}

\def\Tr{\rm Tr}

\newcommand\uot{U(1)_{\text{top}}}

\newcommand{\M}{{\mathfrak M}}

\newcommand{\ra}{\rightarrow}

\newcommand{\lra}{\leftrightarrow}

\newcommand\sigt{\tilde \sigma}
\def\Phit{\tilde \Phi}
\def\Psit{\tilde \Psi}

\newcommand{\CC}{{\mathcal C}}

\newcommand{\CL}{{\mathcal L}}

\newcommand{\cN}{\mathcal{N}}

\newcommand{\cW}{\mathcal{W}}

\newcommand{\bC}{\mathbb{C}}

\newcommand{\bP}{\mathbb{P}}

\newcommand{\bZ}{\mathbb{Z}}

\def\l{\lambda} 

\def\p{\psi}

 \definecolor{mygreen}{RGB}{50,205,50}
 
 \makeatletter
\newcommand*\bigcdot{\mathpalette\bigcdot@{.5}}
\newcommand*\bigcdot@[2]{\mathbin{\vcenter{\hbox{\scalebox{#2}{$\m@th#1\bullet$}}}}}
\makeatother

\tikzset{
    partial ellipse/.style args={#1:#2:#3}{
        insert path={+ (#1:#3) arc (#1:#2:#3)}
    }
}

\tikzset{
    wave amplitude/.initial=0.2cm,
    wave count/.initial=8,
    mirror semicircle/.is if=mirrorsemicircle,
    mirror semicircle=false,
    wavy semicircle/.style={
        to path={
            let \p1 = (\tikztostart),
            \p2 = (\tikztotarget),
            \n1 = {veclen(\x2-\x1,\y2-\y1)},
            \n2 = {atan2(\x2-\x1,\y2-\y1))} in
            plot [
                smooth,
                samples=(\pgfkeysvalueof{/tikz/wave count}+0.5)*8+1, 
                domain=0:1,
                shift={($(\p1)!0.5!(\p2)$)}
            ] ({ 
                (\x*180-\n2 + 180 + \ifmirrorsemicircle 1 \else -1 \fi * 90%
            }:{ 
                (%
                    \n1/2+\pgfkeysvalueof{/tikz/wave amplitude} * %
                    sin(
                        \x * 360 * (\pgfkeysvalueof{/tikz/wave count} + 0.5%
                    )%
                )%
            })
        } (\tikztotarget)
    }
}

\pgfdeclaredecoration{complete sines}{initial}
{
    \state{initial}[
        width=+0pt,
        next state=sine,
        persistent precomputation={\pgfmathsetmacro\matchinglength{
            \pgfdecoratedinputsegmentlength / int(\pgfdecoratedinputsegmentlength/\pgfdecorationsegmentlength)}
            \setlength{\pgfdecorationsegmentlength}{\matchinglength pt}
        }] {}
    \state{sine}[width=\pgfdecorationsegmentlength]{
        \pgfpathsine{\pgfpoint{0.25\pgfdecorationsegmentlength}{0.5\pgfdecorationsegmentamplitude}}
        \pgfpathcosine{\pgfpoint{0.25\pgfdecorationsegmentlength}{-0.5\pgfdecorationsegmentamplitude}}
        \pgfpathsine{\pgfpoint{0.25\pgfdecorationsegmentlength}{-0.5\pgfdecorationsegmentamplitude}}
        \pgfpathcosine{\pgfpoint{0.25\pgfdecorationsegmentlength}{0.5\pgfdecorationsegmentamplitude}}
}
    \state{final}{}
}

\tikzset{ 
    fermion/.style={thick, draw=mygreen, postaction={decorate}, decoration={markings, mark=at position .5 with {\arrow[mygreen]{triangle 45}}}} ,
       scalar/.style={thick, draw=blue, postaction={decorate}, decoration={markings, mark=at position .43 with {\arrow[blue]{triangle 45}}}} ,
          scalarr/.style={thick, draw=blue, postaction={decorate}, decoration={markings, mark=at position .56 with {\arrowreversed[blue]{triangle 45}}}},
       photon/.style={decorate, draw=black,thick,
        decoration={complete sines,amplitude=5pt, segment length=7pt}},
         photon1/.style={decorate, draw=black,thick,
        decoration={complete sines,amplitude=3pt, segment length=5pt}},
          photon2/.style={decorate, draw=black,thick,
        decoration={complete sines,amplitude=3pt, segment length=4pt}}
      }

\title{Easy-plane QED$_3$'s in the large $N_f$ limit}

\preprint{SISSA 02/2019/FISI}

\author[1]{Sergio Benvenuti,}
\author[2]{Hrachya Khachatryan}

\affiliation[1,2]{International School of Advanced Studies (SISSA), Via Bonomea 265, 34136 Trieste, Italy}
\affiliation[1]{INFN, Sezione di Trieste, Via Valerio 2, 34127 Trieste, Italy}
\emailAdd{benve79@gmail.com, hrachya.khachatryan@sissa.it}

\abstract{We consider Quantum Electrodynamics in $2{+}1$ dimensions with $N_f$ fermionic or bosonic flavors, allowing for interactions that respect the global symmetry $U(N_f/2)^2$. There are four bosonic and four fermionic fixed points, which we analyze using  the large $N_f$ expansion. We systematically compute,  at order $O(1/N_f)$, the scaling dimensions of quadratic and quartic mesonic operators. 

We also consider Quantum Electrodynamics  with minimal supersymmetry. In this case the large $N_f$ scaling dimensions, extrapolated at $N_f{=}2$, agree quite well with the scaling dimensions of a dual supersymmetric Gross-Neveu-Yukawa model. This provides a quantitative check of the conjectured duality.

}

\begin{document}

\maketitle
\newpage
\section{Introduction and summary}
Quantum Electrodynamics (QED) in $2{+}1$ dimensions, with fermionic and/or bosonic flavors, is a prime example of interacting Quantum Field Theory, with both theoretical and experimental relevance.  In this paper we study QED's in the limit of large number of flavors, the large $N_f$ limit, where perturbation theory allows to find quantitative results.

Our goal is to define and study models that admit a tractable large $N_f$ expansion but at the same time might be realistic when the number of flavor is small. For this reason we consider an even number of flavors and allow for interactions that respect only $U(N_f/2)^2$ global symmetry, instead of the usual $U(N_f)$. Most of the results derived in this paper have been presented in \cite{Benvenuti:2018cwd}. We use the name "easy plane" QED's because for $N_f{=}2$, one of the bosonic fixed points is the "easy-plane" $\bC\bP^1$ model. Together with $SU(2)$-$\bC\bP^1$ model it describes the N\`eel --- Valence Bond Solid quantum phase transition in the $SU(2)$ and $XY$ magnets \cite{QCP1, QCP2, Motrunich:2003fz}.

We find four bosonic and four fermionic fixed points. The various models differ by the form of the quartic interactions, which in the large $N_f$ limit are modeled introducing one or two Hubbard-Stratonovich scalar fields. In each of the $8$ models we systematically compute the anomalous dimensions of all the scalar operators that at infinite-$N_f$ have small scaling dimension ($\Delta{=}1$ or $\Delta{=}2$). Some operators are quadratic or quartic in the charged fields, some are linear or quadratic in the Hubbard-Stratonovich fields. We work at the leading non trivial order in the large $N_f$ expansion, $O(1/N_f)$, providing many details of the computations, including results for all individual Feynman diagrams.

The results of this paper were used in \cite{Benvenuti:2018cwd} to argue for the following physical picture.

If the number of flavors $N_f$ is large enough, it is well known that the Renormalization Group (RG) flows in the infrared to a Conformal Field Theory (CFT) describing a second order phase transition. However, when $N_f$ is decreased, the large $N_f$ anomalous dimensions suggest that below a certain critical value $N_f^*$ (which is a model dependent quantity), the various fixed points merge pairwise and become complex Conformal Field Theories. This is also called annihilation and merging scenario: varying a parameter, in this case the number of flavors,  two real fixed points of the RG flow (real CFT's) annihilate and become a pair of complex conjugate CFT's. The RG flow preserves unitarity and doesn't hit those complex fixed points, instead it slows down while passing between the complex fixed points (walking) \cite{Gorbenko:2018ncu}. For $N_f < N_f^*$, the infrared physics is not described by a second order phase transition, but by a weakly first order phase transition. 

It is important that in the fermionic case a fixed point with $U(N_f)$ global symmetry ($N_f$ is the number of the two-component Dirac fermions) merges with a  fixed point with $U(N_f/2)$ global symmetry. This provides a rationale for chiral symmetry breaking in fermionic QED's \cite{Pisarski:1984dj, Kubota:2001kk, Kaveh:2004qa, DiPietro:2015taa, Giombi:2015haa, DiPietro:2017kcd, Herbut:2016ide, Gusynin:2016som, Benvenuti:2018cwd, Kotikov:2019rww} \footnote{Recent studies \cite{Li:2018lyb} using conformal bootstrap  suggest merging between $U(N_f)$ QED with $N_f$ two-component fermions minimally coupled to the gauge field and $U(N_f)$ QED-GN. }. In the bosonic case fixed points with the same symmetry merge, so there is no symmetry breaking \cite{MarchRussell:1992ei, Nogueira:2013oza, Nahum:2015jya, Nahum:2015vka, Benvenuti:2018cwd, Serna:2018tct}.

In  \cite{Benvenuti:2018cwd} we showed that the merging pattern of the $4$ bosonic and $4$ bosonic fixed points suggested by large $N_f$ argument is consistent with various boson $\lra$ fermion dualities conjectured to hold when the number of flavors is $N_f{=}2$ \cite{ Karch:2016sxi, Wang:2017txt, Benvenuti:2018cwd} (see also \cite{Aharony:2011jz, Giombi:2011kc, Aharony:2012nh, Son:2015xqa, Aharony:2015mjs, Seiberg:2016gmd, Karch:2016aux, Metlitski:2016dht, Hsin:2016blu, Aharony:2016jvv, Benini:2017dus, Komargodski:2017keh, Benini:2017aed, Jensen:2017bjo, Gomis:2017ixy, Bashmakov:2018wts, Benini:2018umh, Choi:2018ohn, Choi:2018tuh, Senthil:2018cru} for recent progress in $2{+}1d$ dualities). This scenario is also consistent with lattice simulations, that at small $N_f$ \cite{LSK, Kaul:2011dqx, Karthik:2016ppr, DEmidio:2016wwg, Zhang:2018bfc} suggest second order or weakly first order phase transition, and with numerical bootstrap results relevant for $N_f{=}2$ \cite{Nakayama:2016jhq, DSD, Poland:2018epd, IliesiuTALK}.

 \vspace{0.2cm}

 Studying quantum field theories in the large $N_f$ limit has been proved to be useful in different circumstances. In $2{+}1d$ the large $N_f$ limit  has recently been applied to calculate scaling dimensions of monopole operators, $S^3$ partition functions and central charges \cite{Klebanov:2011td, Pufu:2013vpa, Dyer:2015zha, Diab:2016spb, Giombi:2016fct, Chester:2017vdh, Murthy:1989ps, Borokhov:2002ib}. We think that it would be interesting to generalize those computations to the "easy-plane" models described in this paper.
 
\vspace{0.4cm}

After discussing QED's with bosonic flavors in section \ref{sec:sQEDs} and QED's with fermionic flavors in section \ref{sec:fQEDs} we, move to  QED with minimal supersymmetry, $\mathcal{N}=1$. In section \ref{sec:susy} we compute the scaling dimensions of bilinear and quartic mesonic operators. We also include the large $N_f$ dimensions of monopole operators from \cite{Chester:2017vdh}. In this case, lowering $N_f$ there is no evidence of merging or symmetry breaking. Instead, even for $N_f{=}2$ the large $N_f$ results are physical. $\cN=1$ QED with $N_f=2$ is supposed to be dual to a supersymmetric Gross-Neveu-Yukawa model \cite{Gaiotto:2018yjh, Benini:2018bhk}, which can be studied quantitatively in the $4-\varepsilon$ expansion \cite{Benini:2018bhk}. We compare the large $N_f$ results on the gauge theory side of the duality with the $4-\varepsilon$ results on the Gross-Neveu-Yukawa side of the duality, and we find good quantitative agreement, providing a check of the conjectured $\cN=1$ duality.

\vspace{2cm}



\section{Four bosonic QED fixed points in the large $N_f$ limit}\label{sec:sQEDs}
In this section we study bosonic QED with large $N_f$ complex scalar fields, imposing at least $U(N_f/2)^2$ global symmetry. There are four different fixed points, two fixed points have $U(N_f)$ global symmetry, two fixed points have $U(N_f/2)^2$ global symmetry.

We start by considering the following UV (Euclidean) lagrangian 
\begin{align} \nonumber  \CL &=  \frac{1}{4 e^2} F_{\mu \nu} F^{\mu \nu } + \sum_{i=1}^{N_f/2} (|D \Phi_i|^2+|D \tilde{\Phi}_i|^2)  + \lambda \sum_{i,j=1}^{N_f/2}  |\Phi_i|^2 |\tilde{\Phi}_j|^2 \\
&+ \lambda_{ep} \left( (\sum_{i=1}^{N_f/2} |\Phi_i|^2)^2+(\sum_{i=1}^{N_f/2} |\tilde{\Phi}_i|^2)^2\right) 
+\frac{N_f}{32  (1-\xi)} \int d^3 y \frac{\partial_\mu A^\mu(x) \partial _\nu A^ \nu(y)}{2 \pi^2 |x-y|^2} \ .    \label{h1}   \end{align}
 Where $F_{\mu \nu} = \partial_\mu A_\nu - \partial_\nu A_\mu $, and $D_\mu=\partial_\mu +i A_\mu$ is the covariant derivative with respect to the $U(1)$ gauge field $A_\mu$. The complex scalar fields $(\Phi_i, \tilde\Phi_i ) \ (i=1,..,N_f/2)$ carry charge $+1$ under the gauge group. The conformal gauge fixing is defined by the last term in (\ref{h1}). Choosing the gauge fixing parameter to be zero $(\xi=0)$ simplifies the calculations a lot, however we prefer to keep $\xi$ arbitrary (notice that in this parametrization $\xi=1$ is the Landau gauge). Calculating correlation functions of gauge invariant operators, we will see that some Feynman graphs depend on $\xi$, but the sum (at a given order in $1/N_f$) doesn't as expected. This is a useful check of the calculations. In the following, we will always assume conformal gauge fixing for all the QED actions, but will not write it explicitly. 
 
The quartic potential in (\ref{h1}) is a relevant deformation of the free theory. 
As explained in \cite{Benvenuti:2018cwd}, depending on the form of the quartic couplings $\{\lambda_{ep},\lambda\}$ there are four different fixed points \footnote{We tune all the mass terms to zero.}:
\begin{itemize}
\item bQED (tricritical), defined by vanishing quartic potential,
\item bQED$_+$ ($\bC\bP^{N_f-1}$ model), defined by $V \sim (\sum|\Phi_i|^2+|\tilde\Phi_i|^2)^2 $, 
\item ep-bQED ("easy-plane"), defined by $V \sim (\sum|\Phi_i|^2)^2+(\sum|\tilde\Phi_i|^2)^2 $,
\item bQED$_{-}$, defined by  $V \sim (\sum |\Phi_i|^2-|\tilde\Phi_i|^2)^2 $.
\end{itemize}
In appendix (\ref{sec:eps}) we study the RG flow diagram and the fixed points of the model (\ref{h1}) using the epsilon expansion technique. The zeros of the beta functions support the existence of precisely these four RG fixed points.  See \cite{Benvenuti:2018cwd, Calabrese:2002bm} for discussions about the ungauged fixed points and RG flow.

 We study the critical behaviour of the fixed points in the large $N_f$ limit. For this purpose we engineer the quartic interactions in terms of cubic and quadratic interactions via the Hubbard-Stratonovich trick. Introducing two Hubbard-Stratonovich (HS) fields $\sigma$ and $\tilde{\sigma}$, we get an expression equivalent to (\ref{h1})
\begin{align} \nonumber
  \CL = &  \frac{1}{4 e^2} F_{\mu \nu} F^{\mu \nu } + \sum_{i=1}^{N_f/2} (|D \Phi_i|^2+|D \tilde{\Phi}_i|^2)  + \sigma  \sum_{i=1}^{N_f/2} |\Phi_i|^2  +\tilde{\sigma}  \sum_{i=1}^{N_f/2} |\tilde{\Phi}_i|^2  \\
  & - \frac{\eta_1}{2}(\sigma^2 + \sigt^2) - \eta_2 \sigma \sigt\, \ .
  \end{align}
Integrating out $\sigma$ and $\sigt$, one recovers the quartic potential in (\ref{h1})  with couplings $\{\l_{ep}, \l \}$ expressed  in terms of $\{\eta_1, \eta_2\}$:
\begin{align} 
&\lambda_{ep}=\frac{\eta_1}{2(\eta_1^2-\eta_2^2)}  \\
&\lambda=-\frac{\eta_2}{\eta_1^2-\eta_2^2} \ .
\end{align}
It is sometimes convenient to work with the following HS fields 
\begin{align} \nonumber
&\sigma_+=\frac{\sigma+\tilde\sigma}{2} \\
&\sigma_-=\frac{\sigma-\tilde\sigma}{2} \ . \label{h4}
\end{align}
With the choice (\ref{h4}) there is no mixed quadratic term between $\sigma_+$ and $\sigma_-$. 
\begin{align} \nonumber
  \CL =&  \frac{1}{4 e^2} F_{\mu \nu} F^{\mu \nu } + \sum_{i=1}^{N_f/2} (|D \Phi_i|^2+|D \tilde{\Phi}_i|^2)  + \sigma_+  \sum_{i=1}^{N_f/2} (|\Phi_i|^2 +|\tilde\Phi_i|^2) +\sigma_- \sum_{i=1}^{N_f/2} (|\Phi_i|^2 -|\tilde\Phi_i|^2)   \\
  & -(\eta_1+\eta_2) \sigma_+^2-(\eta_1-\eta_2) \sigma_-^2 \  .
 \end{align}


\subsection{bQED (tricritical QED)}\label{sec:bQED}
The  bQED is reached tuning to zero both the mass terms and the quartic interactions. For this reason another name for it is tricritical. The large $N_f$ effective action is described by $N_f$ copies of complex scalars $\Phi_i$ (we collected all the scalars $(\Phi, \tilde\Phi)$ into a single field and denoted it by $\Phi$) minimally coupled to the effective photon 
\begin{align} \label{h2}
\CL _{eff}=   \sum_{i=1}^{N_f} |D_\mu \Phi_i|^2 \ .
\end{align}
Effective photon propagator is obtained by summing geometric series of bubble diagrams such as (\ref{F1}). 
\begin{align}
 \label{photonprop} \langle A_\mu (x) A_\nu(0) \rangle_{\text{eff}}= \frac{8 }{ \pi^2 N_f |x|^2} \Big ( (1-\xi)\delta_{\mu \nu}+2 \xi \frac{x_\mu x_\nu}{|x|^2} \Big)
 \end{align}
The feynman rules for the bQED action (\ref{h2}) are summarised in the table (\ref{F2}). 

The faithful global symmetry is
\begin{align}
  \label{bQED-symmetry} \Big(\frac{SU(N_f)}{\bZ_{N_f}} \times \uot \Big)\rtimes\bZ_2^{\CC} \ .
   \end{align}
Where $\bZ_{N_f}$ is the center of $SU(N_f)$, generated by $e^{2 \pi i /N_f}\mathbb{I} \in SU(N_f)$, which is a gauge transformation, so the symmetry is $PSU(N_f)= \frac{SU(N_f)}{\bZ_{N_f}} $ instead of $SU(N_f)$ (the gauge invariant local operators transform in $SU(N_f)$ representations with zero $N_f$-ality). $\bZ_2^{\CC}$ is the charge-conjugation symmetry $\Phi_i \ra \Phi^*_i, A_\mu \ra - A_\mu$. There is also parity symmetry. 

\begin{figure}[]
  \centering
  \begin{tikzpicture} [scale=0.5]  
  \draw[photon, red] (0,0) to (2,0);
  \node at (2.5,0) {$=$};
  \draw[photon] (3,0) to (5,0);
  \node at (5.5,0) {$+$};
  \draw[photon] (6,0) to (7.1,0);
  \draw[thick, blue] (8.1,0) circle(1cm);
    \draw[photon] (9.1,0) to (10.2,0);
      \node at (10.7,0) {$+$};
      \draw[photon] (11.3,0) to (12.4,0);
  \draw[thick, blue] (13.4,0) circle(1cm);
    \draw[photon] (14.4,0) to (15.5,0);
    \draw[thick,blue] (16.5,0) circle (1cm);
    \draw[photon] (17.5,0) to (18.6,0);
    \node at (19,0) {$ \ \ \  \ \ \ + \  \ ....$};     
  \end{tikzpicture}
\caption{Effective photon propagator (red wavy line). The black wavy line stands for tree level photon propagator.} \label{F1}
  \end{figure}

\begin{figure}[]
  \centering 
  \begin{tikzpicture} [scale=0.8]  
  \draw[black] (-0.5,1) to (8,1);
  \draw[black] (-0.5,1) to (-0.5,-6);
  \draw[black]  (-0.5,-6) to (8,-6);
  \draw[black] (8,1) to (8,-6);
  \draw[photon,red] (0,0) to (2,0); 
  \node at (5,0) {$=\langle A_\mu (x) A_\nu(0) \rangle_{\text{eff}}$};
   \draw[thick, draw=blue, postaction={decorate}, decoration={markings, mark=at position .55 with {\arrow[blue]{triangle 45}}}] (0,-1.5+0.2) to (2,-1.5+0.2); 
   \node at (4-0.2,-1.5+0.2) {$=\frac{1}{4 \pi |x|}$};
   \draw[photon,red] (0,-2.5) to (1,-3);
   \draw[photon,red] (0,-3.5) to (1,-3);
   \draw[thick, draw=blue, postaction={decorate}, decoration={markings, mark=at position .4 with {\arrowreversed[blue]{triangle 45}}}] (1,-3) to (2,-2.5);
    \draw[thick, draw=blue, postaction={decorate}, decoration={markings, mark=at position .65 with {\arrow[blue]{triangle 45}}}] (1,-3) to (2,-3.5);
    \node at (4-0.1,-3) {$=-\delta_{\mu\nu} $};
    \draw[photon,red] (0,-3-1.8) to (1,-3-1.8);
    \draw[thick, draw=blue, postaction={decorate}, decoration={markings, mark=at position .4 with {\arrowreversed[blue]{triangle 45}}}] (1,-3-1.8) to (2,-2.5-1.8);
    \draw[thick, draw=blue, postaction={decorate}, decoration={markings, mark=at position .65 with {\arrow[blue]{triangle 45}}}] (1,-3-1.8) to (2,-3.5-1.8);
   \node at (1,-3-1.8+0.3) {$x$};
\node at (4-0.2,-3-1.8) {$=i \overset{\leftrightarrow}{\partial^x_\mu}$};
    \end{tikzpicture}
\caption{bQED Feynman rules.}  \label{F2}
  \end{figure}
  
Using the Feynman rules (\ref{F2}), we compute anomalous dimensions of gauge-invariant operators at order $O(1/N_f)$. For this purpose, first we calculate the 2-point correlation function for a given operator, then using it we extract anomalous contribution to the scaling. It might happen that for a given model there are several gauge invariant operators that have the same scaling dimensions at the order $O(N_f^0)$ and carry the same quantum numbers. These operators can mix by quantum corrections at order $O(1/N_f)$ and one needs to study the matrix of mixed 2-point correlation functions in order to correctly identify the eigenbasis of mixed operators and their anomalous dimensions. 

\subsection*{Scaling dimension of low-lying scalar operators}

\paragraph{Bilinear mesonic operators}

At the quadratic level, there are $N_f^2$ operators of the form $\Phi^*_i \Phi^j$. They transform in the adjoint plus singlet representations of $SU(N_f)$:
\begin{align} 
& |\Phi|^2_{adj} = \Phi^*_i \Phi^j - \frac{\delta_i^j}{N_f} \sum_k \Phi^*_k \Phi^k \\
& |\Phi|^2_{sing}=\frac{1}{\sqrt{N_f}}\sum_k \Phi^*_k \Phi^k \ .
\end{align}

 \begin{figure}[]
  \centering
\begin{tikzpicture}[scale=0.4]

\draw[black] (-4,3) to (18,3);
\draw[black] (-4,3) to (-4,-30+2+5);
\draw[black] (18,3) to (18,-30+2+5);
\draw[black] (-4,-30+2+5) to (18,-30+2+5);

\draw[thick, blue] (0,0) circle (2cm);
\draw[thick,fill] (-2,0) circle (4pt);
\draw[thick,fill] (2,0) circle (4pt);

\node at (8,0) {$=\Big(\frac{1}{4\pi|x|}\Big)^2 \equiv W$};

\node at (-3,0)  {$A$};
\node at (-3,-5)  {$B$};
\node at (-3,-10)  {$C$};
\node at (-3,-15)  {$D$};
\node at (-3,-20)  {$E$};

\draw[blue,thick] ([shift=(180:2cm)]0,-5)   arc (180:0:2cm);
\draw[blue,thick] ([shift=(0:2cm)]0,-5)   arc (0:-180:2cm);
\draw[photon,red] (-1.66,-3.9) to (1.66,-3.9);
\draw[thick,fill] (-2,-5) circle (4pt);
\draw[thick,fill] (2,-5) circle (4pt);
\node at (9,-5+0.4) {$ =2 \times \frac{ 4\big( 5+3 \xi \big) \log {x^2\Lambda^2} }{ 3 \pi^2 N_f } W$};

\draw[blue,thick] ([shift=(180:2cm)]0,-10)   arc (180:0:2cm);
\draw[blue,thick] ([shift=(0:2cm)]0,-10)   arc (0:-180:2cm);
\draw[photon, red] (0,-10) ([shift=(90:2cm)]0,-10)  to   ([shift=(-90:2cm)]0,-10);
\draw[thick,fill] (-2,-10) circle (4pt);
\draw[thick,fill] (2,-10) circle (4pt);
 \node at (9,-10+0.4) {$=\frac{24\big( 1-\xi \big) \log {x^2\Lambda^2} }{3\pi^2 N_f}W $}; 

\draw[thick,blue]  ([shift=(0:2cm)]0,-15)  arc (0:180:2cm);
\draw[thick,blue]  ([shift=(-180:2cm)]0,-15) arc (-180:0:2cm);
\draw[photon, red] ([shift=(0:2cm)]0,-15)  to   ([shift=(-210:2cm)]6,-15);
\draw[photon, red] ([shift=(0:2cm)]0,-15)  to   ([shift=(-150:2cm)]6,-15);
\draw[thick,blue ] ([shift=(180:2cm)]6,-15) arc (180:0:2cm);
\draw[thick,blue] ([shift=(0:2cm)]6,-15) arc (0:-180:2cm);
\draw[thick,fill] (-2,-15) circle (4pt);
\draw[thick,fill] (8,-15) circle (4pt);
\node at (13,-15) {$= 4 \times\frac{-48\log {x^2\Lambda^2}}{ 3 \pi^2 N_f } W$};

\draw[thick,blue]  ([shift=(0:2cm)]0,-20)  arc (0:180:2cm);
\draw[thick,blue]  ([shift=(-180:2cm)]0,-20) arc (-180:0:2cm); 
\draw[photon, red] ([shift=(40-14:2cm)]0,-20)  to   ([shift=(140+14:2cm)]6,-20);
\draw[photon, red] ([shift=(-40+14:2cm)]0,-20)  to   ([shift=(-140-14:2cm)]6,-20);
\draw[thick,blue ] ([shift=(180:2cm)]6,-20) arc (180:0:2cm);
\draw[thick,blue] ([shift=(0:2cm)]6,-20) arc (0:-180:2cm);
\draw[thick,fill] (-2,-20) circle (4pt);
\draw[thick,fill] (8,-20) circle (4pt);
\node at (10.5,-20) {$=0$};

  \end{tikzpicture}
\caption{ (bQED) Results for individual Feynman graphs appearing in the 2-point correlation function of the scalar-bilinear operators. \protect\footnotemark } \label{F3}
  \end{figure}
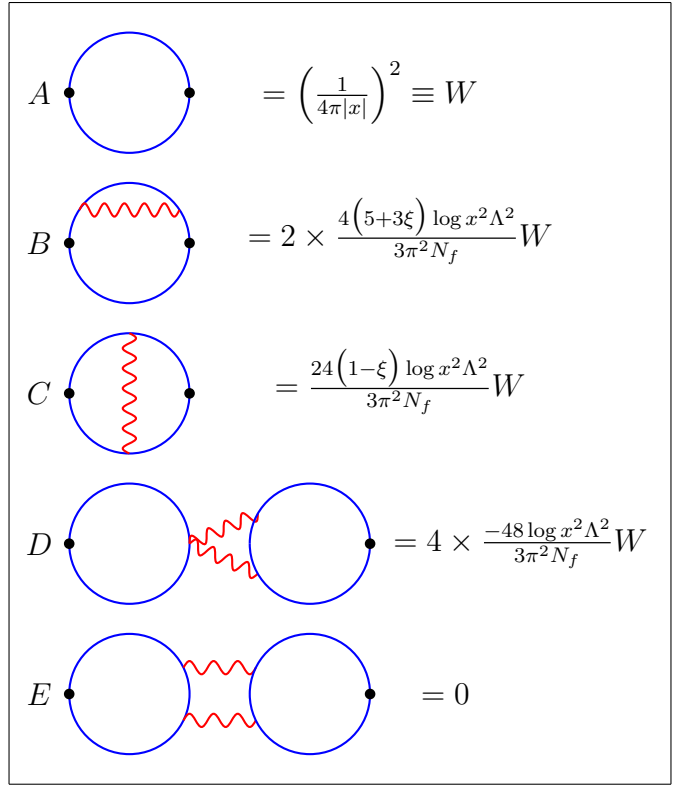
  
  \footnotetext{ When it is not crucial for the graph evaluation, we drop the arrows from propagators. }
  
The 2-point correlation function for the adjoint operator is the sum of the graphs A,B,C (\ref{F3}). For the adjoint operator the graphs D and E are of order $O(1/N_f^2)$ because each photon brings a factor $1/N_f$ and the two blue loops are of order $O(N_f^0)$. All the divergent graphs are regularized by putting an UV cutoff $\Lambda$ on the momentum integrals. Check the appendix (\ref{sec:diag}) for more details of the loop calculations. 
 \begin{align} \nonumber
 \langle |\Phi|^2_{adj}(x) |\Phi|^2_{adj}(0) \rangle &\! =\! \Big(\frac{1}{4\pi|x|}\Big)^2\!\!+ \!\! \frac{ 8\big( 5+3\xi \big) \log {x^2\Lambda^2}}{ 3 \pi^2 N_f } \! \Big(\frac{1}{4\pi|x|}\Big)^2\!+\!\frac{24\big( 1-\xi \big) \log {x^2 \Lambda^2 }}{3 \pi^2 N_f} \! \Big(\frac{1}{4\pi|x|}\Big)^2 \\ \nonumber
 &= \Big(\frac{1}{4\pi|x|}\Big)^2 \Big[1 - \Big( -\frac{64}{3 \pi^2 N_f} \Big)\log {x^2 \Lambda^2} \Big] \\
 &= \Big(\frac{1}{4\pi|x|}\Big)^2 \Big( \frac{1}{x^2\Lambda^2} \Big) ^{ \Delta^{(1)}_{adj}} \ .
 \end{align}
Where we defined anomalous dimension of adjoint operator $\Delta^{(1)}_{adj}$, so $\Delta^{(1)}_{adj}=-\frac{64}{3 \pi^2 N_f}$.  We  extract the anomalous dimension for the singlet operator in a similar way. Notice that for the singlet operator there is an additional order $O(1/N_f)$ contribution coming from the graph D (\ref{F3}) (as opposed to the adjoint bilinear, in the singlet case each loop in (D,E) gives a factor $N_f$). Below we give the final results 
\begin{align} 
& \Delta[ |\Phi|^2_{adj}] = 1 - \frac{64}{3 \pi^2 N_f} + O(1/{N_f^2})    \\
&\Delta[ |\Phi|^2_{sing}] = 1 + \frac{128}{3 \pi^2 N_f} + O(1/{N_f^2}) \ .
\end{align}

\paragraph{Quartic mesonic operators}

\begin{figure}[]
  \centering
\begin{tikzpicture}[scale=0.4]

\draw[black] (-3,3) to (18,3);
\draw[black] (-3,3) to (-3,-23);
\draw[black] (18,3) to (18,-23);
\draw[black] (-3,-23) to (18,-23);
\draw[black] (-3,-17.5) to (18,-17.5);

\draw[scalar] (0,0) (0:2) arc (0:180:2 and 1);
\draw[scalar] (0,0) (-180:2) arc (-180:0:2 and 1);
\draw[scalar] (0,0) (0:2) arc (0:180:2);
\draw[scalar] (0,0) (-180:2) arc (-180:0:2);
\draw[thick,fill] (-2,0) circle (4pt);
\draw[thick,fill] (2,0) circle (4pt);
\node at (5,0) {$=W^2$};

\begin{scope}[shift={(0,-5)}]
\draw[scalar] (0,0) (0:2) arc (0:180:2 and 1);
\draw[scalar] (0,0) (-180:2) arc (-180:0:2 and 1);
\draw[scalar] (0,0) (0:2) arc (0:180:2);
\draw[scalar] (0,0) (-180:2) arc (-180:0:2);
\draw[photon,red] ([shift=(45:2)]0,0) to ([shift=(135:2)]0,0);
\draw[thick,fill] (-2,0) circle (4pt);
\draw[thick,fill] (2,0) circle (4pt);
\node at (9,0) {$=4\times  \frac{ 4\big( 5+3\xi \big) \log {x^2\Lambda^2} }{ 3  \pi^2 N_f } W^2$};
\end{scope}

\begin{scope}[shift={(0,-10)}]
\draw[scalar] (0,0) (0:2) arc (0:180:2 and 1);
\draw[scalar] (0,0) (-180:2) arc (-180:0:2 and 1);
\draw[scalar] (0,0) (0:2) arc (0:180:2);
\draw[scalar] (0,0) (-180:2) arc (-180:0:2);
\draw[photon1,red ] ([shift=(90:2)]0,0) to ([shift=(90:2 and 1)]0,0);
\draw[thick,fill] (-2,0) circle (4pt);
\draw[thick,fill] (2,0) circle (4pt);
\node at (9,0) {$=2\times  \frac{ -24 \big( 1-\xi \big) \log {x^2\Lambda^2} }{ 3 \pi^2 N_f } W^2$};
\end{scope}

\begin{scope}[shift={(0,-15)}]
\draw[scalar] (0,0) (0:2) arc (0:180:2 and 1);
\draw[scalar] (0,0) (-180:2) arc (-180:0:2 and 1);
\draw[scalar] (0,0) (0:2) arc (0:180:2);
\draw[scalar] (0,0) (-180:2) arc (-180:0:2);
\draw[photon,red ] ([shift=(90:2)]0,0) to ([shift=(-90:2 and 1)]0,0);
\draw[thick,fill] (-2,0) circle (4pt);
\draw[thick,fill] (2,0) circle (4pt);
\node at (9,0) {$=4 \times  \frac{24 \big( 1-\xi \big) \log {x^2\Lambda^2} }{ 3 \pi^2 N_f } W^2$};
\end{scope}

\begin{scope}[shift={(0,-20)}]
\draw[blue,thick,postaction={decorate}, decoration={markings, mark=at position .45 with {\arrowreversed[blue]{triangle 45}}}] (0,0) (0:2) arc (0:180:2 and 1);
\draw[blue,thick] (0,0) (-180+100:2 and 1) arc (-180+100:0:2 and 1);
\draw[blue,thick] (-180+100: 2 and 1 ) to (-180+110: 2);
\draw[blue,thick,postaction={decorate}, decoration={markings, mark=at position .53 with {\arrow[blue]{triangle 45}}}] (0,0) (0:2) arc (0:180:2);
\draw[blue,thick,blue,thick,postaction={decorate}, decoration={markings, mark=at position .63 with {\arrow[blue]{triangle 45}}}] (0,0) (-180+110:2) arc (-180+110:0:2);
\draw[blue,thick] ([shift=(-15:1.4)]-1.9,-1.5) arc (-15:100:1.4);
\draw[blue,thick, postaction={decorate}, decoration={markings, mark=at position .44 with {\arrowreversed[blue]{triangle 45}}}] (0,0) (-130+23:2) arc (-130+23:-180:2);
\draw[thick,fill] (-2,0) circle (4pt);
\draw[thick,fill] (2,0) circle (4pt);
\draw[photon1,red] (-130+23:2) to (-180+110:2);
\draw[photon1,red] (-130+23:2) to ([shift=(-180+100:2 and 1)]0,0);
\node at (9,0) {$=4 \times  \frac{ -48 \log {x^2\Lambda^2} }{ 3 \pi^2 N_f } W^2$};
\end{scope}

  \end{tikzpicture}
\caption{(bQED) adjoint-2 and adjoint quartic operator renormalization. For adjoint-2 the last graph contributes at order $O(1/N_f^2)$ .} \label{F4}
  \end{figure}
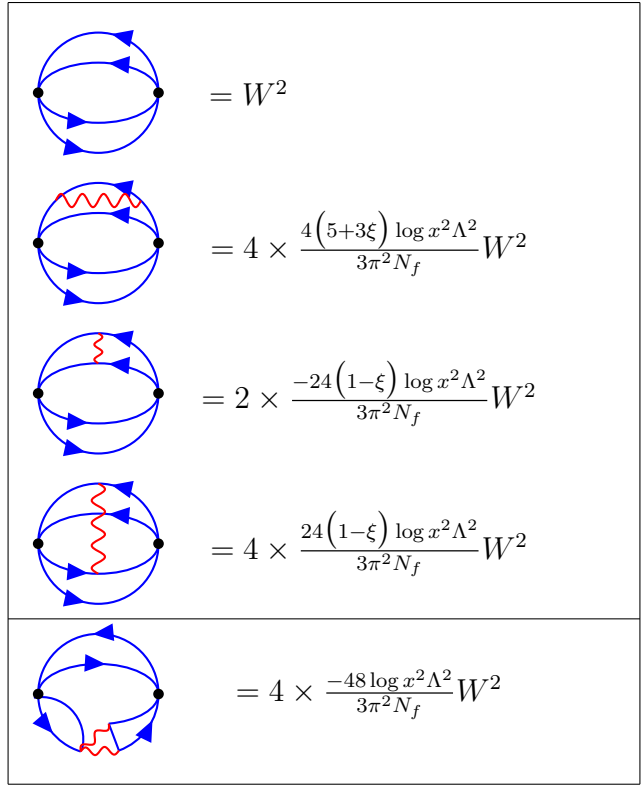

 Next we consider scalar quartic operators 
\begin{align} \label{scalar quartic}
T^{i j}_{k l} \equiv \Phi^{i} \Phi^{ j} \Phi^{*}_k \Phi^{*}_l \ . 
\end{align}
 $T^{i j}_{k l}$ is a gauge invariant operator, symmetric in its upper and lower indices. The following decomposition of $T$ into irreducible representations under the $SU(N_f)$ group is useful for discussion of their scaling dimensions 
\begin{align} \nonumber 
T^{i j}_{k l} =& \frac{1}{N_f(N_f+1)} \big[ \delta^{( i}_k \delta^{j )}_{l} T^{m n}_{ m n} \big] +\frac{1}{N_f+2} \big[ \delta^{(j}_{(l} T^{ i) n}_{ k) n}-\frac{2}{N_f} \delta^{( j}_l \delta^{i )}_{k} T^{ m n}_{m n}   \big ]  \\ 
&+\big[ T^{i j}_{k l}  -\frac{1}{N_f+2}  \delta^{(j}_{(l}T^{ i) n}_{ k)n}+\frac{1}{(N_f+1)(N_f+2)} \delta^{( j}_l \delta^{i )}_{k} T^{m n}_{ m n} \big] \ .  \label{h3}  
\end{align}
The first, second and third terms in the right hand side of (\ref{h3}) are correspondingly singlet, adjoint and adjoint-2 (Dynkin labels $[2,0,\ldots,0,2]$) quartic operators. All of them have scaling dimension 2 at leading order, it remains to calculate order $O(1/N_f)$ corrections. \\
\indent Let us consider quartic adjoint-2 operator defined by the last term of (\ref{h3}). It is enough to study the two-point correlation function for only one component of the adjoint-2 representation, which we choose to be
\begin{align} \label{adj-2}
T^{1 2}_{3 4}= \Phi^{1} \Phi^{ 2} \Phi^{*}_3 \Phi^{*}_4 \ .
\end{align}
All the relevant graphs for extracting the anomalous dimension of the operator (\ref{adj-2}) are collected in the table (\ref{F4}) (the last graph comes at sub-leading order). It receives contribution from the anomalous dimensions of the $\Phi_i$ fields (there are 4 such graphs) plus graphs with a photon connecting two different legs ("kite"-graphs, there are 6 "kite"-graphs). In 2 "kite"-graphs the photon connects the scalar propagators with arrows going in the same direction, while in the other 4 "kite"-graphs the photon connects propagators with arrows going in the opposite direction. The contribution of a "kite"-graph where the photon connects arrows going in the same direction is equal to minus the contribution of a "kite"-graph where the photon connects  arrows going in the opposite direction. So effectively we are left with the contribution of 2 such "kite"-graphs.  \footnote{One can consider degree-$2k$ operators which transform in the adjoint-k representation (Dynkin labels $[k,0,\ldots,0,k]$). These operators do not mix with other operators. The anomalous dimension of a degree-$2k$ adjoint-$k$ operator, at order $O(1/N_f)$, receives contribution from the anomalous dimensions of the $\Phi_i$ fields (there are $2k$ such graphs) plus the contribution of "kite" graphs (there are $\binom{2k}{2}=2k^2-k$  "kite"-graphs). In $2\cdot \binom{k}{2}=k^2-k$ "kite"-graphs the photon connects fields with arrows going in the same direction, while in the other $k^2$ "kite"-graphs the photon connects fields with arrows going in the opposite direction. These two groups of "kite"-graphs contribute with opposite signs, so effectively we are left with the contribution of $k^2-(k^2-k)=k$ such "kite"-graphs. Therefore the scaling dimension of the degree-$2k$ adjoint-$k$ operator is \be \nonumber \Delta[|\Phi|^{2k}_{adj-k}]  = k  \Delta[ |\Phi|^2_{adj}]+ O(1/N_f^2) = k - \frac{64k}{3 \pi^2 N_f} + O(1/N_f^2)  \ . \ee
 }

For the quartic adjoint operator the last graph in (\ref{F4}) contributes at order $O(1/N_f)$. For the singlet quartic operator the last graph contributes twice as much as for the quartic adjoint operator. We list the quartic operators and their scaling dimensions 
\begin{align}
& \Delta[ |\Phi|^4_{adj-2}] = 2\Delta[ |\Phi|^2_{adj}] + O(1/{N_f^2})= 2 - \frac{128}{3 \pi^2 N_f} + O(1/{N_f^2})   \\
&\Delta[ |\Phi|^4_{adj}] = \Delta[ |\Phi|^2_{adj}] +\Delta[ |\Phi|^2_{sing}] + O(1/{N_f^2})= 2 + \frac{64}{3 \pi^2 N_f} + O(1/{N_f^2})  \\
 &\Delta[ |\Phi|^4_{sing}] =2\Delta[ |\Phi|^2_{sing}] + O(1/{N_f^2})=2 + \frac{256}{3 \pi^2 N_f} + O(1/{N_f^2}) \ .
 \end{align}

\subsection{bQED$_+$ ($\bC\bP^{N_f-1}$ model)}\label{sec:scQED}
\begin{figure}[H]
  \centering
  \begin{tikzpicture} [scale=0.5]  
  \draw[dashed, red, thick] (-0.8,0) to (1.2,0);
  \node at (2.1,-0.05) {$=$};
  \draw[dashed,thick] (3,0) to (5,0);
  \node at (5.5,0) {$+$};
  \draw[dashed,thick] (6,0) to (7.1,0);
  \draw[thick, blue] (8.1,0) circle(1cm);
   \draw[dashed,thick] (9.1,0) to (10.2,0);
  \node at (10.7,0) {$+$};
   \draw[dashed,thick] (11.3,0) to (12.4,0);
  \draw[thick, blue] (13.4,0) circle(1cm);
   \draw[dashed, thick] (14.4,0) to (15.5,0);
   \draw[thick,blue] (16.5,0) circle (1cm);
   \draw[dashed,thick] (17.5,0) to (18.6,0);
   \node at (19,0) {$ \ \ \  \ \ \ + \  \ ....$};     
  \end{tikzpicture}
 \caption{(bQED$_+$) HS field $\sigma_+$ effective propagator (red dashed line). The black dashed line stands for tree level HS field propagator. } \label{F5}
  \end{figure}
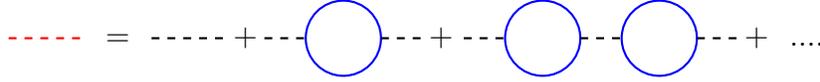

The bQED$_+$ fixed point is reached with $SU(N)_f$ invariant quartic deformation $V \sim (\sum|\Phi_i|^2+|\tilde\Phi_i|^2)^2 $ and by tuning the mass term to zero. In the literature this model is also known as Abelian Higgs model or $\bC\bP^{N_f-1}$ model. The large $N_f$ effective action is described by $N_f$ copies of complex scalars $\Phi_i$ (we collected all the scalars $(\Phi, \tilde\Phi)$ into a single field and denoted it by $\Phi$), minimally coupled to an effective photon and interacting with a single Hubbard-Stratonovich field $\sigma_+$ via a cubic interaction: 
\be \CL_{eff} =   \sum_{i=1}^{N_f} |D_\mu \Phi_i|^2 +\sigma_+ \sum_{i=1}^{N_f}  |\Phi_i|^2  \,. \ee
The effective photon propagator is the same as in (\ref{photonprop}) and the effective propagator for the HS field is obtained from summing geometric series of the bubble diagrams in (\ref{F5}).
\begin{align}
 \langle \sigma_+(x) \sigma_+(0) \rangle_{eff} = \frac{8}{\pi^2 N_f |x|^4} \ .
\end{align}
The global symmetry is the same as in bQED (\ref{bQED-symmetry}).

\subsection*{Scaling dimension of low-lying scalar operators}

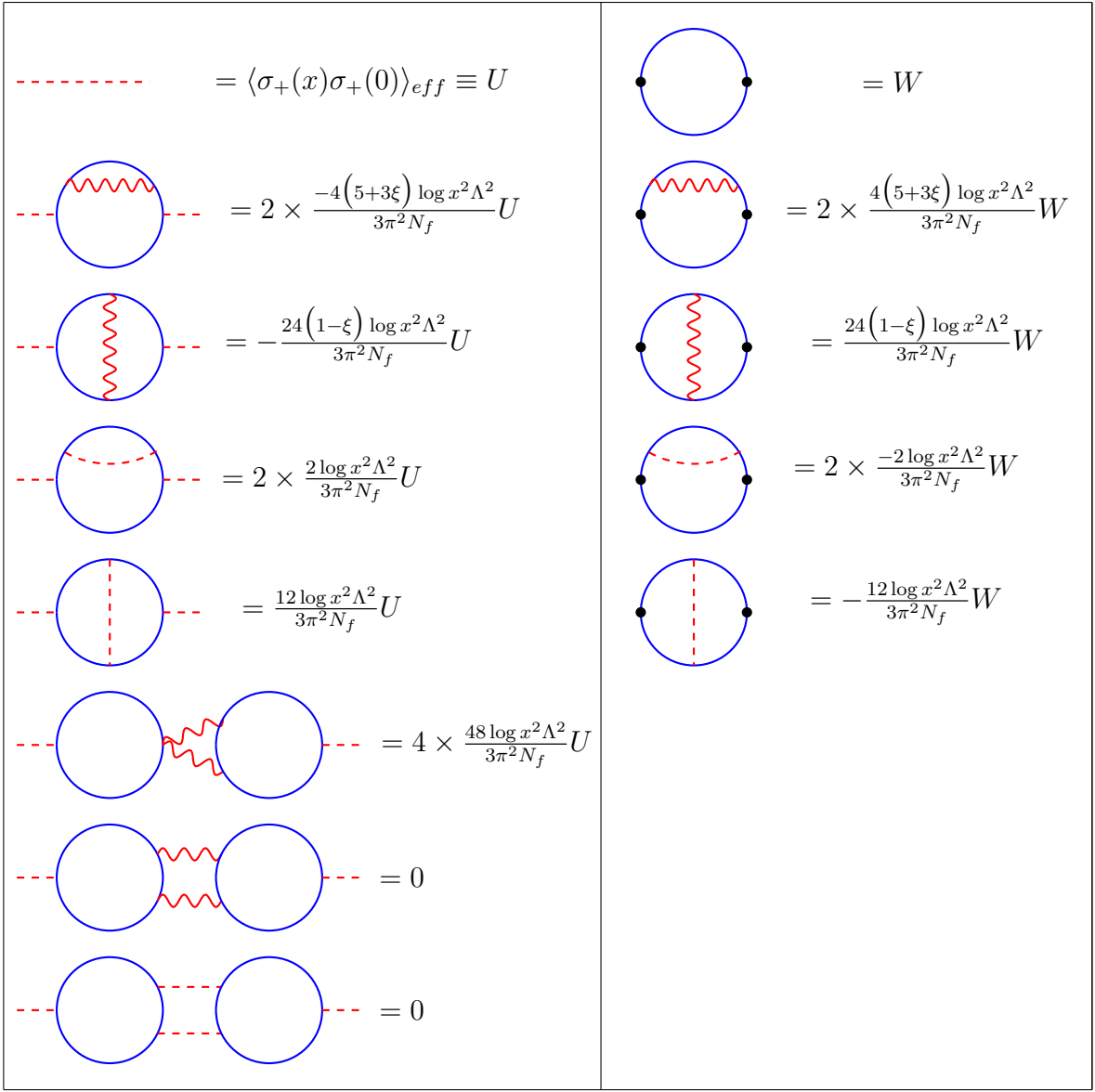
\begin{figure}[H]
  \centering
\begin{tikzpicture}[scale=0.38]

\draw[black] (-4,3) to (37,3);
\draw[black] (-4,3) to (-4,-30+2-10);
\draw[black] (37,3) to (37,-30+2-10);
\draw[black] (37,3) to (37,-30+2-10);
\draw[black] (-4,-30+2-10) to (37,-30+2-10);
\draw[black] (-4+22.5,3) to (-4+22.5,-30+2-10);

\draw[dashed,red,thick] (-3.5,0) to (1.5,0);

\node at (10-0.5,0) {$=\langle \sigma_+(x) \sigma_+(0) \rangle_{eff} \equiv U$};

\draw[blue,thick] ([shift=(180:2cm)]0,-5)   arc (180:0:2cm);
\draw[blue,thick] ([shift=(0:2cm)]0,-5)   arc (0:-180:2cm);
\draw[dashed,red,thick] (-3.5,-5) to (-2,-5);
\draw[dashed,red,thick] (2,-5) to (3.5,-5);
\draw[photon,red] (-1.66,-3.9) to (1.66,-3.9);

\node at (10,-5+0.4) {$ =2\times \frac{ -4\big( 5+3 \xi \big) \log{x^2 \Lambda^2}}{ 3  \pi^2 N_f } U$};

\draw[blue,thick] ([shift=(180:2cm)]0,-10)   arc (180:0:2cm);
\draw[blue,thick] ([shift=(0:2cm)]0,-10)   arc (0:-180:2cm);
\draw[photon, red] (0,-10) ([shift=(90:2cm)]0,-10)  to   ([shift=(-90:2cm)]0,-10);
\draw[dashed,red,thick] (-3.5,-10) to (-2,-10);
\draw[dashed,red,thick] (2,-10) to (3.5,-10);  
 \node at (9,-10+0.4) {$=-\frac{24\big( 1-\xi \big) \log{x^2 \Lambda^2} }{3 \pi^2 N_f}U $};

\draw[thick, blue] ([shift=(180:2cm)]0,-15) arc (180:135:2cm);
\draw[thick,blue] ([shift=(135:2cm)]0,-15) arc (135:45:2cm);
\draw[thick, blue] ([shift=(45:2cm)]0,-15) arc (45:0:2cm);
\draw[thick,blue] ([shift=(0:2cm)]0,-15)   arc (0:-180:2cm);
\draw[thick, red, dashed] (0,-12+0.1) [partial ellipse=-40-14:-140+12:3cm and 2.5cm];
\draw[dashed,red,thick] (-3.5,-15) to (-2,-15);
\draw[dashed,red,thick] (2,-15) to (3.5,-15);  
\node at (8,-15) {$=2\times\frac{2\log {x^2\Lambda^2}}{3  \pi^2 N_f } U$};

\draw[thick,blue] ([shift=(180:2cm)]0,-20)   arc (180:0:2cm);
\draw[thick,blue] ([shift=(0:2cm)]0,-20)   arc (0:-180:2cm);
\draw[dashed,red,thick] (0,-20) ([shift=(90:2cm)]0,-20)  to   ([shift=(-90:2cm)]0,-20);
\draw[dashed,red,thick] (-3.5,-20) to (-2,-20);
\draw[dashed,red,thick] (2,-20) to (3.5,-20);  

\node at (8,-20+0.1) {$= \frac{12\log{x^2\Lambda^2}}{3 \pi^2 N_f } U$};

\draw[thick,blue]  ([shift=(0:2cm)]0,-25)  arc (0:180:2cm);
\draw[thick,blue]  ([shift=(-180:2cm)]0,-25) arc (-180:0:2cm);
\draw[photon, red] (0,-25) ([shift=(0:2cm)]0,-25)  to   ([shift=(-210:2cm)]6,-25);
\draw[photon, red] (0,-25) ([shift=(0:2cm)]0,-25)  to   ([shift=(-150:2cm)]6,-25);
\draw[thick,blue ] ([shift=(180:2cm)]6,-25) arc (180:0:2cm);
\draw[thick,blue] ([shift=(0:2cm)]6,-25) arc (0:-180:2cm);
\draw[dashed,red,thick] (-3.5,-25) to (-2,-25);
\draw[dashed,red,thick] (8,-25) to (9.5,-25);  

\node at (14.2,-25) {$= 4\times \frac{48\log{x^2 \Lambda^2}}{3  \pi^2 N_f } U$};

\draw[thick,blue]  ([shift=(0:2cm)]0,-30)  arc (0:180:2cm);
\draw[thick,blue]  ([shift=(-180:2cm)]0,-30) arc (-180:0:2cm); 
\draw[photon, red] ([shift=(40-14:2cm)]0,-30)  to   ([shift=(140+14:2cm)]6,-30);
\draw[photon, red] ([shift=(-40+14:2cm)]0,-30)  to   ([shift=(-140-14:2cm)]6,-30);
\draw[thick,blue ] ([shift=(180:2cm)]6,-30) arc (180:0:2cm);
\draw[thick,blue] ([shift=(0:2cm)]6,-30) arc (0:-180:2cm);
\node at (11,-30) {$=0$};
\draw[dashed,red,thick] (-3.5,-30) to (-2,-30);
\draw[dashed,red,thick] (8,-30) to (9.5,-30);  

\draw[thick,blue]  ([shift=(0:2cm)]0,-35)  arc (0:180:2cm);
\draw[thick,blue]  ([shift=(-180:2cm)]0,-35) arc (-180:0:2cm); 
\draw[dashed,thick, red] ([shift=(40-14:2cm)]0,-35)  to   ([shift=(140+14:2cm)]6,-35);
\draw[dashed,thick, red] ([shift=(-40+14:2cm)]0,-35)  to   ([shift=(-140-14:2cm)]6,-35);
\draw[thick,blue ] ([shift=(180:2cm)]6,-35) arc (180:0:2cm);
\draw[thick,blue] ([shift=(0:2cm)]6,-35) arc (0:-180:2cm);
\node at (11,-35) {$=0$};
\draw[dashed,red,thick] (-3.5,-35) to (-2,-35);
\draw[dashed,red,thick] (8,-35) to (9.5,-35);

\begin{scope}[shift={(22,0)}]

\draw[blue,thick] (0,0) circle (2cm);
\draw[fill]  (-2,0) circle (5pt);
\draw[fill]  (2,0) circle (5pt);
\node at (7.5,0) {$=W$};

\draw[blue,thick] ([shift=(180:2cm)]0,-5)   arc (180:0:2cm);
\draw[blue,thick] ([shift=(0:2cm)]0,-5)   arc (0:-180:2cm);
\draw[fill]  (-2,-5) circle (5pt);
\draw[fill]  (2,-5) circle (5pt);
\draw[photon,red] (-1.66,-3.9) to (1.66,-3.9);
\node at (9-0.2,-5+0.4) {$ = 2\times \frac{ 4\big( 5+3 \xi \big) \log{x^2 \Lambda^2}}{ 3  \pi^2 N_f} W$};

\draw[blue,thick] ([shift=(180:2cm)]0,-10)   arc (180:0:2cm);
\draw[blue,thick] ([shift=(0:2cm)]0,-10)   arc (0:-180:2cm);
\draw[photon, red] (0,-10) ([shift=(90:2cm)]0,-10)  to   ([shift=(-90:2cm)]0,-10);
\draw[fill]  (-2,-10) circle (5pt);
\draw[fill]  (2,-10) circle (5pt);
 \node at (9-0.2,-10+0.4) {$=\frac{24\big( 1-\xi \big) \log {x^2 \Lambda^2} }{ 3 \pi^2 N_f}W $}; 

\draw[thick, blue] ([shift=(180:2cm)]0,-15) arc (180:135:2cm);
\draw[thick,blue] ([shift=(135:2cm)]0,-15) arc (135:45:2cm);
\draw[thick, blue] ([shift=(45:2cm)]0,-15) arc (45:0:2cm);
\draw[thick,blue] ([shift=(0:2cm)]0,-15)   arc (0:-180:2cm);
\draw[thick, red, dashed] (0,-12+0.1) [partial ellipse=-40-14:-140+12:3cm and 2.5cm];
\draw[fill]  (-2,-15) circle (5pt);
\draw[fill]  (2,-15) circle (5pt);
\node at (8,-15+0.4) {$=2 \times \frac{-2\log {x^2 \Lambda^2}}{3   \pi^2 N_f } W$};

\draw[thick,blue] ([shift=(180:2cm)]0,-20)   arc (180:0:2cm);
\draw[thick,blue] ([shift=(0:2cm)]0,-20)   arc (0:-180:2cm);
\draw[dashed,red,thick] (0,-20) ([shift=(90:2cm)]0,-20)  to   ([shift=(-90:2cm)]0,-20);
\draw[fill]  (-2,-20) circle (5pt);
\draw[fill]  (2,-20) circle (5pt);
\node at (8,-20+0.4) {$= -\frac{12\log {x^2 \Lambda^2}}{ 3\pi^2 N_f} W$};
\end{scope}

  \end{tikzpicture}
\caption{ (bQED$_+$)  Results for the individual Feynman graphs appearing in the 2-point correlation functions  $\langle \sigma_+ (x) \sigma_+(0) \rangle $ (left column)\protect\footnotemark \  and $\langle|\Phi|^2_{adj}(x) |\Phi|^2_{adj}(0) \rangle$ (right column).}  \label{F6}
  \end{figure}

The $N_f^2$ gauge invariant operators $\Phi^*_i \Phi^j$ transform in the adjoint plus singlet of $SU(N_f)$. The singlet operator is set to zero by the equation of motion of the Hubbard-Stratonovich field $\sigma_+$.\footnote{As a simple check of this statement, one can explicitly check that the two point function $\langle |\Phi|^2_{sing} (x) |\Phi|^2_{sing} (0) \rangle$ is zero at order $O(N_f^0)$. 
\begin{align} \nonumber
  \begin{tikzpicture} [scale=0.5]  
  \draw[blue,thick] (0,0) circle (1cm);
    \draw[fill]  (-1,0) circle (3pt);
        \draw[fill]  (1,0) circle (3pt);
  \node at (2,0) {$+$};
  \draw[blue,thick]  (4,0) circle (1cm);
  \draw[dashed,red,thick] (5,0) to (6,0);
    \draw[blue,thick]  (7,0) circle (1cm);
      \draw[fill]  (3,0) circle (3pt);
        \draw[fill]  (8,0) circle (3pt);
    \node at (9.4,0) {$= \ \ 0$};
\end{tikzpicture}
  \end{align}
The $1$-loop diagram cancels with a $2$-loop diagram given by two bubbles connected by a $\sigma_+$ propagator (normalizing the singlet operator as $\frac{1}{\sqrt{N_f}}\sum_k \Phi^*_k \Phi^k $, both such graphs are of order $1$ at large $N_f$). We thank to Silviu Pufu for clarifying this point.} So we consider the scaling dimension of $\sigma_+$ instead. The scaling dimensions  of these operators can be read from the table (\ref{F6}) 
\begin{align} \label{hh4}
& \Delta[|\Phi|^2_{adj}] = 1 - \frac{48}{3 \pi^2 N_f} +O(1/{N_f^2})  \\
 &\Delta[\sigma_+] = 2 - \frac{144}{3 \pi^2 N_f} + O(1/{N_f^2}) \ .\label{hh5}
\end{align}
The formulas above have already been discussed in \cite{Halperin:1973jh, Hikami:1979ih, Vas:1983, Kaul:2008xw}. The scaling dimensions (\ref{hh4}, \ref{hh5}) are related to traditional critical exponents by
\begin{align}
&\eta_N = 2\Delta[|\Phi|^2_{adj}]-1 = 1- \frac{96}{3 \pi^2 N_f} +O(1/{N_f^2}) \\
&\nu^{-1}=3-\Delta[\sigma_+]=1+\frac{144}{3 \pi^2 N_f} + O(1/{N_f^2}) \ .
\end{align} 
Where $\eta_N$ is the anomalous scaling dimension of the adjoint scalar-bilinear operator also known as Néel field \cite{Kaul:2008xw}. 

  \footnotetext{The last two graphs have no logarithmic divergences.}
  
  Next we discuss scaling dimension of the quartic adjoint-2 operator (with Dynkin labels $[2,0,...,,0,2]$). This operator is in the spectrum and has scaling dimension $2$ at order $O(N_f^0)$. The graphs that contribute to its 2-point correlation function at the order $O(1/N_f)$ are the ones in (\ref{F4}) (already discussed in the context of bQED), supplemented with the list of graphs in the table (\ref{F7}). There are 4 graphs with HS field connecting a leg with itself and 6 kite graphs with the HS field joining two different legs. Summing all the contributions we can extract anomalous dimension of the adjoint-2 quartic operator
  \begin{align}
  \Delta[|\Phi|^4_{adj-2}]=2-\frac{48}{3\pi^2 N_f}+O(1/{N_f^2}) \ .
  \end{align}
  
 \begin{figure}[]
  \centering
\begin{tikzpicture}[scale=0.45]

\draw[black] (3,3) to (17,3);
\draw[black] (3,3) to (3,-8);
\draw[black] (3,-8) to (17,-8);
\draw[black] (17,3) to (17,-8);

\draw[scalar] ([shift=(0:2)]5.5,0) arc (0:180:2 and 1);
\draw[scalar] ([shift=(-180:2)]5.5,0) arc (-180:0:2 and 1);
\draw[scalar] ([shift=(0:2)]5.5,0) arc (0:180:2 );
\draw[scalar] ([shift=(-180:2)]5.5,0) arc (-180:0:2 );
\draw[thick, red, dashed] (5.5,4-0.2) [partial ellipse=-40-27:-140+12+15:3.4cm and 2.5cm];
\draw[fill]  (3.5,0) circle (4pt);
\draw[fill]  (7.5,0) circle (4pt);
\node at (12,0) {$=4 \times  \frac{ -2 \log {x^2\Lambda^2} }{ 3 \pi^2 N_f } W^2$};

\begin{scope}[shift={(-11,-5)}]
\draw[scalar] ([shift=(0:2)]16.5,0) arc (0:180:2 and 1);
\draw[scalar] ([shift=(-180:2)]16.5,0) arc (-180:0:2 and 1);
\draw[scalar] ([shift=(0:2)]16.5,0) arc (0:180:2 );
\draw[scalar] ([shift=(-180:2)]16.5,0) arc (-180:0:2 );
\draw[dashed,red , thick] ([shift=(90:2)]16.5,0) to ([shift=(-90:2 and 1)]16.5,0);
\draw[fill]  (14.5,0) circle (4pt);
\draw[fill]  (18.5,0) circle (4pt);
\node at (23,0) {$=6 \times  \frac{ -12 \log {x^2\Lambda^2} }{ 3  \pi^2 N_f } W^2$};
\end{scope}

  \end{tikzpicture}
\caption{(bQED$_+$) quartic adjoint-2 renormalization. (contribution from graphs with HS prop.)} \label{F7}
  \end{figure}


\subsection{ep-bQED ("easy-plane" QED)}\label{sec:epscQED}
 The ep-bQED fixed point is reached with the quartic potential $V \sim (\sum _{i=1}^{N_f/2}|\Phi_i|^2)^2+(\sum_{i=1}^{N_f/2}|\tilde\Phi_i|^2)^2 $ and by tuning the mass terms to zero.  The large $N_f$  effective action is described by complex scalar fields $(\Phi_i, \tilde\Phi_i)$ minimally coupled to the effective photon and interacting with two HS fields via cubic interactions 
\be \label{epsQED} \CL _{eff}=  \sum_{i=1}^{N_f/2} (|D \Phi_i|^2+|D \tilde{\Phi}_i|^2)  + \sigma  \sum_{i=1}^{N_f/2} |\Phi_i|^2  +\tilde{\sigma}  \sum_{i=1}^{N_f/2} |\tilde{\Phi}_i|^2  \,.  \ee
The effective propagator for the photon is the same as in (\ref{photonprop}), and the effective propagators for the HS fields are
\begin{align}
 \label{effpropbqedpm} \langle \sigma(x) \sigma(0) \rangle = \langle \sigt(x) \sigt(0) \rangle  = \frac{8  }{\pi^2 (N_f/2) |x|^4} \ .
 \end{align}
The photon "sees" all the $N_f$ flavors, $\sigma$ and $\sigt$ only "see" $N_f/2$ flavors. In the Feynman graphs, we use red dashed (double dashed) line for $\sigma$ ($\tilde\sigma$) and blue (double blue) line for $\Phi$ ($\tilde\Phi$).

The global symmetry of the effective action (\ref{epsQED}) is 
\begin{align} \label{symepQED}
\Big( \frac{SU(N_f/2) \times SU(N_f/2)  \times U(1)_b \rtimes\bZ_2^e }{\bZ_{N_f}}\times U(1)_{top} \Big ) \rtimes \bZ_2^C \ .
\end{align}
The $U(1)_b$ acts: $\{ \Phi_i \rightarrow e^{i\alpha}\Phi_i, \tilde\Phi_i \rightarrow e^{-i\alpha}\tilde\Phi_i \}$. The $\bZ_2^e $ acts: $\{\Phi_i \leftrightarrow \tilde\Phi_i,\ \sigma \leftrightarrow \tilde\sigma \}$. There is also parity invariance. 

\subsection*{Scaling dimension of low-lying scalar operators}
The $N_f^2$ quadratic gauge invariant operators transform as two adjoints, two singlets and two bifundamentals of $SU(N_f/2)^2$. More precisely, in the reducible representation
\begin{align}
 \label{decomp1} ( \bf{adj}, \bf{1} ) \oplus ( \bf{1}, \bf{adj}) \oplus (\bf{\bar F}, F ) \oplus ( F , \bar{F}) \oplus 2 \cdot (\bf{1},\bf{1}) \ ,
 \end{align}
where by $\bf{F}$ we denoted the fundamental representation of $SU(N_f/2)$.

 Feynman graphs that contribute to the anomalous scaling dimension of $|\Phi|^2_{adj}$  are the graphs in the right column of the table (\ref{F6}). One has to keep in mind that the photon "sees" all the flavors, while each sigma field "sees" only half of them, therefore the contribution of graphs that involve an HS propagator is twice as big as the contribution of the corresponding graphs in bQED$_+$. For the adjoint operator $|\Phit|^2_{adj}$ one has the same set of graphs, but the blue lines are exchanged by blue double lines  and red dashed lines are exchanged by red dashed double lines. On the other hand, the scaling dimension of the bifundamental operators $(\Phi_i \Phit_j^*, \Phi^*_i\Phit_j)$ is corrected by graphs similar to those in the right column of (\ref{F6}), except  that the last graph is absent. The two scalar-bilinear singlets are set to zero by the equations of motion of the HS fields $\sigma$ and $\tilde\sigma$. 
 
 The 2-point correlation function  $\langle \sigma(x) \sigma(0) \rangle$ is corrected by the left column graphs (\ref{F6}), and similar graphs stand for $\langle \tilde\sigma(x) \tilde\sigma(0) \rangle$. It is preferable to denote by $U$ the effective propagator of the HS field $\sigma$ (\ref{effpropbqedpm}), then the graphs involving single photon contribute as in bQED$_+$, the graphs involving HS propagator contribute 2 times the corresponding graphs in bQED$_+$, the graph involving two photons contributes twice less than the same graph in bQED$_+$. So we conclude that the $O(1/N_f)$ corrected propagator for the HS $\sigma$ field is 
 \begin{align} \label{h8}
  \langle \sigma(x) \sigma(0) \rangle =\Big (1+\frac{64 \log {x^2 \Lambda^2}}{3\pi^2 N_f} \Big) \Big( \frac{8 }{\pi^2 (N_f/2) |x|^4} \Big) \ .
 \end{align} 
  It turns out that already at order $O(1/N_f)$ there is a mixing between HS fields $\sigma$ and $\tilde\sigma$ (\ref{F8}).  
   \begin{align} \label{h9}
  \langle \sigma(x) \tilde\sigma(0) \rangle = \frac{96 \log {x^2 \Lambda^2}}{3\pi^2 N_f} \Big( \frac{8 }{\pi^2 (N_f/2) |x|^4} \Big) \ .
 \end{align} 
  The HS fields $\sigma_\pm$ defined in (\ref{h4}) are the eigenvectors of the mixing matrix. Using (\ref{h8}, \ref{h9}) one readily extracts anomalous dimensions of those fields. 

   \begin{figure}[]
  \centering
\begin{tikzpicture}[scale=0.4]
\draw[blue,thick] (0,0) (0:2cm) arc (0:180:2cm);
\draw[blue,thick] (0,0)  (-180:2cm) arc (-180:0:2cm);
\draw[red, thick,dashed] (-4,0) to (-2,0);
\draw[photon, red] (0,0) (0:2cm) to   ([shift=(-210:2cm)]6,0);
\draw[photon, red] (0,0) (0:2cm) to   ([shift=(-150:2cm)]6,0);
\draw[thick, blue, double ] ([shift=(180:2cm)]6,0) arc (180:0:2cm);
\draw[thick, blue, double] ([shift=(0:2cm)]6,0) arc (0:-180:2cm);
\draw[red,double,thick,dashed] (8,0) to (10,0);
\node at (17,0.2) {$= 4\times \frac{24 \log{x^2 \Lambda^2}}{3\pi^2 N_f} \Big( \frac{8 }{ \pi^2 (N_f/2) |x|^4} \Big)$};
  \end{tikzpicture}
\caption{Diagram responsible for mixing $\langle \sigma(x) \tilde\sigma(0) \rangle$ .} \label{F8}
  \end{figure}
  
  The $N_f^4$ quartic gauge invariant operators transform as reducible representation of $SU(N_f/2)^2$ with the following decomposition into irreducible blocks
  \begin{align}  \nonumber 
 & (\bf{adj_2},\bf{1})\oplus (\bf{1}, \bf{adj_2}) \oplus (\overline{sym},sym) \oplus (sym,\overline{sym})\\
 \nonumber      \oplus & (\bf{adj},adj) \oplus (R,\bar{F})\oplus (\bar{R},F) \oplus (\bar{F},R) \oplus (F,\bar{R}) \\
        \oplus & \bf{2} \cdot (\bf{adj},1) \oplus 2 \cdot  (1,adj) \oplus 2\cdot(F,\bar{F}) \oplus 2\cdot (\bar{F},F) \oplus 3\cdot (1,1) \ . \label{decomp2}
  \end{align}
Where by $\bf{R}$ we denote the representation of $SU(N_f/2)$ with Dynkin labels $[2,0,...,0,1]$.   All the irreducible blocks in the third row of (\ref{decomp2}) contain a singlet quadratic factor and therefore they are out of spectrum. In the table (\ref{F15}) we collected all the relevant graphs for extracting the anomalous scaling dimension of the operator $({\bf \overline{ sym},sym}) = \Phi_i^* \Phi_j^* \tilde\Phi_k \tilde\Phi_l$. One can make similar tables for the other quartic operators which are in the spectrum. 

 \begin{figure}[]
  \centering
\begin{tikzpicture}[scale=0.45]

\draw[black] (-3,3) to (29,3);
\draw[black] (-3,3) to (-3,-13);
\draw[black] (-3,-13) to (29,-13);
\draw[black] (29,3) to (29,-13);

\draw[scalar] (0,0) (0:2) arc (0:180:2 and 1);
\draw[scalar, double] (0,0) (-180:2) arc (-180:0:2 and 1);
\draw[scalar] (0,0) (0:2) arc (0:180:2);
\draw[scalar,double] (0,0) (-180:2) arc (-180:0:2);
\draw[fill,thick] (-2,0) circle (4pt);
\draw[fill,thick] (2,0) circle (4pt);
\node at (4.5,0) {$=W^2$};

\begin{scope}[shift={(15,0)}]
\draw[scalar] (0,0) (0:2) arc (0:180:2 and 1);
\draw[scalar, double] (0,0) (-180:2) arc (-180:0:2 and 1);
\draw[scalar] (0,0) (0:2) arc (0:180:2);
\draw[scalar,double] (0,0) (-180:2) arc (-180:0:2);
\draw[photon,red] (50:2) to (130:2);
\draw[fill,thick] (-2,0) circle (4pt);
\draw[fill,thick] (2,0) circle (4pt);
\node at (7.5,0.5) {$=4\times \frac{4 \Big(5+3 \xi \Big) \log{x^2\Lambda^2} }{3 \pi^2 N_f}W^2$};
\end{scope}

\begin{scope}[shift={(0,-5)}]
\draw[scalar] (0,0) (0:2) arc (0:180:2 and 1);
\draw[scalar, double] (0,0) (-180:2) arc (-180:0:2 and 1);
\draw[scalar] (0,0) (0:2) arc (0:180:2);
\draw[scalar,double] (0,0) (-180:2) arc (-180:0:2);
\draw[photon1,red] (90:2) to (90:2 and 1);
\draw[fill,thick] (-2,0) circle (4pt);
\draw[fill,thick] (2,0) circle (4pt);
\node at (7.5,0) {$=2\times \frac{-24(1-\xi) \log{x^2\Lambda^2}}{3 \pi^2 N_f}W^2$};
\end{scope}

\begin{scope}[shift={(15,-5)}]
\draw[scalar] (0,0) (0:2) arc (0:180:2 and 1);
\draw[scalar, double] (0,0) (-180:2) arc (-180:0:2 and 1);
\draw[scalar] (0,0) (0:2) arc (0:180:2);
\draw[scalar,double] (0,0) (-180:2) arc (-180:0:2);
\draw[photon,red] (90:2 and 1) to (-90:2 and 1);
\draw[fill,thick] (-2,0) circle (4pt);
\draw[fill,thick] (2,0) circle (4pt);
\node at (7.5,0) {$=4\times \frac{24(1-\xi) \log{x^2\Lambda^2}}{3 \pi^2 N_f}W^2$};
\end{scope}

\begin{scope}[shift={(0,-10)}]
\draw[scalar] (0,0) (0:2) arc (0:180:2 and 1);
\draw[scalar, double] (0,0) (-180:2) arc (-180:0:2 and 1);
\draw[scalar] (0,0) (0:2) arc (0:180:2);
\draw[scalar,double] (0,0) (-180:2) arc (-180:0:2);
\draw[red,dashed,thick] (0,4-0.2) [partial ellipse=-40-27:-140+12+15:3.4cm and 2.5cm];
\draw[fill,thick] (-2,0) circle (4pt);
\draw[fill,thick] (2,0) circle (4pt);
\node at (7.5,0) {$=4\times \frac{-4 \log{x^2\Lambda^2}}{3\pi^2 N_f}W^2$};
\end{scope}

\begin{scope}[shift={(15,-10)}]
\draw[scalar] (0,0) (0:2) arc (0:180:2 and 1);
\draw[scalar, double] (0,0) (-180:2) arc (-180:0:2 and 1);
\draw[scalar] (0,0) (0:2) arc (0:180:2);
\draw[scalar,double] (0,0) (-180:2) arc (-180:0:2);
\draw[red, dashed,thick] (90:2) to (90:2 and 1);
\draw[fill,thick] (-2,0) circle (4pt);
\draw[fill,thick] (2,0) circle (4pt);
\node at (7,0) {$=2\times \frac{-24\log{x^2\Lambda^2}}{3\pi^2 N_f}W^2$};
\end{scope}

  \end{tikzpicture}
\caption{Renormalization of the $\bf{(\overline{sym},sym)}$ quartic operator. }  \label{F15}
  \end{figure}

  The scaling dimensions are as follows
\begin{align} 
\label{epBQEDsc}
 &\Delta[|\Phi|^2_{adj}] =  \Delta[|\tilde\Phi|^2_{adj}] = 1 - \frac{32}{3 \pi^2 N_f} + O(1/{N_f^2})    \\
 &\Delta[\Phi_i \Phit_j^*] = \Delta[\Phi^*_i\Phit_j] = 1 - \frac{56}{3 \pi^2 N_f} + O(1/{N_f^2}) \\
 &\Delta[\sigma_-] = 2 + \frac{32}{3 \pi^2 N_f} + O(1/{N_f^2})     \\
 &\Delta[\sigma_+] = 2 - \frac{160}{3 \pi^2 N_f}+O(1/{N_f^2})  \\
 &\Delta[|\Phi|^4_{adj-2}]=\Delta[|\tilde\Phi|^4_{adj-2}]=2+\frac{32}{3\pi^2 N_f}+O(1/{N_f^2}) \\
  & \Delta[\tilde\Phi_l^*(\Phi_i \Phi_j \Phi_k^*)_{[2,0,...,0,1]} ]= \Delta[\tilde\Phi_l( \Phi_k \Phi^*_i \Phi^*_j)_{[1,0,...,0,2]} ]=2-\frac{40}{3\pi^2 N_f} +O(1/{N_f^2}) \\
   & \Delta[\Phi_l^*(\tilde\Phi_i \tilde\Phi_j \tilde\Phi_k^*)_{[2,0,...,0,1]} ]= \Delta[\Phi_l( \tilde\Phi_k \tilde\Phi^*_i \tilde\Phi^*_j)_{[1,0,...,0,2]} ]=2-\frac{40}{3\pi^2 N_f} +O(1/{N_f^2}) \\
 &\Delta[|\Phi|^2_{adj} |\tilde\Phi|^2_{adj} ]=2-\frac{64}{3\pi^2 N_f} +O(1/{N_f^2}) \\
  &\Delta[\Phi^*_{i} \Phi^*_{j} \tilde\Phi_{k} \tilde\Phi_{l}]=\Delta[\Phi_{i} \Phi_{j} \tilde\Phi^*_{k} \tilde\Phi^*_{l}]=2-\frac{64}{3 \pi^2 N_f}+O(1/{N_f^2}) \ .
\end{align}

\subsection{bQED$_-$}

The bQED$_-$ is reached with quartic deformation $V \sim (\sum|\Phi_i|^2-|\tilde\Phi_i|^2)^2 $ and by tuning mass terms to zero. Large $N_f$ effective action is described by complex scalar fields $(\Phi_i,\tilde\Phi_i)$ minimally coupled to the effective photon and interacting with single HS field via cubic interaction. 
\begin{align}
 \label{bsigma-} \CL_{eff} =  \sum_{i=1}^{N_f/2} (|D \Phi_i|^2+|D \tilde{\Phi}_i|^2)  + \sigma_- ( \sum_{i=1}^{N_f/2} |\Phi_i|^2  -   \sum_{i=1}^{N_f/2} |\tilde{\Phi}_i|^2 ) \,.  
 \end{align}
Effective propagator for the photon is the same as in (\ref{photonprop}), and the effective propagators for the HS field $\sigma_-$ is as follows
\begin{align}
 \langle \sigma_-(x) \sigma_-(0) \rangle  = \frac{8  }{\pi^2 N_f |x|^4} 
 \end{align}
In the Feynman graphs we will use red dashed line for the effective propagator of $\sigma_-$. The global symmetry of the bQED$_-$ action is the same as for ep-bQED (\ref{symepQED}).
\subsection*{Scaling dimension of low-lying scalar operators}
 The $N_f^2$ quadratic gauge invariant operators are  decomposed into irreducible representations of $SU(N_f/2)^2$ as in (\ref{decomp1}). Feynman graphs that contribute to the scaling dimensions of the operators $\{ |\Phi|^2_{adj}, |\tilde\Phi|^2_{adj}\}$ are those in the left column of (\ref{F6}). The same graphs can be used to calculate scaling dimension of the bifundamental operators $\{\Phi_i \tilde\Phi_j^*, \Phi_i^* \tilde\Phi_j\}$, however the graph with HS field $\sigma_-$ joining propagators $\Phi$ and $\tilde\Phi$ contributes with the opposite sign compared to the similar graph in the bQED$_+$. This is because the cubic vertices with HS field coupled to the scalar flavors $(\Phi, \tilde\Phi)$ have different signs as it follows from the effective action (\ref{bsigma-}). Notice that EOM of the HS field $\sigma_-$ sets to zero the operator $( \sum|\Phi_i|^2  -  \sum |\tilde{\Phi}_i|^2 )$. Therefore that operator is out of the spectrum, while the plus combination is in the spectrum and has a dimension 1 at leading order.  
 
\begin{figure}[H]
  \centering
\begin{tikzpicture}[scale=0.35]

\draw[black] (-3.5,2.5) to (29,2.5) ; 
\draw[black] (-3.5,2.5) to (-3.5,-38);
\draw[black] (-3.5,-38) to (29,-38);
\draw[black] (29,2.5) to (29,-38);

\draw[blue,thick,double, postaction={decorate}, decoration={markings, mark=at position .43 with {\arrow[scale=0.75,blue]{triangle 45}}}] (0,0) (0:2) arc (0:180:2 and 1);
\draw[scalar] (0,0) (-180:2) arc (-180:0:2 and 1);
\draw[scalarr] (0,0) (0:2) arc (0:180:2);
\draw[scalarr] (0,0) (-180:2) arc (-180:0:2);
\draw[thick,fill] (-2,0) circle (4pt);
\draw[thick,fill] (2,0) circle (4pt);

\begin{scope}[shift={(6,0)}]
\draw[blue,thick,double,  postaction={decorate}, decoration={markings, mark=at position .43 with {\arrow[scale=0.75,blue]{triangle 45}}}] (0,0) (0:2) arc (0:180:2 and 1);
\draw[blue,thick,double, postaction={decorate}, decoration={markings, mark=at position .43 with {\arrow[scale=0.75,blue]{triangle 45}}}] (0,0) (-180:2) arc (-180:0:2 and 1);
\draw[scalarr] (0,0) (0:2) arc (0:180:2);
\draw[blue,thick,double, postaction={decorate}, decoration={markings, mark=at position .6 with {\arrowreversed[scale=0.75,blue]{triangle 45}}}] (0,0) (-180:2) arc (-180:0:2);
\draw[thick,fill] (-2,0) circle (4pt);
\draw[thick,fill] (2,0) circle (4pt);
\end{scope}

\node at (15,0) {$=W^2$};

\begin{scope}[shift={(0,-5)}]
\draw[blue,thick,double,  postaction={decorate}, decoration={markings, mark=at position .43 with {\arrow[scale=0.75,blue]{triangle 45}}}] (0,0) (0:2) arc (0:180:2 and 1);
\draw[scalar] (0,0) (-180:2) arc (-180:0:2 and 1);
\draw[scalarr] (0,0) (0:2) arc (0:180:2);
\draw[scalarr] (0,0) (-180:2) arc (-180:0:2);
\draw[photon1,red] ([shift=(45:2)]0,0) to ([shift=(135:2)]0,0);
\draw[thick,fill] (-2,0) circle (4pt);
\draw[thick,fill] (2,0) circle (4pt);
\end{scope}

\begin{scope}[shift={(5,-5)}]
\draw[blue,thick,double,  postaction={decorate}, decoration={markings, mark=at position .43 with {\arrow[scale=0.75,blue]{triangle 45}}}] (0,0) (0:2) arc (0:180:2 and 1);
\draw[blue,thick,double, postaction={decorate}, decoration={markings, mark=at position .43 with {\arrow[scale=0.75,blue]{triangle 45}}}] (0,0) (-180:2) arc (-180:0:2 and 1);
\draw[scalarr] (0,0) (0:2) arc (0:180:2);
\draw[blue,thick,double, postaction={decorate}, decoration={markings, mark=at position .6 with {\arrowreversed[scale=0.75,blue]{triangle 45}}}] (0,0) (-180:2) arc (-180:0:2);
\draw[photon1,red] ([shift=(45:2)]0,0) to ([shift=(135:2)]0,0);
\draw[thick,fill] (-2,0) circle (4pt);
\draw[thick,fill] (2,0) circle (4pt);
\end{scope}

\node at (14,-5) {$=4\times  \frac{ 4\big( 5+3 \xi \big) \log {x^2\Lambda^2} }{ 3  \pi^2 N_f } W^2$};

\begin{scope}[shift={(0,-10)}]
\draw[blue,thick,double,  postaction={decorate}, decoration={markings, mark=at position .43 with {\arrow[scale=0.75,blue]{triangle 45}}}] (0,0) (0:2) arc (0:180:2 and 1);
\draw[scalar] (0,0) (-180:2) arc (-180:0:2 and 1);
\draw[scalarr] (0,0) (0:2) arc (0:180:2);
\draw[scalarr] (0,0) (-180:2) arc (-180:0:2);
\draw[photon2,red ] ([shift=(90:2)]0,0) to ([shift=(90:2 and 1)]0,0);
\draw[thick,fill] (-2,0) circle (4pt);
\draw[thick,fill] (2,0) circle (4pt);
\end{scope}

\begin{scope}[shift={(5,-10)}]
\draw[blue,thick,double,  postaction={decorate}, decoration={markings, mark=at position .43 with {\arrow[scale=0.75,blue]{triangle 45}}}] (0,0) (0:2) arc (0:180:2 and 1);
\draw[blue,thick,double, postaction={decorate}, decoration={markings, mark=at position .43 with {\arrow[scale=0.75,blue]{triangle 45}}}] (0,0) (-180:2) arc (-180:0:2 and 1);
\draw[scalarr] (0,0) (0:2) arc (0:180:2);
\draw[blue,thick,double, postaction={decorate}, decoration={markings, mark=at position .6 with {\arrowreversed[scale=0.75,blue]{triangle 45}}}] (0,0) (-180:2) arc (-180:0:2);
\draw[photon2,red ] ([shift=(90:2)]0,0) to ([shift=(90:2 and 1)]0,0);
\draw[thick,fill] (-2,0) circle (4pt);
\draw[thick,fill] (2,0) circle (4pt);
\end{scope}

\node at (14,-10) {$=4\times  \frac{ 24\big( 1-\xi \big) \log {x^2\Lambda^2} }{ 3  \pi^2 N_f } W^2$};

\begin{scope}[shift={(0,-15)}]
\draw[blue,thick,double,  postaction={decorate}, decoration={markings, mark=at position .43 with {\arrow[scale=0.75,blue]{triangle 45}}}] (0,0) (0:2) arc (0:180:2 and 1);
\draw[scalar] (0,0) (-180:2) arc (-180:0:2 and 1);
\draw[scalarr] (0,0) (0:2) arc (0:180:2);
\draw[scalarr] (0,0) (-180:2) arc (-180:0:2);
\draw[photon,red ] ([shift=(90:2)]0,0) to ([shift=(-90:2 and 1)]0,0);
\draw[thick,fill] (-2,0) circle (4pt);
\draw[thick,fill] (2,0) circle (4pt);
\end{scope}

\begin{scope}[shift={(5,-15)}]
\draw[blue,thick,double,  postaction={decorate}, decoration={markings, mark=at position .43 with {\arrow[scale=0.75,blue]{triangle 45}}}] (0,0) (0:2) arc (0:180:2 and 1);
\draw[blue,thick,double, postaction={decorate}, decoration={markings, mark=at position .43 with {\arrow[scale=0.75,blue]{triangle 45}}}] (0,0) (-180:2) arc (-180:0:2 and 1);
\draw[scalarr] (0,0) (0:2) arc (0:180:2);
\draw[blue,thick,double, postaction={decorate}, decoration={markings, mark=at position .6 with {\arrowreversed[scale=0.75,blue]{triangle 45}}}] (0,0) (-180:2) arc (-180:0:2);
\draw[photon,red ] ([shift=(90:2)]0,0) to ([shift=(-90:2 and 1)]0,0);
\draw[thick,fill] (-2,0) circle (4pt);
\draw[thick,fill] (2,0) circle (4pt);
\end{scope}

\node at (14,-15) {$=2 \times  \frac{- 24 \big( 1-\xi \big) \log {x^2\Lambda^2} }{ 3 \pi^2 N_f } W^2$};

\begin{scope}[shift={(0,-20)}]
\draw[blue,thick,double,  postaction={decorate}, decoration={markings, mark=at position .43 with {\arrow[scale=0.75,blue]{triangle 45}}}] (0,0) (0:2) arc (0:180:2 and 1);
\draw[blue,thick] (0,0) (-180+100:2 and 1) arc (-180+100:0:2 and 1);
\draw[blue,thick] (-180+100: 2 and 1 ) to (-180+110: 2);
\draw[scalarr] (0,0) (0:2) arc (0:180:2);
\draw[blue,thick] (0,0) (-180+110:2) arc (-180+110:0:2);
\draw[blue,thick] ([shift=(-15:1.4)]-1.9,-1.5) arc (-15:100:1.4);
\draw[blue,thick] (0,0) (-130+23:2) arc (-130+23:-180:2);
\draw[thick,fill] (-2,0) circle (4pt);
\draw[thick,fill] (2,0) circle (4pt);
\draw[photon1,red] (-130+23:2) to (-180+110:2);
\draw[photon1,red] (-130+23:2) to ([shift=(-180+100:2 and 1)]0,0);
\end{scope}

\begin{scope}[shift={(5,-20)}]
\draw[blue,thick,double,  postaction={decorate}, decoration={markings, mark=at position .43 with {\arrow[scale=0.75,blue]{triangle 45}}}] (0,0) (0:2) arc (0:180:2 and 1);
\draw[blue,thick,double] (0,0) (-180+100:2 and 1) arc (-180+100:0:2 and 1);
\draw[blue,thick,double] (-180+100: 2 and 1 ) to (-180+110: 2);
\draw[scalarr] (0,0) (0:2) arc (0:180:2);
\draw[blue,thick,double] (0,0) (-180+110:2) arc (-180+110:0:2);
\draw[blue,thick,double] ([shift=(-15:1.4)]-1.9,-1.5) arc (-15:100:1.4);
\draw[blue,thick,double] (0,0) (-130+23:2) arc (-130+23:-180:2);
\draw[thick,fill] (-2,0) circle (4pt);
\draw[thick,fill] (2,0) circle (4pt);
\draw[photon1,red] (-130+23:2) to (-180+110:2);
\draw[photon1,red] (-130+23:2) to ([shift=(-180+100:2 and 1)]0,0);
\end{scope}

\begin{scope}[shift={(10,-20)}]
\draw[blue,thick,double,  postaction={decorate}, decoration={markings, mark=at position .43 with {\arrow[scale=0.75,blue]{triangle 45}}}] (0,0) (0:2) arc (0:180:2 and 1);
\draw[blue,thick,double] (0,0) (-180+100:2 and 1) arc (-180+100:0:2 and 1);
\draw[blue,thick,double] (-180+100: 2 and 1 ) to (-180+110: 2);
\draw[scalarr] (0,0) (0:2) arc (0:180:2);
\draw[blue,thick,double] (0,0) (-180+110:2) arc (-180+110:0:2);
\draw[blue,thick] ([shift=(-15:1.4)]-1.9,-1.5) arc (-15:100:1.4);
\draw[blue,thick] (0,0) (-130+23:2) arc (-130+23:-180:2);
\draw[thick,fill] (-2,0) circle (4pt);
\draw[thick,fill] (2,0) circle (4pt);
\draw[photon1,red] (-130+23:2) to (-180+110:2);
\draw[photon1,red] (-130+23:2) to ([shift=(-180+100:2 and 1)]0,0);
\end{scope}

\begin{scope}[shift={(15,-20)}]
\draw[blue,thick,double,  postaction={decorate}, decoration={markings, mark=at position .43 with {\arrow[scale=0.75,blue]{triangle 45}}}] (0,0) (0:2) arc (0:180:2 and 1);
\draw[blue,thick] (0,0) (-180+100:2 and 1) arc (-180+100:0:2 and 1);
\draw[blue,thick] (-180+100: 2 and 1 ) to (-180+110: 2);
\draw[scalarr] (0,0) (0:2) arc (0:180:2);
\draw[blue,thick] (0,0) (-180+110:2) arc (-180+110:0:2);
\draw[blue,thick,double] ([shift=(-15:1.4)]-1.9,-1.5) arc (-15:100:1.4);
\draw[blue,thick,double] (0,0) (-130+23:2) arc (-130+23:-180:2);
\draw[thick,fill] (-2,0) circle (4pt);
\draw[thick,fill] (2,0) circle (4pt);
\draw[photon1,red] (-130+23:2) to (-180+110:2);
\draw[photon1,red] (-130+23:2) to ([shift=(-180+100:2 and 1)]0,0);
\end{scope}

\node at (23,-20) {$=4 \times  \frac{ -48 \log {x^2\Lambda^2} }{3  \pi^2 N_f } W^2$};

\begin{scope}[shift={(-5,-25)}]
\draw[blue,thick,double,  postaction={decorate}, decoration={markings, mark=at position .43 with {\arrow[scale=0.75,blue]{triangle 45}}}] ([shift=(0:2)]5.5,0) arc (0:180:2 and 1);
\draw[scalar] ([shift=(-180:2)]5.5,0) arc (-180:0:2 and 1);
\draw[scalarr] ([shift=(0:2)]5.5,0) arc (0:180:2 );
\draw[scalarr] ([shift=(-180:2)]5.5,0) arc (-180:0:2 );
\draw[thick, red, dashed] (5.5,4-0.2) [partial ellipse=-40-27:-140+12+15:3.4cm and 2.5cm];
\draw[fill]  (3.5,0) circle (4pt);
\draw[fill]  (7.5,0) circle (4pt);
\end{scope}

\begin{scope}[shift={(1,-25)}]
\draw[blue,thick,double,  postaction={decorate}, decoration={markings, mark=at position .43 with {\arrow[scale=0.75,blue]{triangle 45}}}] ([shift=(0:2)]5.5,0) arc (0:180:2 and 1);
\draw[blue,thick,double, postaction={decorate}, decoration={markings, mark=at position .43 with {\arrow[scale=0.75,blue]{triangle 45}}}] ([shift=(-180:2)]5.5,0) arc (-180:0:2 and 1);
\draw[scalarr] ([shift=(0:2)]5.5,0) arc (0:180:2 );
\draw[blue,thick,double, postaction={decorate}, decoration={markings, mark=at position .6 with {\arrowreversed[scale=0.75,blue]{triangle 45}}}] ([shift=(-180:2)]5.5,0) arc (-180:0:2 );
\draw[thick, red, dashed] (5.5,4-0.2) [partial ellipse=-40-27:-140+12+15:3.4cm and 2.5cm];
\draw[fill]  (3.5,0) circle (4pt);
\draw[fill]  (7.5,0) circle (4pt);
\end{scope}

\node at (14,-25) {$=4 \times  \frac{ -2 \log {x^2\Lambda^2} }{ 3 \pi^2 N_f } W^2$};

\draw[black] ([shift= (-90:1 and 4.5)]8.5,-32.5) arc (-90:90:1 and 4.5);
\draw[black] ([shift= (90:1 and 4.5)]-1.5,-32.5) arc (90:180:1 and 4.5);
\draw[black] ([shift= (-90:1 and 4.5)]-1.5,-32.5) arc (-90:-180:1 and 4.5);

\begin{scope}[shift={(-5,-30)}]
\draw[blue,thick,double,  postaction={decorate}, decoration={markings, mark=at position .43 with {\arrow[scale=0.75,blue]{triangle 45}}}] ([shift=(0:2)]5.5,0) arc (0:180:2 and 1);
\draw[scalar] ([shift=(-180:2)]5.5,0) arc (-180:0:2 and 1);
\draw[scalarr] ([shift=(0:2)]5.5,0) arc (0:180:2 );
\draw[scalarr] ([shift=(-180:2)]5.5,0) arc (-180:0:2 );
\draw[thick,red,dashed] ([shift=(90:2)]5.5,0) to ([shift=(-90:2 and 1 )]5.5,0);
\draw[fill]  (3.5,0) circle (4pt);
\draw[fill]  (7.5,0) circle (4pt);
\end{scope}

\begin{scope}[shift={(1,-30)}]
\draw[blue,thick,double,  postaction={decorate}, decoration={markings, mark=at position .43 with {\arrow[scale=0.75,blue]{triangle 45}}}] ([shift=(0:2)]5.5,0) arc (0:180:2 and 1);
\draw[blue,thick,double, postaction={decorate}, decoration={markings, mark=at position .43 with {\arrow[scale=0.75,blue]{triangle 45}}}] ([shift=(-180:2)]5.5,0) arc (-180:0:2 and 1);
\draw[scalarr] ([shift=(0:2)]5.5,0) arc (0:180:2 );
\draw[blue,thick,double, postaction={decorate}, decoration={markings, mark=at position .6 with {\arrowreversed[scale=0.75,blue]{triangle 45}}}] ([shift=(-180:2)]5.5,0) arc (-180:0:2 );
\draw[thick,red,dashed] ([shift=(-90:2)]5.5,0) to ([shift=(90:2 and 1 )]5.5,0);
\draw[fill]  (3.5,0) circle (4pt);
\draw[fill]  (7.5,0) circle (4pt);
\end{scope}

\begin{scope}[shift={(-5,-35)}]
\draw[blue,thick,double,  postaction={decorate}, decoration={markings, mark=at position .43 with {\arrow[scale=0.75,blue]{triangle 45}}}] ([shift=(0:2)]5.5,0) arc (0:180:2 and 1);
\draw[scalar] ([shift=(-180:2)]5.5,0) arc (-180:0:2 and 1);
\draw[scalarr] ([shift=(0:2)]5.5,0) arc (0:180:2 );
\draw[scalarr] ([shift=(-180:2)]5.5,0) arc (-180:0:2 );
\draw[thick,red,dashed] ([shift=(90:2)]5.5,0) to ([shift=(90:2 and 1 )]5.5,0);
\draw[fill]  (3.5,0) circle (4pt);
\draw[fill]  (7.5,0) circle (4pt);
\end{scope}

\begin{scope}[shift={(1,-35)}]
\draw[blue,thick,double,  postaction={decorate}, decoration={markings, mark=at position .43 with {\arrow[scale=0.75,blue]{triangle 45}}}] ([shift=(0:2)]5.5,0) arc (0:180:2 and 1);
\draw[blue,thick,double, postaction={decorate}, decoration={markings, mark=at position .43 with {\arrow[scale=0.75,blue]{triangle 45}}}] ([shift=(-180:2)]5.5,0) arc (-180:0:2 and 1);
\draw[scalarr] ([shift=(0:2)]5.5,0) arc (0:180:2 );
\draw[blue,thick,double, postaction={decorate}, decoration={markings, mark=at position .6 with {\arrowreversed[scale=0.75,blue]{triangle 45}}}] ([shift=(-180:2)]5.5,0) arc (-180:0:2 );
\draw[thick,red,dashed] ([shift=(90:2)]5.5,0) to ([shift=(90:2 and 1 )]5.5,0);
\draw[fill]  (3.5,0) circle (4pt);
\draw[fill]  (7.5,0) circle (4pt);
\end{scope}

\node at (12,-32.5) {$=0$};

  \end{tikzpicture}
\caption{(bQED$_-$) quartic bifundamental operator renormalization. Each graph has the flavor index $k$ running in its bottom loop.} \label{F16}
  \end{figure}
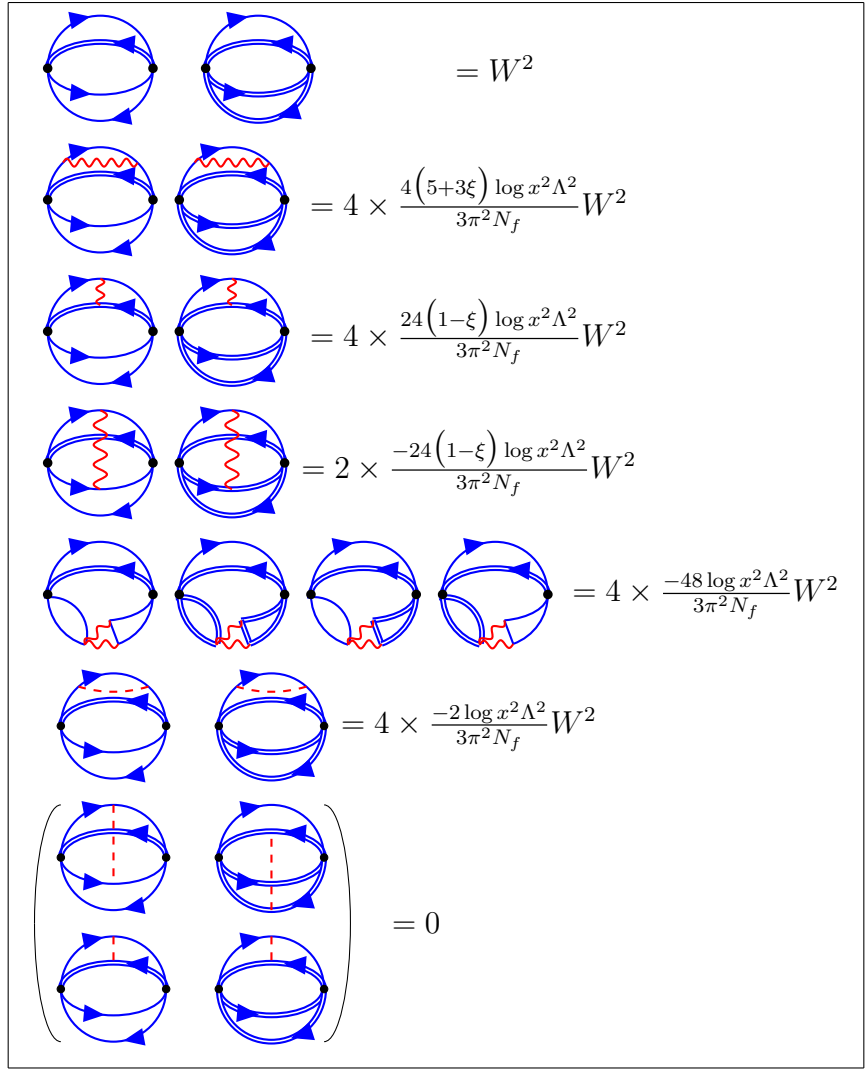

  The $N_f^4$ quartic gauge invariant operators are decomposed into irreducible representations of $SU(N_f/2)^2$ as in (\ref{decomp2}). Notice that in the last line of (\ref{decomp2}) not all the operators are excluded from the spectrum:  the quartic operators which are a product of a quadratic operator $( \sum|\Phi_i|^2  + \sum |\tilde{\Phi}_i|^2 )$ and a quadratic adjoint or bifundamental operator, as well as the quartic operator $( \sum|\Phi_i|^2  + \sum |\tilde{\Phi}_i|^2 )^2$ are in the spectrum and have scaling dimension equal to 2 in the leading order. In the table (\ref{F16}) we collected all the graphs that contribute to the anomalous scaling dimension of the quartic bifundamental operator $\frac{\big ( \sum|\Phi_k|^2  +  \sum |\tilde{\Phi}_k|^2 \big ) \Phi_i \tilde\Phi_j^* }{\sqrt{N_f}}$. Similar computations can be done for the other operators. Below we give the list of operators and their scaling dimensions.

\begin{align}
 &\Delta[|\Phi|^2_{adj}] =  \Delta[|\tilde\Phi|^2_{adj}] = 1 - \frac{48}{3 \pi^2 N_f} + O(1/{N_f^2})    \\
 &\Delta[\Phi_i \Phit_j^*] = \Delta[\Phi^*_i\Phit_j]= 1 - \frac{72}{3 \pi^2 N_f} + O(1/{N_f^2})  \\
  & \Delta \big [ \big ( \sum|\Phi_i|^2  +  \sum |\tilde{\Phi}_i|^2 \big )  \big ]= 1+ \frac{144}{3 \pi^2 N_f} + O(1/{N_f^2})  \\
  &\Delta[|\Phi|^4_{adj-2}]=\Delta[|\tilde\Phi|^4_{adj-2}]=2-\frac{48}{3\pi^2 N_f}+O(1/{N_f^2}) \\
    & \Delta[\tilde\Phi_l^*(\Phi_i \Phi_j \Phi_k^*)_{[2,0,...,0,1]} ]= \Delta[\tilde\Phi_l( \Phi_k \Phi^*_i \Phi^*_j)_{[1,0,...,0,2]} ]=2-\frac{120}{3\pi^2 N_f} +O(1/{N_f^2}) \\
   & \Delta[\Phi_l^*(\tilde\Phi_i \tilde\Phi_j \tilde\Phi_k^*)_{[2,0,...,0,1]} ]= \Delta[\Phi_l( \tilde\Phi_k \tilde\Phi^*_i \tilde\Phi^*_j)_{[1,0,...,0,2]} ]=2-\frac{120}{3\pi^2 N_f} +O(1/{N_f^2}) \\
       &\Delta[|\Phi|^2_{adj} |\tilde\Phi|^2_{adj} ]=2-\frac{144}{3\pi^2 N_f}+O(1/{N_f^2})  \\
  &\Delta[\Phi^*_{i} \Phi^*_{j} \tilde\Phi_{k} \tilde\Phi_{l}]=\Delta[\Phi_{i} \Phi_{j} \tilde\Phi^*_{k} \tilde\Phi^*_{l}]=2-\frac{144}{3 \pi^2 N_f}+O(1/{N_f^2})  \\
  & \Delta \big [ |\Phi|^2_{adj}\!  \sum \! \big ( |\Phi_i|^2  \!+\!  |\tilde{\Phi}_i|^2 \big )  \big ]\!\!=\! \Delta \big [ |\tilde\Phi|^2_{adj}  \! \sum \!  \big ( |\Phi_i|^2  +  |\tilde{\Phi}_i|^2 \big )  \big ]\!\!=\! 2\!+\! \frac{96}{3\pi^2 N_f}\!+\!O(1/{N_f^2})  \\
    & \Delta \big [  \Phi_i \tilde\Phi_j^* \! \sum \! \big (  |\Phi_i|^2 \! +\!  |\tilde{\Phi}_i|^2 \big )  \big ] \!\!=\!\Delta \big [ \Phi_i^* \tilde\Phi_j  \! \sum \! \big (  |\Phi_i|^2  \!+\!  |\tilde{\Phi}_i|^2 \big )\big ]  \!\!=\! 2\!+\! \frac{72}{3\pi^2 N_f} \!+\!O(1/{N_f^2}) \\
   &\Delta[\sigma_-] = 2 + \frac{48}{3 \pi^2 N_f} + O(1/{N_f^2})  \\
   & \Delta \big [ \big ( \sum|\Phi_i|^2  +  \sum |\tilde{\Phi}_i|^2 \big ) ^2 \big ] = 2+\frac{288}{3\pi^2 N_f}+O(1/{N_f^2}) 
 \end{align}


\section{Four fermionic QED fixed points in the large $N_f$ limit}\label{sec:fQEDs}
In this section we study fermionic QED, with large $N_f$ complex fermionic flavors, imposing at least $U(N_f/2)^2$ global symmetry. There are four different fixed points, two fixed points have $U(N_f)$ global symmetry, two fixed points have $U(N_f/2)^2$ global symmetry.

Let us consider the following UV (Euclidean) lagrangian 
\begin{align} \nonumber 
\CL = & \frac{1}{4 e^2} F_{\mu \nu} F^{\mu \nu } + \sum_{i=1}^{N_f/2} ( \bar\Psi_i \slashed D \Psi^i + \bar{\tilde\Psi}_i \slashed D \tilde\Psi^i ) + \rho_+\sum_{i=1}^{N_f/2} ( \bar\Psi_i \Psi^i + \bar{\tilde\Psi}_i  \tilde\Psi^i ) \\
&+\rho_- \sum_{i=1}^{N_f/2} ( \bar\Psi_i \Psi^i - \bar{\tilde\Psi}_i  \tilde\Psi^i ) + m_+^2 \rho_+^2+ m_-^2 \rho_-^2+ ... \label{h5} \ .
\end{align}
Where the dots stand for kinetic terms and quartic interactions of the Hubbard-Stratonovich fields $\rho_+$ and $\rho_-$. We choose the gamma matrices to be equal to the Pauli matrices: $\gamma^0 = \sigma_2 , \ \gamma^1=  \sigma_1, \ \gamma^2 =  \sigma _3$, and $\slashed D=\gamma^\mu D_\mu$. The two-component Dirac fermions $(\Psi_i, \tilde\Psi_i) \ (i=1,...,N_f/2)$ carry charge $+1$ under the gauge group. We also implicitly assume a conformal gauge fixing term.  These type of theories (\ref{h5}) have been studied using various techniques, e.g. epsilon expansion and functional RG flow \cite{Pisarski:1984dj,DiPietro:2015taa, DiPietro:2017kcd, Braun:2014wja, Janssen:2017eeu, Ihrig:2018ojl, Zerf:2018csr}.

As explained in \cite{Benvenuti:2018cwd}, depending on the form of the Yukawa interactions, there are four different fixed points:
\begin{itemize}
\item   fQED, both HS fields are massive and the Yukawa interactions are absent, 
\item QED-GN$_+$, the Yukawa interaction involving HS field $\rho_+$ is turned on and the HS field $\rho_-$ is massive, 
\item QED-NJL (gauged Nambu–Jona-Lasinio), both HS fields $\rho_\pm$ are massless, and both Yukawa interactions are turned on,  
\item QED-GN$_-$, the Yukawa interaction involving HS field $\rho_-$ is turned on and the HS field $\rho_+$ is massive.
\end{itemize}

\subsection{fQED}\label{sec:fQED}
\begin{figure}[]
  \centering
  \begin{tikzpicture} [scale=0.8]  
  
 \draw[black] (-0.5,1) to (8,1);
 \draw[black] (-0.5,1) to (-0.5,-4);
  \draw[black]  (-0.5,-4) to (8,-4);
  \draw[black] (8,1) to (8,-4);
        
  \draw[photon,red] (0,0) to (2,0); 
  \node at (5,0) {$=\langle A_\mu (x) A_\nu(0) \rangle_{\text{eff}}$};
  
   \draw[thick, draw=mygreen, postaction={decorate}, decoration={markings, mark=at position .55 with {\arrow[mygreen]{triangle 45}}}] (0,-1.5+0.2) to (2,-1.5+0.2); 
   \node at (4-0.2,-1.5+0.2) {$=\frac{\slashed x}{4 \pi |x|^3}$};
  \begin{scope}[shift={(0,2)}]
   \draw[photon,red] (0,-3-1.8) to (1,-3-1.8);
   \draw[thick, draw=mygreen, postaction={decorate}, decoration={markings, mark=at position .4 with {\arrowreversed[mygreen]{triangle 45}}}] (1,-3-1.8) to (2,-2.5-1.8);
    \draw[thick, draw=mygreen, postaction={decorate}, decoration={markings, mark=at position .65 with {\arrow[mygreen]{triangle 45}}}] (1,-3-1.8) to (2,-3.5-1.8);
\node at (4-0.2,-3-1.8) {$=-i \gamma_\mu$};
\end{scope}
 \end{tikzpicture}
 
\caption{(fQED) Feynman rules for propagators and vertices.}  \label{F9}
  \end{figure}
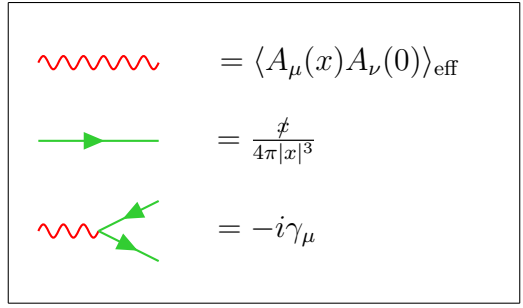

In fQED both HS fields massive and therefore decoupled from the IR spectrum. The large $N_f$ effective action for the fQED fixed point is described by $N_f$ copies of Dirac fermions $\Psi_i$ (we collected all the fermions $(\Psi, \tilde\Psi)$ into a single field and denoted it by $\Psi$) minimally coupled to the effective photon
\begin{align}
 \CL_{eff} = \sum_{i=1}^{N_f} \bar{\Psi^i}    \slashed{D} \Psi_i  \ . 
    \end{align}
The effective photon propagator is obtained summing geometric series of bubble diagrams  (\ref{F1}), where all the scalar (blue) loops are exchanged with fermion (green) loops. 
\begin{align}
 \label{photonpropf} \langle A_\mu (x) A_\nu(0) \rangle_{\text{eff}}= \frac{8 }{ \pi^2 N_f |x|^2} \Big ( (1-\xi)\delta_{\mu \nu}+2 \xi \frac{x_\mu x_\nu}{x^2} \Big)\ .
 \end{align}
We notice that the effective photon propagator in the fQED coincides with the effective photon propagator in the bosonic QED's. This is because the fermion and boson loops that appear in the geometric sums are equal to each other. 
 Feynman rules for the vertices and for the propagators are given in the table (\ref{F9}).
 
The faithful global symmetry is 
\begin{align}
 \label{h6} \frac{SU(N_f) \times \uot}{\bZ_{N_f}} \rtimes \bZ_2^{\CC} \ .
 \end{align}
Where $\bZ_{N_f}$ is generated by $\left( e^{2 \pi i /N_f}\mathbb{I} , -1 \right) \in SU(N_f) \times \uot$ (this fact comes from a careful treatment of the monopoles operators, which are dressed with fermionic zero-modes). $\bZ_2^{\CC}$ is the charge-conjugation symmetry. There is also symmetry under parity.\footnote{It is crucial to have even number of Dirac fermions, otherwise the theory suffers from parity anomaly.}

\subsection*{Scaling dimension of low-lying scalar operators \footnote{Check \cite{Xu:2008, Chester:2016ref} for scaling dimensions of quartic operators, which at infinite $N_f$ have $\Delta=4$.}}


 \begin{figure}[]
  \centering
\begin{tikzpicture}[scale=0.4]

\draw[black] (-4,3) to (18,3);
\draw[black] (-4,3) to (-4,-30+2+5+5);
\draw[black] (18,3) to (18,-30+2+5+5);
\draw[black] (-4,-30+2+5+5) to (18,-30+2+5+5);

\draw[thick, mygreen] (0,0) circle (2cm);
\draw[thick,fill] (-2,0) circle (4pt);
\draw[thick,fill] (2,0) circle (4pt);

\node at (8,0) {$=\frac{1}{8 \pi^2 |x|^4} \equiv \tilde W$};

\node at (-3,0)  {$A$};
\node at (-3,-5)  {$B$};
\node at (-3,-10)  {$C$};
\node at (-3,-15)  {$D$};

\draw[mygreen,thick] ([shift=(180:2cm)]0,-5)   arc (180:0:2cm);
\draw[mygreen,thick] ([shift=(0:2cm)]0,-5)   arc (0:-180:2cm);
\draw[photon,red] (-1.66,-3.9) to (1.66,-3.9);
\draw[thick,fill] (-2,-5) circle (4pt);
\draw[thick,fill] (2,-5) circle (4pt);
\node at (10,-5+0.4) {$ =2\times \frac{ -4\big(1-3\xi \big) \log x^2\Lambda^2}{ 3 \pi^2 N_f } \tilde W$};

\draw[mygreen,thick] ([shift=(180:2cm)]0,-10)   arc (180:0:2cm);
\draw[mygreen,thick] ([shift=(0:2cm)]0,-10)   arc (0:-180:2cm);
\draw[photon, red] (0,-10) ([shift=(90:2cm)]0,-10)  to   ([shift=(-90:2cm)]0,-10);
\draw[thick,fill] (-2,-10) circle (4pt);
\draw[thick,fill] (2,-10) circle (4pt);
 \node at (9,-10+0.4) {$=\frac{24\big( 3-\xi \big) \log x^2 \Lambda^2 }{3 \pi^2 N_f} \tilde W $}; 

\draw[thick,mygreen]  ([shift=(0:2cm)]0,-15)  arc (0:180:2cm);
\draw[thick,mygreen]  ([shift=(-180:2cm)]0,-15) arc (-180:0:2cm); 
\draw[photon, red] ([shift=(40-14:2cm)]0,-15)  to   ([shift=(140+14:2cm)]6,-15);
\draw[photon, red] ([shift=(-40+14:2cm)]0,-15)  to   ([shift=(-140-14:2cm)]6,-15);
\draw[thick,mygreen ] ([shift=(180:2cm)]6,-15) arc (180:0:2cm);
\draw[thick,mygreen] ([shift=(0:2cm)]6,-15) arc (0:-180:2cm);
\draw[thick,fill] (-2,-15) circle (4pt);
\draw[thick,fill] (8,-15) circle (4pt);
\node at (13,-15) {$=2\times  \frac{-96 \log x^2 \Lambda^2  }{3 \pi^2 N_f} \tilde W$};

  \end{tikzpicture}
\caption{ (fQED) Results for individual Feynman graphs appearing in the 2-point correlation function for the fermion-bilinear operators.  } \label{F10}
  \end{figure}
  
The $N_f^2$ gauge invariant operators $\bar{\Psi}_i \Psi^j$ transform in the adjoint plus singlet of $SU(N_f)$. Their scaling dimensions at large $N_f$ can be read from (\ref{F10}) and have been already discussed in \cite{Rantner:2002zz, Hermele:2005dkq,Erratum}
\begin{align}
&\Delta[|\Psi|^2_{adj}] = 2 - \frac{64}{3 \pi^2 N_f} + O(1/{N_f^2})  \\
 &\Delta[|\Psi|^2_{sing}] = 2 + \frac{128}{3 \pi^2 N_f} + O(1/{N_f^2}) \ .
  \end{align}


\subsection{QED-GN$_+$}\label{sec:QEDGN}
In the QED-GN$_+$ fixed point the action is (\ref{h5}), with Yukawa interaction involving HS field $\rho_+$, while the HS field $\rho_-$ is massive and is decoupled from the IR spectrum. The large $N_f$ effective action is described by $N_f$ copies of Dirac fermions $\Psi_i$ (we collected all the fermions $(\Psi, \tilde\Psi)$ into a single field and denoted it by $\Psi$) minimally coupled to the effective photon and interacting with HS field $\rho_+$ via Yukawa interaction 
\begin{align}
 \CL_{eff} = \sum_{i=1}^{N_f} \bar{\Psi_i}    \slashed{D} \Psi^i + \rho_+ \sum_{i=1}^{N_f} \bar{\Psi_i} \Psi^i \ .
  \end{align}
The effective propagator for the photon is the same as in the fQED (\ref{photonpropf}). The effective propagator for the HS field $\rho_+$ follows from summing geometric series of bubble diagrams as in (\ref{F5})  with all the scalar (blue) loops exchanged with fermion(green) loops 
\begin{align}
\langle \rho_+(x) \rho_+(0) \rangle_{eff} = \frac{4}{\pi^2 N_f |x|^2} \ .
\end{align}
In the Feynman graphs we use a red dashed line in order to represent the $\rho_+$ propagator. The global symmetry is the same as in fQED (\ref{h6}). There is also parity symmetry ($\rho_+$ is parity-odd). 

\subsection*{Scaling dimension of low-lying scalar operators}
\begin{figure}[H]
  \centering
\begin{tikzpicture}[scale=0.38]

\draw[black] (-4,3) to (36,3);
\draw[black] (-4,3) to (-4,-30+2-5);
\draw[black] (36,3) to (36,-30+2-5);
\draw[black] (17,3) to (17,-30+2-5);
\draw[black] (-4,-30+2-5) to (36,-30+2-5);

\draw[dashed,red,thick] (-3.5,0) to (3.5,0);
\node at (9.7,0.1) {$=\langle \rho_+(x) \rho_+(0) \rangle_{eff} \equiv \tilde U$};

\draw[mygreen,thick] ([shift=(180:2cm)]0,-5)   arc (180:0:2cm);
\draw[mygreen,thick] ([shift=(0:2cm)]0,-5)   arc (0:-180:2cm);
\draw[dashed,red,thick] (-3.5,-5) to (-2,-5);
\draw[dashed,red,thick] (2,-5) to (3.5,-5);
\draw[photon,red] (-1.66,-3.9) to (1.66,-3.9);
\node at (10,-5) {$ = 2\times \frac{ 4\big( 1-3\xi \big) \log x^2 \Lambda^2}{3  \pi^2 N_f} \tilde U$};

\draw[mygreen,thick] ([shift=(180:2cm)]0,-10)   arc (180:0:2cm);
\draw[mygreen,thick] ([shift=(0:2cm)]0,-10)   arc (0:-180:2cm);
\draw[photon, red] (0,-10) ([shift=(90:2cm)]0,-10)  to   ([shift=(-90:2cm)]0,-10);
\draw[dashed,red,thick] (-3.5,-10) to (-2,-10);
\draw[dashed,red,thick] (2,-10) to (3.5,-10);  
\node at (9.5,-10) {$=-\frac{24\big( 3-\xi \big) \log x^2 \Lambda^2 }{3 \pi^2 N_f} \tilde U $}; 

\draw[thick, mygreen] ([shift=(180:2cm)]0,-15) arc (180:135:2cm);
\draw[thick,mygreen] ([shift=(135:2cm)]0,-15) arc (135:45:2cm);
\draw[thick, mygreen] ([shift=(45:2cm)]0,-15) arc (45:0:2cm);
\draw[thick,mygreen] ([shift=(0:2cm)]0,-15)   arc (0:-180:2cm);
\draw[thick, red, dashed] (0,-12+0.1) [partial ellipse=-40-14:-140+12:3cm and 2.5cm];
\draw[dashed,red,thick] (-3.5,-15) to (-2,-15);
\draw[dashed,red,thick] (2,-15) to (3.5,-15);  
\node at (9.5,-15) {$=2 \times \frac{2\log x^2 \Lambda^2}{3  \pi^2 N_f } \tilde U$};

\draw[thick,mygreen] ([shift=(180:2cm)]0,-20)   arc (180:0:2cm);
\draw[thick,mygreen] ([shift=(0:2cm)]0,-20)   arc (0:-180:2cm);
\draw[dashed,red,thick] (0,-20) ([shift=(90:2cm)]0,-20)  to   ([shift=(-90:2cm)]0,-20);
\draw[dashed,red,thick] (-3.5,-20) to (-2,-20);
\draw[dashed,red,thick] (2,-20) to (3.5,-20);  
\node at (8,-20) {$= \frac{12\log x^2\Lambda^2}{3  \pi^2 N_f} \tilde U$};

\draw[mygreen,thick]  ([shift=(0:2cm)]0,-25)  arc (0:180:2cm);
\draw[mygreen,thick]  ([shift=(-180:2cm)]0,-25) arc (-180:0:2cm);
\draw[photon, red] (0,-25) ([shift=(31:2cm)]0,-25)  to   ([shift=(149:2cm)]5,-25);
\draw[photon, red] (0,-25) ([shift=(-31:2cm)]0,-25)  to   ([shift=(-149:2cm)]5,-25);
\draw[mygreen,thick ] ([shift=(180:2cm)]5,-25) arc (180:0:2cm);
\draw[mygreen,thick] ([shift=(0:2cm)]5,-25) arc (0:-180:2cm);
\draw[dashed,red,thick] (-3.5,-25) to (-2,-25);
\draw[dashed,red,thick] (7,-25) to (8.5,-25);  
\node at (12.7,-25) {$= 2\times  \frac{96\log x^2 \Lambda^2}{ 3 \pi^2 N_f } \tilde U$};

\draw[mygreen,thick]  ([shift=(0:2cm)]0,-30)  arc (0:180:2cm);
\draw[mygreen,thick]  ([shift=(-180:2cm)]0,-30) arc (-180:0:2cm);
\draw[dashed,thick, red] (0,-25) ([shift=(31:2cm)]0,-30)  to   ([shift=(149:2cm)]5,-30);
\draw[dashed,thick, red] (0,-25) ([shift=(-31:2cm)]0,-30)  to   ([shift=(-149:2cm)]5,-30);
\draw[mygreen,thick ] ([shift=(180:2cm)]5,-30) arc (180:0:2cm);
\draw[mygreen,thick] ([shift=(0:2cm)]5,-30) arc (0:-180:2cm);
\draw[dashed,red,thick] (-3.5,-30) to (-2,-30);
\draw[dashed,red,thick] (7,-30) to (8.5,-30);  
\node at (11,-30) {$= 0$};

\begin{scope}[shift={(21,0)}]
\draw[mygreen,thick] (0,0) circle (2cm);

\node at (7.5,0) {$= \tilde W$};
\draw[fill]  (-2,0) circle (5pt);
\draw[fill]  (2,0) circle (5pt);

\draw[mygreen,thick] ([shift=(180:2cm)]0,-5)   arc (180:0:2cm);
\draw[mygreen,thick] ([shift=(0:2cm)]0,-5)   arc (0:-180:2cm);
\draw[fill]  (-2,-5) circle (5pt);
\draw[fill]  (2,-5) circle (5pt);
\draw[photon,red] (-1.66,-3.9) to (1.66,-3.9);
\node at (9-0.2,-5+0.4) {$  = 2\times  \frac{ -4\big( 1-3\xi \big) \log x^2 \Lambda^2}{ 3 \pi^2 N_f } \tilde W$};

\draw[mygreen,thick] ([shift=(180:2cm)]0,-10)   arc (180:0:2cm);
\draw[mygreen,thick] ([shift=(0:2cm)]0,-10)   arc (0:-180:2cm);
\draw[photon, red] (0,-10) ([shift=(90:2cm)]0,-10)  to   ([shift=(-90:2cm)]0,-10);
\draw[fill]  (-2,-10) circle (5pt);
\draw[fill]  (2,-10) circle (5pt);
 \node at (8,-10+0.4) {$=\frac{24\big( 3-\xi \big) \log x^2 \Lambda^2 }{ 3 \pi^2 N_f} \tilde W $}; 

\draw[thick, mygreen] ([shift=(180:2cm)]0,-15) arc (180:135:2cm);
\draw[thick,mygreen] ([shift=(135:2cm)]0,-15) arc (135:45:2cm);
\draw[thick, mygreen] ([shift=(45:2cm)]0,-15) arc (45:0:2cm);
\draw[thick,mygreen] ([shift=(0:2cm)]0,-15)   arc (0:-180:2cm);
\draw[thick, red, dashed] (0,-12+0.1) [partial ellipse=-40-14:-140+12:3cm and 2.5cm];
\draw[fill]  (-2,-15) circle (5pt);
\draw[fill]  (2,-15) circle (5pt);
\node at (8,-15+0.4) {$=2\times \frac{-2\log x^2 \Lambda^2}{3   \pi^2 N_f } \tilde W$};

\draw[thick,mygreen] ([shift=(180:2cm)]0,-20)   arc (180:0:2cm);
\draw[thick,mygreen] ([shift=(0:2cm)]0,-20)   arc (0:-180:2cm);
\draw[dashed,red,thick] (0,-20) ([shift=(90:2cm)]0,-20)  to   ([shift=(-90:2cm)]0,-20);
\draw[fill]  (-2,-20) circle (5pt);
\draw[fill]  (2,-20) circle (5pt);
\node at (8,-20+0.4) {$= -\frac{12\log x^2\Lambda^2}{ 3\pi^2 N_f} \tilde W$};

\end{scope}

\end{tikzpicture}
\caption{ (QED-GN$_+$) Results for individual Feynman graphs appearing in the 2-point correlation functions  $\langle \rho_+ (x) \rho_+(0) \rangle $(left column)\protect\footnotemark  \ and $\langle |\Psi|^2_{adj}(x) |\Psi|^2_{adj}(0) \rangle$ (right column).}  \label{F11}  \end{figure}
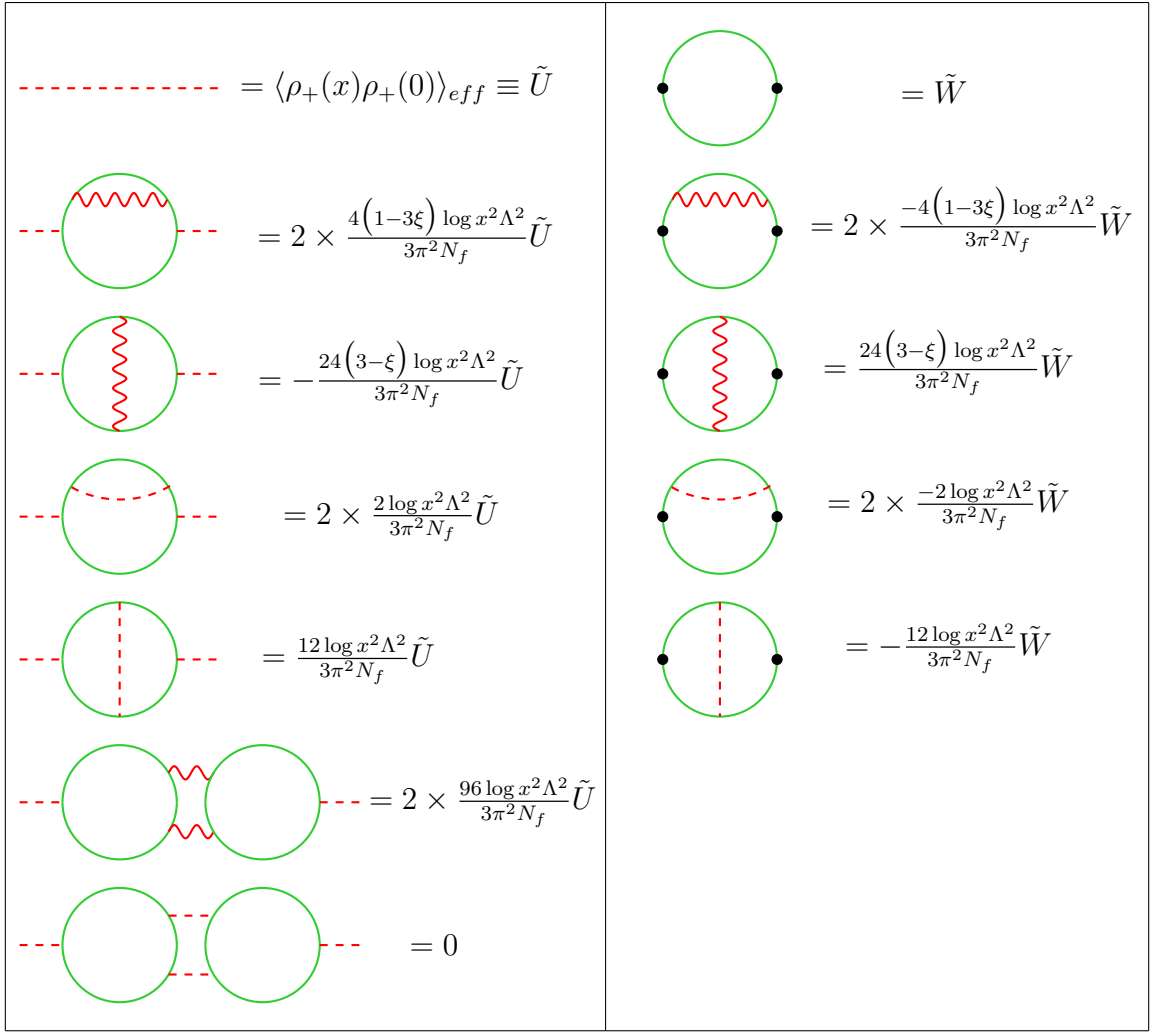

\begin{figure}[H]
 \centering
 \begin{tikzpicture}[scale=0.38]
\draw[black] (-3,3) to (17,3);
 \draw[black] (-3,3) to (-3,-27);
 \draw[black] (-3,-27) to (17,-27);
 \draw[black] (17,3) to (17,-27);
\node at (10,0) {$=2 \times \big(\frac{4}{\pi^2 N_f  |x|^2} \big)^2 \equiv Z$};
 
\draw[dashed, red, thick] (0,0) circle (2cm);
\draw[fill, red] (-2,0) circle (5pt);
\draw[fill, red] (2,0) circle (5pt);

\draw[dashed, red, thick] (0,-6) circle (2cm);
\draw[fill,black] (0,-4.1) ellipse (0.8cm and 0.5cm);
\draw[fill, red] (-2,-6) circle (5pt);
\draw[fill, red] (2,-6) circle (5pt);

\node at (10,-5.5) {$= 2\times \frac{144 \log x^2\Lambda^2 }{3  \pi^2 N_f} Z$};

\begin{scope}[shift={(-2,-12)}]
 \draw[dashed ,thick, red] (0,0) to (2,2);
  \draw[dashed ,thick, red] (0,0) to (2,-2);
  \draw[fill, red] (0,0) circle (5pt);
   \draw[mygreen, thick] (2,2) to (2,-2);
   \draw[mygreen, thick] (2,-2) to (5,-2);
 \draw[mygreen, thick] (2,2) to (5,2);
  \draw[mygreen, thick] (5,2) to (5,-2);
   \draw[dashed ,thick, red] (5,2) to (7,0);
 \draw[dashed ,thick, red] (5,-2) to (7,0);
   \draw[fill, red] (7,0) circle (5pt);
\end{scope}

\node at (10,-11.5) {$= 4\times \frac{-6\log x^2 \Lambda^2 }{3 \pi^2 N_f} Z$};

 \node at (10,-17.5) {$= 2\times \frac{-12\log x^2 \Lambda^2 }{ 3\pi^2 N_f} Z$};
 
  \begin{scope}[shift={(-2,-18)}]

  \draw[dashed ,thick, red] (0,0) to (2,2);
  \draw[dashed ,thick, red] (0,0) to (2,-2);
  \draw[fill, red] (0,0) circle (5pt);
   \draw[mygreen, thick] (2,-2) to (5,2);
   \draw[mygreen, thick] (2,-2) to (5,-2);
 \draw[mygreen, thick] (2,2) to (5,2);
  \draw[mygreen, thick] (2,2) to (3.5-0.2,0+0.2);
    \draw[mygreen, thick] (5,-2) to (3.5+0.2,0-0.2);
   \draw[dashed ,thick, red] (5,2) to (7,0);
 \draw[dashed ,thick, red] (5,-2) to (7,0);
   \draw[fill, red] (7,0) circle (5pt);
   \draw[mygreen,thick]     ([shift=(-27:0.36cm)]3.4,-0.1)  arc (-27:120-5:0.36);
 \end{scope}

  \begin{scope}[shift={(-2,-24)}]

  \draw[dashed ,thick, red] (0,0) to (2.5,2.5);
  \draw[dashed ,thick, red] (0,0) to (2.5,-2.5);
  \draw[fill, red] (0,0) circle (5pt);
   \draw[mygreen, thick] (2.5,2.5) to (4.5,2.5);
   \draw[mygreen,thick] (2.5,2.5) to (3.5,1);
      \draw[mygreen,thick] (4.5,2.5) to (3.5,1);
\draw[red,dashed,thick] (3.5,1) to (3.5,-1);   
      \draw[mygreen, thick] (2.5,-2.5) to (4.5,-2.5);
     \draw[mygreen,thick] (2.5,-2.5) to (3.5,-1);
      \draw[mygreen,thick] (4.5,-2.5) to (3.5,-1);
   \draw[dashed ,thick, red] (4.5,2.5) to (7,0);
 \draw[dashed ,thick, red] (4.5,-2.5) to (7,0);
   \draw[fill, red] (7,0) circle (5pt);
 \end{scope}
 \node at (8,-24) {$= 0$};

  \end{tikzpicture} 
\caption{(QED-GN$_+$) Feynman graphs appearing in the 2-point correlation function of the composite operator $\rho_+^2$ \protect\footnotemark .   The black ellipse in the second diagram means dressing HS field propagator with graphs in the left column of (\ref{F11}) .}  \label{F12}
  \end{figure}
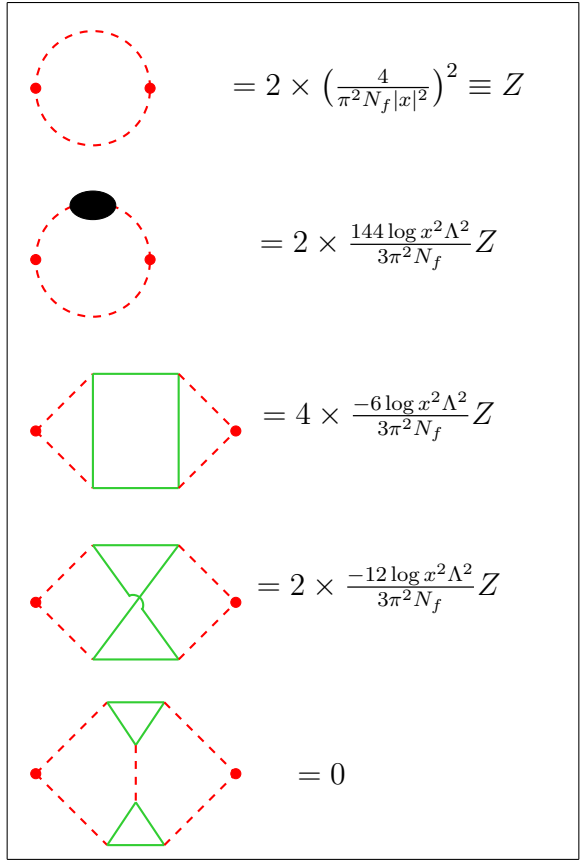

As in fQED, the $N_f^2$ gauge invariant operators $\bar{\Psi}_i \Psi^j$ transform in the adjoint plus singlet representation of $SU(N_f)$. However, the  singlet operator is set to zero by the equation of motion of the HS field $\rho_+$.  Order $O(1/N_f)$ scaling dimensions for the adjoint operator and for $\rho_+$ can be read using table (\ref{F11}), for $\rho_+^2$ using table (\ref{F12}) \footnote{Note added: soon after  we presented these results in \cite{Benvenuti:2018cwd}, also \cite{Boyack:2018zfx} computed the scaling dimensions (\ref{hh1}, \ref{hh2}, \ref{hh3}). Their results agree with ours.}
\begin{align} \label{hh1}
& \Delta[|\Psi|^2_{adj}] = 2 - \frac{48}{3 \pi^2 N_f} + O\left(1/N_f^2\right) \\
 &\Delta[\rho_+] =  1-  \frac{144}{3 \pi^2 N_f}   + O\left(1/N_f^2\right) \label{hh2} \\
& \Delta[\rho_+^2] =  2 -  \frac{240}{3 \pi^2 N_f}   + O\left(1/N_f^2\right) \label{hh3} \ .
   \end{align}
The scaling dimension of the order parameter $\rho_+^2$ is related to the critical exponent $\nu$:
   \begin{align}
   \nu^{-1}=3-\Delta[\rho^2_+]=1+ \frac{240}{3 \pi^2 N_f}   + O\left(1/N_f^2\right)
   \end{align}

\footnotetext{The last graph is vanishing (both the divergent and finite parts are zero). This is because parity invariance forbids single parity odd HS field to decay into 2 HS fields.} 

  \footnotetext{The last Feynman graph is vanishing because the triangle subgraphs made by fermion propagators are identically  zero.}


\subsection{QED-NJL}\label{sec:epqedgn}
In the QED-NJL fixed point, the action is (\ref{h5}). It involves Yukawa interactions and the masses of the HS fields are tuned to zero. The large $N_f$ effective action is described by $N_f$ Dirac fermions $(\Psi_i, \tilde\Psi_i)$ minimally  coupled to the effective photon and interacting with the HS fields $(\rho, \tilde\rho)$ via Yukawa interactions 
\begin{align}
 \CL_{eff} =   \sum_{i=1}^{N_f/2} \bar{\Psi}_i    \slashed{D} \Psi^i+  \sum_{i=1}^{N_f/2} {\bar\Psit}_i \slashed{D} \Psit^i + \rho \sum_{i=1}^{N_f/2} {\bar\Psi}_i \Psi^i +  \tilde\rho \sum_{i=1}^{N_f/2} {\bar\Psit}_i \Psit^i 
 \end{align}
where  
\begin{align}
&\rho=\rho_+ + \rho_- \\ 
&\tilde\rho=\rho_+ - \rho_- \ .
\end{align}
The photon "sees" all the flavors, therefore effective photon propagator is the same as in fQED (\ref{photonpropf}). The effective propagators for the HS fields are 
\begin{align}
\langle \rho(x) \rho(0) \rangle =\langle \tilde\rho(x) \tilde\rho(0) \rangle = \frac{4}{ \pi^2 (N_f/2) |x|^2} \ .
\end{align}
The continuous global symmetry is:
\begin{align}
 \label{h7} \frac{(SU(N_f/2) \times SU(N_f/2) \times U(1)_b \times \uot )   \rtimes \bZ_2^e } {\bZ_{N_f}}  \rtimes \bZ_2^C \ .
 \end{align}
Parity is preserved, provided $(\rho,\tilde\rho)$ and $\rho_\pm$ are odd under parity transformation. The other global symmetries act as follows. $U(1)_b$: $\{\Psi \rightarrow e^{i\alpha}\Psi, \  \tilde\Psi \rightarrow e^{-i\alpha}\tilde\Psi \}$, $\bZ_2^e$: $\{\Psi \leftrightarrow \tilde\Psi, \rho \leftrightarrow \tilde\rho\}$.

\subsection*{Scaling dimension of low-lying scalar operators}

The gauge invariant fermion bilinear operators are classified as irreducible representations (\ref{decomp1}) under $SU(N_f/2)^2$ symmetry group. The calculation of the scaling dimensions for the adjoint and the bifundamental operators is parallel to the calculation of the scaling dimensions of the similar operators in the ep-bQED and can be done using the graphs (\ref{F11}). 

 The quadratic singlet operators are out of the spectrum, they are set to zero by the EOM of the HS fields $\rho, \tilde\rho$. The two-point correlation function for the $\rho$ field can be calculated using the left column diagrams of the table (\ref{F13}). Taking into account the necessary changes we get
\begin{align} \label{h10} 
  \langle \rho(x) \rho(0) \rangle =\Big (1+\frac{64 \log {x^2 \Lambda^2}}{3\pi^2 N_f} \Big) \Big( \frac{4 }{\pi^2 (N_f/2) |x|^2} \Big) \ .
\end{align}
Notice that at order $O(1/N_f)$ there is a mixing between $\rho$ and $\tilde\rho$ (\ref{F13}). 
\begin{align} \label{h11}
  \langle \rho(x) \tilde\rho(0) \rangle = \frac{96 \log {x^2 \Lambda^2}}{3\pi^2 N_f} \Big( \frac{4 }{\pi^2 (N_f/2) |x|^2} \Big) \ .
\end{align}
Instead, the fields $\rho_+, \rho_-$ do not mix, they are the eigenvectors of the mixing matrix. Using (\ref{h10}, \ref{h11}) one can calculate anomalous dimensions of these fields. 

In the table (\ref{F14}) we collected all the graphs that contribute to the mixing of operators quadratic in HS fields: $\{\rho^2(x), \tilde\rho^2(x), \sqrt{2}\rho\tilde\rho(x) \}$. We get the following mixing matrix 
\begin{align} \label{mixmat}
\begin{pmatrix} 
1+\frac{32 \log{x^2\Lambda^2}}{3\pi^2 N_f} & 0 & \frac{96\sqrt{2} \log{x^2\Lambda^2}}{\pi^2 N_f} \\
0 & 1+\frac{32 \log{x^2\Lambda^2}}{3\pi^2 N_f} &  \frac{96\sqrt{2} \log{x^2\Lambda^2}}{\pi^2 N_f} \\
\frac{96\sqrt{2} \log{x^2\Lambda^2}}{\pi^2 N_f} &  \frac{96\sqrt{2} \log{x^2\Lambda^2}}{\pi^2 N_f} & 1+\frac{128 \log{x^2\Lambda^2}}{3\pi^2 N_f}
 \end{pmatrix} 
 \times \tilde Z .
\end{align}
Where $\tilde Z$ is defined in (\ref{F14}). Using (\ref{mixmat}) it is straightforward to pass to the eigenbasis and find the scaling dimension for each of the eigenbasis operators. Below we give the list of operators and their scaling dimensions 
\begin{align}
 &\Delta[|\Psi|^2_{adj}] =  \Delta[|\Psit|^2_{adj}]= 2 - \frac{32}{3 \pi^2 N_f} + O(1/{N_f^2})  \\
&\Delta[\bar\Psit_i\Psi_j] = \Delta[\bar\Psi_j\Psit_i]  = 2 - \frac{56}{3 \pi^2 N_f} + O(1/{N_f^2})  \\
 &\Delta[\rho_+] =  1-  \frac{160}{3 \pi^2 N_f}   + O(1/{N_f^2}) \\
 &\Delta[\rho_-] =  1 + \frac{32}{3 \pi^2 N_f}   + O(1/{N_f^2}) \\
 &\Delta[\rho_+ \rho_-] =  2 -  \frac{32}{3 \pi^2 N_f}   + O(1/{N_f^2})  \\
 &\Delta[\rho_+^2+ (4+\sqrt{17})\rho_-^2] =  2 -  \frac{16(5-3\sqrt{17})}{3 \pi^2 N_f}   + O(1/{N_f^2})  \\
 &\Delta[\rho_+^2+(4-\sqrt{17})\rho_-^2] =  2 -  \frac{16(5+3\sqrt{17})}{3 \pi^2 N_f}   +O(1/{N_f^2}) \ .
 \end{align}
 
 A similar model with two HS scalars was studied in \cite{Gracey:1993ka}. Their model seems to be different from QED-NJL we discuss, in particular the anomalous dimensions of HS fields are different from ours. 

  \begin{figure}[H]
  \centering
\begin{tikzpicture}[scale=0.38]

\draw[mygreen,thick]  ([shift=(0:2cm)]0,0)  arc (0:180:2cm);
\draw[mygreen,thick]  ([shift=(-180:2cm)]0,0) arc (-180:0:2cm);
\draw[photon, red] (0,0) ([shift=(31:2cm)]0,0)  to   ([shift=(149:2cm)]5,0);
\draw[photon, red] (0,0) ([shift=(-31:2cm)]0,0)  to   ([shift=(-149:2cm)]5,0);
\draw[mygreen,double,thick ] ([shift=(180:2cm)]5,0) arc (180:0:2cm);
\draw[mygreen,double, thick] ([shift=(0:2cm)]5,0) arc (0:-180:2cm);
\draw[dashed,red,thick] (-3.5,0) to (-2,0);
\draw[dashed,red,double,thick] (7,0) to (8.5,0);  
\node at (14.9,0) {$= 2\times  \frac{48\log x^2 \Lambda^2}{ 3 \pi^2 N_f } \frac{4}{ \pi^2 (N_f/2)|x|^2}  $};

\end{tikzpicture}
\caption{Diagram responsible for mixing $\langle \rho(x) \tilde\rho(0) \rangle$ .} \label{F13}
  \end{figure}

 \begin{figure}[H]
 \centering
\begin{tikzpicture}[scale=0.34]

 \draw[black] (-5,3) to (-5,-24);
 \draw[black] (-5,-24) to (42,-24);
 \draw[black] (-5,3) to (42,3);
 \draw[black] (-5,-9) to (42,-9);
 \draw[black] (42,3) to (42,-24);
 
 \draw[dashed, red, thick] (6,0) circle (2cm);
 \node at (-1,-3) {$\langle \rho^2(x)\rho^2(0)\rangle :$};
 \draw[fill, red] (4,0) circle (5pt);
 \draw[fill, red] (8,0) circle (5pt);
 \node at (15.5,0) {$=2 \times \big(\frac{4}{\pi^2 (N_f/2)  |x|^2} \big)^2 \equiv \tilde Z$};
 
 \begin{scope}[shift={(27,5.8)}]
 \draw[dashed, red, thick] (0,-6) circle (2cm);
 \draw[fill,black] (0,-4.1) ellipse (0.8cm and 0.5cm);
 \draw[fill, red] (-2,-6) circle (5pt);
 \draw[fill, red] (2,-6) circle (5pt);
 \node at (7,-6) {$=2\times \frac{64 \log x^2 \Lambda^2}{3\pi^2N_f} \tilde Z$};
 \end{scope}
 
 \begin{scope}[shift={(4.5,-6)}]
 \draw[dashed ,thick, red] (0,0) to (2,2);
 \draw[dashed ,thick, red] (0,0) to (2,-2);
 \draw[fill, red] (0,0) circle (5pt);
 \draw[mygreen, thick] (2,2) to (2,-2);
 \draw[mygreen, thick] (2,-2) to (5,-2);
 \draw[mygreen, thick] (2,2) to (5,2);
 \draw[mygreen, thick] (5,2) to (5,-2);
 \draw[dashed ,thick, red] (5,2) to (7,0);
 \draw[dashed ,thick, red] (5,-2) to (7,0);
 \draw[fill, red] (7,0) circle (5pt);
 \node at (12.5,0) {$=4\times \frac{-12 \log x^2 \Lambda^2}{3 \pi^2N_f} \tilde Z$};
\end{scope}

 \begin{scope}[shift={(23,-6)}]
 \draw[dashed ,thick, red] (0,0) to (2,2);
 \draw[dashed ,thick, red] (0,0) to (2,-2);
\draw[fill, red] (0,0) circle (5pt);
 \draw[mygreen, thick] (2,-2) to (5,2);
 \draw[mygreen, thick] (2,-2) to (5,-2);
 \draw[mygreen, thick] (2,2) to (5,2);
 \draw[mygreen, thick] (2,2) to (3.5-0.2,0+0.2);
 \draw[mygreen, thick] (5,-2) to (3.5+0.2,0-0.2);
 \draw[dashed ,thick, red] (5,2) to (7,0);
 \draw[dashed ,thick, red] (5,-2) to (7,0);
 \draw[fill, red] (7,0) circle (5pt);
 \draw[mygreen,thick]     ([shift=(-27:0.36cm)]3.4,-0.1)  arc (-27:120-5:0.36);
 \node at (13,0) {$=2\times \frac{-24 \log x^2 \Lambda^2}{3\pi^2N_f} \tilde Z$};
 \end{scope}
 
 \node at (,-12.5) {$\langle \sqrt{2}\rho\tilde\rho(x) \sqrt{2}\rho\tilde\rho(0)\rangle :$};
 
  \draw[red,dashed,thick]   ([shift=(0:2cm)]10,-13)  arc   (0:180:2cm);
  \draw[red,dashed,double,thick]   ([shift=(-180:2cm)]10,-13)  arc   (-180:0:2cm);
  \draw[fill,red]  (8,-13) circle  (5pt);
  \draw[fill,red]  (12,-13) circle  (5pt);
  \node at (14,-13) {$=\tilde Z$};
  
 \begin{scope}[shift={(13,0)}]
\draw[red,dashed,thick]   ([shift=(0:2cm)]8,-13)  arc   (0:180:2cm);
  \draw[red,dashed,double,thick]   ([shift=(-180:2cm)]8,-13)  arc   (-180:0:2cm);
  \draw[fill,red]  (6,-13) circle  (5pt);
 \draw[fill,red]  (10,-13) circle  (5pt);
  \draw[fill,black] (8,-11.1) ellipse (0.8cm and 0.5cm);
  \node at (15.5,-13) {$=2\times \frac{64 \log x^2 \Lambda^2}{3 \pi^2N_f} \tilde Z$};
   \end{scope}

\draw[black] (-5,-16) to (42,-16); 

\node at (0,-20) {$\langle \rho^2(0) \sqrt{2} \rho\tilde\rho(0):$};

\draw[red,dashed,double,thick]   ([shift=(0:4 and 2.7)]12,-20)  arc   (0:50:4 and 2.7);
 \draw[red,dashed,thick]   ([shift=(130:4 and 2.7)]12,-20)  arc   (130:180:4 and 2.7);
  \draw[red,dashed,thick]   ([shift=(-180:4 and 2.7)]12,-20)  arc   (-180:0:4 and 2.7);
   \draw[fill,red]  (8,-20) circle  (5pt);
   \draw[fill,red]  (16,-20) circle  (5pt);
                    
 \begin{scope}[shift={(10.5,-18+0.2)}]
  \draw[mygreen,thick]  ([shift=(0:1 and 0.6)]0,0)  arc (0:180:1 and 0.7);
 \draw[mygreen,thick]  ([shift=(-180:1 and 0.6)]0,0) arc (-180:0:1 and 0.6);
 \draw[photon1, red] (0,0) ([shift=(31:1 and 0.6)]0,0)  to   ([shift=(149:1 and 0.6)]3,0);
 \draw[photon1, red] (0,0) ([shift=(-31:1 and 0.6)]0,0)  to   ([shift=(-149:1 and 0.6)]3,0);
 \draw[mygreen,double,thick ] ([shift=(180:1 and 0.6)]3,0) arc (180:0:1 and 0.6);
 \draw[mygreen,double, thick] ([shift=(0:1 and 0.6)]3,0) arc (0:-180:1 and 0.6);
   \end{scope}
                     
\node at (21.5,-20) {$=\frac{96\sqrt{2} \log x^2 \Lambda^2 }{3 \pi^2 N_f} \tilde Z$};

\end{tikzpicture}
\caption{(QED-NJL) mixing of quadratic in HS field operators.} \label{F14}
  \end{figure}

\subsection{QED-GN$_-$}
In the QED-GN$_-$ fixed point the action is (\ref{h5}), with Yukawa interaction involving the HS field $\rho_-$, while the HS field $\rho_+$ is massive and is decoupled from the IR spectrum. The large $N_f$ effective action is described by Dirac fermions $(\Psi_i, \tilde\Psi_i)$ minimally  coupled to the effective photon and interacting with the HS field $\rho_-$ via Yukawa interaction
\begin{align}
 \label{rho-} \CL_{eff} =  \sum_{i=1}^{N_f/2} ( \bar{\Psi}_i    \slashed{D} \Psi^i+  {\bar\Psit}_i \slashed{D} \Psit^i)  + \rho_- ( \sum_{i=1}^{N_f/2} \bar\Psi_i \Psi^i -   \sum_{i=1}^{N_f/2} \bar{\tilde\Psi}_i \tilde\Psi^i) \,. 
  \end{align}
 The effective photon propagator is as in (\ref{photonpropf}) because the photon "sees" all the flavors. The effective propagator for the HS field $\rho_-$ is 
\begin{align}
\langle \rho_-(x) \rho_-(0) \rangle =\frac{4}{ \pi^2 N_f |x|^2} \ .
\end{align}
The continuous global symmetry is the same as that of QED-NJL (\ref{h7}). Parity is preserved provided HS field $\rho_-$ is odd under parity transformation. The discrete symmetry $\bZ_2^e$ acts: $\{\Psi \leftrightarrow \tilde\Psi, \rho_- \leftrightarrow - \rho_-\}$.
\subsection*{Scaling dimension of low-lying scalar operators}
The fermion bilinear operators are classified according to irreducible representations of $SU(N_f/2)^2$, like in (\ref{decomp1}). Notice that $\rho_-$ takes the operator $ \sum_{i=1}^{N_f/2}  (\bar\Psi_i \Psi^i -   \bar{\tilde\Psi}_i \tilde\Psi^i)$ out from the spectrum, while the plus combination remains in the spectrum and has dimension 2 at the leading  order. 

 The scaling dimension of the HS field $\rho_-$ is calculated using the graphs in the left column of the table (\ref{F11}). The contributions of the first 5 graphs remain unchanged, while there are 4 three-loop graphs and they are canceling each other. This is due to the fact that $\rho_-$ field couples to the fermion flavors $(\Psi,\tilde\Psi)$ with different signs and therefore the three loop graph which has fermions $\Psi$ running in one of its loops and fermions $\tilde\Psi$ running in the other loop comes with an opposite sign with respect to the three loop graph made solely by fermions $\Psi$ (or $\tilde\Psi$). The scaling dimensions for the other operators can be calculated easily. 
\begin{align}
 &\Delta[|\Psi|^2_{adj}] =  \Delta[|\Psit|^2_{adj}] = 2 - \frac{48}{3 \pi^2 N_f} + O(1/{N_f^2})   \\
 &\Delta[\bar\Psit_i\Psi_j] = \Delta[\bar\Psi_j\Psit_i] = 2 - \frac{72}{3 \pi^2 N_f} + O(1/{N_f^2})  \\
 & \Delta \big [ \frac{1}{\sqrt{N_f}} \big ( \sum|\Psi_i|^2  +  \sum |\tilde{\Psi}_i|^2 \big )  \big ]= 2+ \frac{144}{3 \pi^2 N_f} + O(1/{N_f^2})  \\
 &\Delta[\rho_-] = 1 + \frac{48}{3 \pi^2 N_f} +O(1/{N_f^2})     \\
 &\Delta[\rho_-^2]=2+\frac{144}{3\pi^2 N_f}+O(1/{N_f^2}) \ .
 \end{align}
 Some of these results have been obtained in \cite{Gracey:1991ry, Gracey:1993zn, Gracey:2018fwq}, which also include some scaling dimensions at order $O(1/N_f^2)$.

 \section{Super-QED in the large $N_f$ limit}\label{sec:susy}
 In this section we compute the large $N_f$ scaling dimension of mesonic operators in QED with minimal supersymmetry, and then compare the results, at $N_f=2$, with a dual Gross-Neveu-Yukawa model. At the end we also consider an $\cN=2$ superQED.

 The UV action of  $2+1d$ QED with minimal supersymmetry, $\mathcal{N}=1 $ (i.e. 2 supercharges), $N_f$ flavors and zero superpotential
 \be \cW_{\cN=1}=0 \,,\ee
  has the form 
\begin{align} \nonumber
S_{UV}= \int d^3 x  & \Big( - \frac{1}{4 e^2} F_{\mu \nu} F^{\mu \nu } + \frac{1}{2 e^2} \bar\lambda i  \slashed{\partial} \lambda  + \bar \Psi_j i  \gamma^\mu D_\mu \Psi ^j  + \overline{D_\mu \Phi _j  } D^\mu \Phi ^j  \\ 
& + i \bar \Psi_j \lambda \Phi^j - i \bar \lambda \Psi^j \Phi ^ {* }_j  - \frac{N_f}{32 i (1-\xi)} \int d^3 y \frac{\partial_\mu A^\mu(x) \partial _\nu A^ \nu(y)}{2 \pi^2 |x-y|^2}  \Big ) \ . \label{h12}
\end{align}
The action (\ref{h12}) is written in the Minkowski metric. Our convention for the Minkowski metric is $(+,-,-)$.  The kinetic terms for the photon and for the gaugino are non canonically normalised, the covariant derivative is $D_\mu = \partial _ \mu + i A_\mu$. We have $N_f$ flavors of Dirac fermions and complex scalars: $ \Psi ^j,  \ \Phi^j ,  \  j=1,...,N_f$.  Our conventions for the gamma matrices are: $\gamma^0 = \sigma_2 , \ \gamma^1= i \sigma_1, \ \gamma^2 = i \sigma _3$, where $\sigma_{i}$ are the Pauli matrices. We define $\bar \Psi = \Psi ^{\dag} \gamma^0$. Notice that  gaugino is a Majorana fermion, with our conventions for the gamma matrices it has two real components.

The action (\ref{h12}) is written in the Wess-Zumino gauge, which explicitly breaks supersymmetry, the remaining gauge symmetry is fixed by adding the conformal gauge fixing term in the action. The $\mathcal{N}=1$ supersymmetry of the action (\ref{h12}) becomes obvious when one constructs it using superspace integrals and superfields, for more details check \cite{Gates:1983nr}. The fields are organized in $\mathcal{N}=1$ super-multiplets: a vector multiplet $\{ \lambda, A_\mu \}$ and $N_f$ scalar matter multiplets $\{ \Phi^i, \Psi^i, F^i \}$. Going on-shell one sets $F^i=0$.  The global symmetry of the action is 
\begin{align}
\frac{SU(N_f) \times \uot}{\bZ_{N_f}} \rtimes \bZ_2^{\CC} \ .
 \end{align}
 Additionally there is parity invariance. These symmetries prevent the generation of additional interactions (quadratic or quartic superpotential interactions would break parity invariance), therefore there is no need of tuning interactions to zero. 

The large $N_f$ effective action of the $\mathcal{N}=1$ SQED is described by $N_f$ scalar and $N_f$ fermion flavors minimally coupled to the effective photon and interacting with the effective gaugino via a Yukawa interaction  
\begin{align}
S_{IR}= \int d^3 x  \big(    \bar \Psi_j i  \gamma^\mu D_\mu \Psi ^j  + \overline{D_\mu \Phi _j  } D^\mu \Phi ^j   + i \bar \Psi_j \lambda \Phi^j - i \bar \lambda \Psi^j \Phi ^ {\dag }_j   \big ) \ . \label{h13}
\end{align}
The effective photon propagator is obtained by summing a geometric series of bubble diagrams with fermion and scalar loops. We give the effective photon propagator after Wick rotation 
\begin{align} 
\langle A_\mu (x) A_\nu(0) \rangle_{\text{eff}}= -\frac{4 i }{ \pi^2 N_f |x|^2} \big ( (1-\xi)\delta_{\mu \nu}+2\xi \frac{x_\mu x_\nu}{|x|^2} \big) \ .
\end{align}
 The effective gaugino propagator is obtained by summing a geometric series of bubble diagrams with each bubble made by one fermion and one boson propagators, after Wick rotation we have following expression 
\begin{align} 
 \big \langle \lambda(x) \lambda^T (0) \big \rangle_{\text{eff}}= \frac{8i (\slashed{x} \gamma^0)}{\pi^2 N_f |x|^4} \ .
\end{align}
We use red dotted line to represent effective gaugino propagator in the Feynman graphs. 

\subsection{Scaling dimension of low-lying mesonic operators}
The following three quadratic operators sit inside the same $\cN=1$ supermultiplet  
\begin{align} \label{h14}
  \begin{pmatrix} 
  \Phi ^* \Phi \\
  \Phi^* \Psi_\alpha + \Phi \Psi_\alpha^*   \\
    \bar \Psi \Psi \ 
      \end{pmatrix} \ .
\end{align}
where $\alpha$ is a spinor index. Depending how the flavor indices are contracted we can construct a singlet and an adjoint representation of the global symmetry $SU(N_f)$.

\begin{figure}[H]
  \centering
\begin{tikzpicture}[scale=0.33]

\draw[black] (-4,3) to (18.5,3);
\draw[black] (-4,3) to (-4,-30+2);
\draw[black] (18.5,3) to (18.5,-30+2);
\draw[black] (-4,-30+2) to (18.5,-30+2);

\node at (-3,0)  {$A$};
\node at (-3,-5)  {$B$};
\node at (-3,-10)  {$C$}; 
\node at (-3,-15)  {$D$};
\node at (-3,-20)  {$E$}; 
\node at (-3,-25)  {$F$};  

\draw[thick, blue] (0,0) circle (2cm);
\draw[thick,fill] (-2,0) circle (4pt);
\draw[thick,fill] (2,0) circle (4pt);
\node at (8,0) {$= \frac{-i}{16 \pi^2 |x|^2}=  V$};

\draw[blue,thick] ([shift=(180:2cm)]0,-5)   arc (180:0:2cm);
\draw[blue,thick] ([shift=(0:2cm)]0,-5)   arc (0:-180:2cm);
\draw[photon,red] (-1.66,-3.9) to (1.66,-3.9);
\draw[thick,fill] (-2,-5) circle (4pt);
\draw[thick,fill] (2,-5) circle (4pt);
\node at (10.5,-5) {$= 2\times \frac{2(5+3\xi)\log{x^2\Lambda^2}}{3 \pi^2 N_f} V$};

\begin{scope}[shift={(0,-5)}]
\draw[thick, blue] ([shift=(180:2cm)]0,-5) arc (180:135:2cm);
\draw[thick,mygreen] ([shift=(135:2cm)]0,-5) arc (135:45:2cm);
\draw[thick, blue] ([shift=(45:2cm)]0,-5) arc (45:0:2cm);
\draw[thick,blue] ([shift=(0:2cm)]0,-5)   arc (0:-180:2cm);
\draw[thick, red, densely dotted] (0,-2+0.6) [partial ellipse=-61:-140+20:3cm and 2.5cm];
\draw[fill]  (-2,-5) circle (5pt);
\draw[fill]  (2,-5) circle (5pt);
\node at (9.5,-5) {$= 2\times \frac{-4\log{x^2\Lambda^2}}{3\pi^2 N_f} V$};
\end{scope}

 \begin{scope}[shift={(0,-5)}]

\draw[blue,thick] ([shift=(180:2cm)]0,-10)   arc (180:0:2cm);
\draw[blue,thick] ([shift=(0:2cm)]0,-10)   arc (0:-180:2cm);
\draw[photon, red] (0,-10) ([shift=(90:2cm)]0,-10)  to   ([shift=(-90:2cm)]0,-10);
\draw[thick,fill] (-2,-10) circle (4pt);
\draw[thick,fill] (2,-10) circle (4pt);
\node at (9.5,-10) {$= \frac{12(1-\xi)\log{x^2\Lambda^2}}{3 \pi^2 N_f} V$};

\draw[thick,blue]  ([shift=(0:2cm)]0,-15)  arc (0:180:2cm);
\draw[thick,blue]  ([shift=(-180:2cm)]0,-15) arc (-180:0:2cm);
\draw[photon, red] ([shift=(0:2cm)]0,-15)  to   ([shift=(-210:2cm)]6,-15);
\draw[photon, red] ([shift=(0:2cm)]0,-15)  to   ([shift=(-150:2cm)]6,-15);
\draw[thick,blue ] ([shift=(180:2cm)]6,-15) arc (180:0:2cm);
\draw[thick,blue] ([shift=(0:2cm)]6,-15) arc (0:-180:2cm);
\draw[thick,fill] (-2,-15) circle (4pt);
\draw[thick,fill] (8,-15) circle (4pt);
\node at (13.5,-15) {$=4\times  \frac{-12\log{x^2\Lambda^2}}{3 \pi^2 N_f} V$};

\draw[blue,thick]  ([shift=(31:2cm)]0,-20) arc (31:180:2cm);
\draw[blue,thick]  ([shift=(-180:2cm)]0,-20) arc (-180:-31:2cm);
\draw[thick, mygreen] ([shift=(31:2cm)]0,-20) arc (31:-31:2cm);
\draw[blue,thick] ([shift=(149:2cm)]6,-20) arc (149:0:2cm);
\draw[thick, mygreen] ([shift=(-149:2cm)]6,-20) arc (-149:-180:2cm);
\draw[thick, mygreen] ([shift=(149:2cm)]6,-20) arc (149:180:2cm);
\draw[thick,blue] ([shift=(0:2cm)]6,-20) arc (0:-149:2cm);
\draw[densely dotted, red,thick]   ([shift=(31:2cm)]0,-20) to ([shift=(149:2cm)]6,-20);
\draw[densely dotted, red,thick]   ([shift=(-31:2cm)]0,-20) to ([shift=(-149:2cm)]6,-20);
\draw[thick,fill] (-2,-20) circle (4pt);
\draw[thick,fill] (8,-20) circle (4pt);
\node at (13.5,-20) {$=2\times  \frac{12\log{x^2\Lambda^2}}{3\pi^2 N_f} V$};
\end{scope}

  \end{tikzpicture}
\caption{ ($\mathcal{N}=1$ SQED) Results for individual Feynman graphs appearing in the 2-point correlation function for the scalar-bilinear operators.  } \label{F17}
  \end{figure}
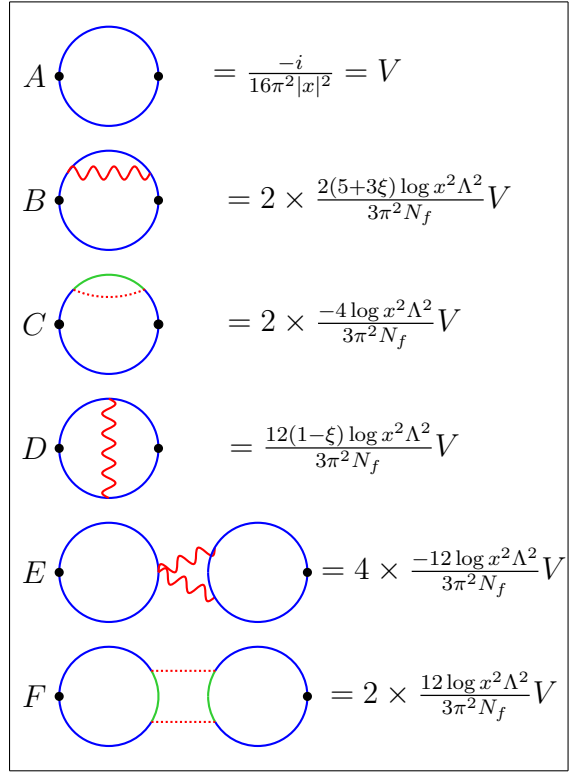

Let us first discuss the adjoint supermultiplet. Using the graphs A,B,C,D from the table (\ref{F17}) we can extract the scaling dimension of the adjoint scalar-bilinear operator 
 \begin{align} \nonumber
 \langle |\Phi|^2_{adj}(x) |\Phi|^2_{adj}(0) \rangle &=\frac{-i}{16\pi^2|x|^2}+\frac{-i}{16\pi^2|x|^2}  \Big( \frac{4(5+3\xi)}{3\pi^2 N_f} -\frac{8}{3\pi^2 N_f} +\frac{12(1-\xi)}{3\pi^2 N_f} \Big)\log{x^2 \Lambda^2}  \\
 &= \frac{-i}{16\pi^2|x|^2}\Big[1 - \Big( -\frac{24}{3 \pi^2 N_f} \Big)\log {x^2 \Lambda^2} \Big] \nonumber \\
 &= \frac{-i}{16\pi^2|x|^2} \Big( \frac{1}{x^2\Lambda^2} \Big) ^{ \Delta^{(1)}_{adj}} \ .
 \end{align}
The anomalous dimension of the adjoint operator is  $\Delta^{(1)}_{adj}=-\frac{24}{3 \pi^2 N_f}$. Due to supersymmetry  the scaling dimensions of the components in (\ref{h14}) are related to each other.
 \begin{align}  \label{h23}
  &\Delta[(\Phi ^* \Phi) _{adj}]=1-\frac{24}{3\pi^2 N_f} +O(1/N_f^2)\\ 
  & \Delta[(\Phi^* \Psi_\alpha + \Phi \Psi_\alpha^*) _{adj}]=\frac{3}{2}-\frac{24}{3\pi^2 N_f}+O(1/N_f^2) \\
  & \Delta[(\bar \Psi \Psi) _{adj}]=2-\frac{24}{3 \pi^2 N_f}+ O(1/N_f^2)  \ .
  \end{align}
  It is also possible to construct another scalar-fermion bilinear 
\begin{align} \label{h15}
(\Phi^* \Psi_\alpha - \Phi \Psi_\alpha^* )_{adj} \ .
\end{align}
We checked that anomalous dimension of (\ref{h15}) is vanishing at order $O(1/N_f)$. This is not surprising since operator (\ref{h15}) sits  in the same supermultiplet with the gauge invariant flavor current operator $\big (\bar \Psi \gamma^\mu \Psi+ i(\Phi^* D^\mu\Phi-\overline{D^\mu\Phi} \cdot \Phi)\big )_{adj}$ which is conserved and has a scaling dimension exactly equal to $2$ for any $N_f$. 

 In order to compute the anomalous dimension of the singlet scalar-bilinear operator, we use all the graphs in the table (\ref{F17}), since for this operator all of them contribute at order $O(1/N_f)$. It turns out that the anomalous scaling dimension vanishes at that order (there seems to be no reason to think that at higher orders in the $1/N_f$ expansion the anomalous corrections are going to be absent). Also the singlet supermultiplet (\ref{h14}) has the dimensions of its components related to each other:
\begin{align} 
&\Delta[(\Phi ^* \Phi) _{sing}]=1+ O(1/N_f^2) \\  
  & \Delta[(\Phi^* \Psi_\alpha + \Phi \Psi_\alpha^*) _{sing}]=\frac{3}{2}+ O(1/N_f^2)\\
  & \Delta[(\bar \Psi \Psi)_{sing}]=2+ O(1/N_f^2)   \ .  \label{h24}
\end{align}
Notice that the singlet counterpart of (\ref{h15}) is out of spectrum. This is precisely the operator that couples to gaugino in the effective action (\ref{h13}) and it is set to zero by the EOM of the gaugino field. 
 \begin{figure}[]
  \centering
\begin{tikzpicture}[scale=0.38]

\draw[black] (-3,3) to (18,3);
\draw[black] (-3,3) to (-3,-23);
\draw[black] (18,3) to (18,-23);
\draw[black] (-3,-23) to (18,-23);

\draw[scalar] (0,0) (0:2) arc (0:180:2 and 1);
\draw[scalar] (0,0) (-180:2) arc (-180:0:2 and 1);
\draw[scalar] (0,0) (0:2) arc (0:180:2);
\draw[scalar] (0,0) (-180:2) arc (-180:0:2);
\draw[thick,fill] (-2,0) circle (4pt);
\draw[thick,fill] (2,0) circle (4pt);
\node at (8,0) {$=-\frac{i}{(4\pi|x|)^4}=H$};

\begin{scope}[shift={(0,-5)}]
\draw[scalar] (0,0) (0:2) arc (0:180:2 and 1);
\draw[scalar] (0,0) (-180:2) arc (-180:0:2 and 1);
\draw[scalar] (0,0) (0:2) arc (0:180:2);
\draw[scalar] (0,0) (-180:2) arc (-180:0:2);
\draw[photon,red] ([shift=(45:2)]0,0) to ([shift=(135:2)]0,0);
\draw[thick,fill] (-2,0) circle (4pt);
\draw[thick,fill] (2,0) circle (4pt);
\node at (9.5,0) {$=4\times  \frac{ 2\big( 5+3 \xi \big) \log {x^2\Lambda^2} }{ 3 \pi^2 N_f } H$};
\end{scope}

\begin{scope}[shift={(0,-10)}]
\draw[scalar] (0,0) (0:2) arc (0:180:2 and 1);
\draw[scalar] (0,0) (-180:2) arc (-180:0:2 and 1);
\draw[thick,blue] (0,0) (0:2) arc (0:45:2);
\draw[fermion] (0,0) (45:2) arc (45:135:2);
\draw[thick,blue] (0,0) (135:2) arc (135:180:2);
\draw[scalar] (0,0) (-180:2) arc (-180:0:2);
\draw[thick, red, densely dotted] (0,3.35) [partial ellipse=-69:-112:3.9cm and 2.1cm];
\draw[thick,fill] (-2,0) circle (4pt);
\draw[thick,fill] (2,0) circle (4pt);
\node at (9,0) {$=4\times  \frac{ -4\log {x^2\Lambda^2} }{  3 \pi^2 N_f } H$};
\end{scope}

\begin{scope}[shift={(0,-15)}]
\draw[scalar] (0,0) (0:2) arc (0:180:2 and 1);
\draw[scalar] (0,0) (-180:2) arc (-180:0:2 and 1);
\draw[scalar] (0,0) (0:2) arc (0:180:2);
\draw[scalar] (0,0) (-180:2) arc (-180:0:2);
\draw[photon1,red ] ([shift=(90:2)]0,0) to ([shift=(90:2 and 1)]0,0);
\draw[thick,fill] (-2,0) circle (4pt);
\draw[thick,fill] (2,0) circle (4pt);
\node at (9.5,0) {$=2\times  \frac{ -12 \big( 1-\xi \big) \log {x^2\Lambda^2} }{ 3  \pi^2 N_f } H$};
\end{scope}

\begin{scope}[shift={(0,-20)}]
\draw[scalar] (0,0) (0:2) arc (0:180:2 and 1);
\draw[scalar] (0,0) (-180:2) arc (-180:0:2 and 1);
\draw[scalar] (0,0) (0:2) arc (0:180:2);
\draw[scalar] (0,0) (-180:2) arc (-180:0:2);
\draw[photon,red ] ([shift=(90:2)]0,0) to ([shift=(-90:2 and 1)]0,0);
\draw[thick,fill] (-2,0) circle (4pt);
\draw[thick,fill] (2,0) circle (4pt);
\node at (9.5,0) {$=4 \times  \frac{ 12 \big( 1-\xi \big) \log {x^2\Lambda^2} }{ 3 \pi^2 N_f } H$};
\end{scope}

  \end{tikzpicture}
\caption{($\mathcal{N}=1$ SQED) quartic adjoint-2 operator renormalization. } \label{F18}
  \end{figure}

Next we consider scalar quartic operators (\ref{scalar quartic}). In the equation (\ref{h3}) we decomposed this operator into irreducible representations of $SU(N_f)$ group: singlet, adjoint, adjoint-2. For the quartic adjoint-2 operator we need the Feynman graphs of table (\ref{F18}) to extract the scaling dimension. Similar calculations can be done for the other two operators. We skip the details and give the final result below
\begin{align} \label{h25}
& \Delta[ |\Phi|^4_{adj-2}] = 2\Delta[ |\Phi|^2_{adj}] + O(1/{N_f^2})  = 2 - \frac{48}{3 \pi^2 N_f} + O(1/{N_f^2})    \\  \label{h26}
&\Delta[ |\Phi|^4_{adj}] = \Delta[ |\Phi|^2_{adj}] +\Delta[ |\Phi|^2_{sing}] + O(1/{N_f^2})  = 2 - \frac{24}{3 \pi^2 N_f} + O(1/{N_f^2})   \\  \label{h27}
 &\Delta[ |\Phi|^4_{sing}] =2\Delta[ |\Phi|^2_{sing}]+ O(1/{N_f^2})   =2 + O(1/{N_f^2})    \ .
 \end{align}

\subsection{The duality $\mathcal{N}{=}1$ SQED with $N_f{=}2$ $\leftrightarrow$ $7$-field Wess-Zumino model: a quantitative check}
The $\mathcal{N}=1$ super-QED with two flavors ($N_f=2$) has been argued to be dual to a cubic $\mathcal{N}=1$ supersymmetric Wess-Zumino model with $SU(2) \times U(1)$ global symmetry(\cite{Gaiotto:2018yjh, Benini:2018bhk}). The field content of the WZ model is given by $7$ real $\cN=1$ supermultiplets: a real triplet $\mu_I$ and a complex doublet $M_\alpha$. The superpotential of the WZ model is dictated by the $SU(2) \times U(1)$ global symmetry and by parity invariance:
\begin{align}
\mathcal{W}_{\cN=1}=\mu_I M_\alpha (\sigma_I)_{\alpha \beta} M_\beta^\dag \ .
\end{align}
The real fields $\mu_I$ map to the quadratic mesons on the gauge theory side. The complex fields $M_\alpha$ map to the monopoles with minimal topological charge. This duality can be obtained starting from the $\mathcal{N}=4$ mirror symmetry (\cite{Intriligator:1996ex, Kapustin:1999ha}), which in the IR relates abelian gauge theory with one hypermultiplet flavor to a free massless hypermultiplet. The $\cN=1$ duality also has a description in terms of S-duality of Type IIB brane setups \cite{Gremm:1999su}.

In table (\ref{F21}) we collect the basic gauge invariant operators. On the left side we list the operators which belong to the spectrum of $\mathcal{N}=1$ SQED, their approximate scaling dimensions are calculated using the large $N_f$ formulas obtained in the previous two sections. We also include the scaling dimension of the monopole operators $\mathfrak{M}^{\pm 1}$ in the large $N_f$ limit ($\Delta[\mathfrak{M}^{\pm 1}] = 0.3619 N_f+O(1)$), which we extract from the results of \cite{Chester:2017vdh} in appendix \ref{N=1mon}. On the right side we list the operators of the dual WZ model, their scaling dimensions are calculated using $4 -\epsilon$ expansion in \cite{Benini:2018bhk}. Using the map discussed in detail in \cite{Benini:2018bhk}, on each row the two operators map into each other under the duality. We notice a quite good agreement between the dimensions of the corresponding operators, providing a nice quantitative check of the $\cN=1$ duality.

\begin{figure}[H]
  \centering
\begin{tikzpicture}[scale=0.38]

\draw[black] (-9,1.5) to (27,1.5);
\draw[black] (-9,1.5) to (-9,-10);
\draw[black]  (-9,-10) to (27,-10);
\draw[black] (27,1.5) to (27,-10);
\draw[black] (10,1.5) to (10,-10);

\node at (0,0) {$\Delta[\mathfrak{M}^{\pm 1}] \sim 0.724$};
\node at (0,-2) {$\Delta[(\Phi^* \Phi)_{spin-1}] \sim \Big( 1- \frac{24}{3\pi^2 2}\Big)=0.595$};
\node at (0,-4) {$\Delta[(\Phi^* \Phi)_{sing}] \sim 1 $};
\node at (0,-6) {$\Delta[|\Phi|^4_{spin-2}] \sim \Big( 2- \frac{48}{3\pi^2 2}\Big)=1.19$};
\node at (0,-8) {$\Delta[|\Phi|^4_{sing}] \sim 2$};

\begin{scope}[shift={(19,0)}]
\node at (0,0) {$\Delta[M_\alpha] \sim 0.76$};
\node at (0,-2) {$\Delta[\mu_I] \sim 0.66$};
\node at (0,-4) {$\Delta[-2\sum \mu_I^2 + \sum |M_\alpha|^2] \sim 1 $};
\node at (0,-6) {$\Delta[\mu_I \mu_J -\frac{\delta_{IJ}}{3} \sum \mu_K^2 ]\sim 1.33$};
\node at (0,-8) {$\Delta[2\sum \mu_I^2 +3 \sum |M_\alpha|^2] \sim 2.33$};

\end{scope}
\end{tikzpicture}
\caption{Operator mapping across the duality and the scaling dimensions of the operators .} \label{F21}
\end{figure}

\subsection{The $\cN=1$ supersymmetric $O(N)$ sigma model and $\mathcal{N}=2$ SQED}
For completeness, we also discuss the large $N_f$ limit of the "chiral" $\cN=2$ QED, with $N_f$ flavors and $0$ anti-flavors (also denoted as $(N_f,0)$ flavors).

In $2+1$ dimensions, the $\mathcal{N}=2$ chiral multiplet has the field content 
\begin{align} \label{h16}
\Phi \ \ \Psi \ \ F \ .
\end{align}
Where the $\Phi$ is a complex scalar, and $\Psi$ is a two-component Dirac fermion and $F$ is an auxiliary complex field. The vector multiplet has the field content 
\begin{align} \label{h17}
A_\mu \ \  \sigma \ \ \lambda_1 \ \ \lambda_2 \ \ D \ .
\end{align}
Where $\sigma$ and $D$ are real scalars, the $\lambda_{1,2}$ are real two-component Majorana fermions, which usually are combined into a single two-component Dirac fermion $(\lambda_1+i \lambda_2)$. $(N_f,0)$ flavored $\mathcal{N}=2$ SQED has $N_f$ chiral multiplets (\ref{h16}) minimally coupled to a vector multiplet (\ref{h17}) with charge $+1$. One can write the action of this theory in $\mathcal{N}=1$ language. For this purpose we regroup the fields (\ref{h16}, \ref{h17}) into the following $\mathcal{N}=1$ multiplets 
\begin{align} \nonumber
&V: \ \ A_\mu \ \ \lambda_1 \\ \label{h18}
&H:  \ \sigma \ \ \lambda_2 \ \ D \\ 
& Q_i: \ \Phi^i \ \  \Psi^i \ \ F^i \ . \nonumber 
\end{align}
The $\mathcal{N}=2$ SQED action can be written as a $\mathcal{N}=1$ SQED action (\ref{h12}), plus a kinetic term for $H$ and interaction  from the superpotential 
\be \cW_{\cN=1} = H  \bar Q_i Q^i \,.\ee
Written in Lorentzian metric, the full action in components become
\begin{align} \nonumber 
S^{\mathcal{N}=2}=& \  S^{\mathcal{N}=1}+\frac{1}{2e^2} \int d^3 x \big( \partial^\mu \sigma \partial_\mu \sigma + \bar\lambda_2 i \gamma^\mu \partial _\mu \lambda_2 + D^2 \big) \\
& \ \ \ + \int d^3 x \big( -\sigma^2 \Phi^{* }_j \Phi^j + \sigma \bar\Psi_j \Psi^j + (\bar\Psi_j \lambda_2 \Phi^j + \bar\lambda_2 \Psi^j \Phi^{* }_j )+ D \Phi^{* }_j \Phi^j\big ) \ . \label{h19}
\end{align}
where the first term in the right hand side of (\ref{h19}) is defined in (\ref{h12}). The gaugino $\lambda$ in (\ref{h12}) is replaced by $\lambda_1$. We have a quartic term in the second line because we have integrated out auxiliary fields $F^j$:
\begin{align} \label{h20}
F^{*}_j F^j- \sigma (\Phi^{* }_j F^j+ \Phi^j F^{* }_j)  \ \rightarrow \ -\sigma^2 \Phi^{*}_j \Phi^j \ .
\end{align}
Performing a $1/N_f$ expansion with quartic vertix is usually more involved task than working with the cubic vertex, therefore one usually doesn't integrate out $F^j$  (\ref{h20}). However at order $O(1/N_f)$ this is not a problem, and we work with the action (\ref{h19}), in order to have less fields. The scaling dimensions of the fields sitting in the chiral multiplet $H$ at the IR fixed point are  
\begin{align} 
\Delta[\sigma]=1, \ \ \Delta[\lambda_2]=3/2,  \ \  \Delta[D]=2 \ .
\end{align}
Due to supersymmetry, these dimensions are exact in $1/N_f$ expansion. This follows from the fact that the dimension of the gauge field $A_\mu$ is exactly $1$ and the fields $(A_\mu, \sigma, \lambda_1, \lambda_2, D)$ sit in the same vector multiplet. The operator $(\Phi^{* } \Phi_{adj})$ has a scaling dimension exactly equal to $1$, since it sits in the same $\cN=2$ supermultiplet of the flavor $SU(N_f)$ currents. 

These observations allow us to check our results for the singlet and adjoint operator dimensions obtained in $\mathcal{N}=1$ SQED. For this purpose first we notice that the action (\ref{h19}) without the first term is the $\mathcal{N}=1$ supersymmetric $O(N)$ sigma model \cite{Gracey:1990ac}. The large $N_f$ scaling dimensions of the field $\sigma$ and of the bilinear adjoint operator for this model model have been computed in \cite{Gracey:1990ac} (see \cite{Benini:2018umh} for a finite-$N_f$ study in the $4-\epsilon$ expansion):
\begin{align} \label{h21}
&\Delta[|\Phi|^2_{adj}]_{\cN=1 \, O(N)}=1+\frac{24}{3\pi^2 N_f }+O(1/{N_f^2}) \\
&\Delta[\sigma]_{\cN=1 \, O(N)}=1+O(1/{N_f^2})  \ .  \label{h22} 
\end{align}
 For both of these operators the list of possible diagrams contributing to the 2-point correlation functions in $\mathcal{N}=2$ SQED are exhausted by the lists given in the context of $\mathcal{N}=1$ SQED and $\mathcal{N}=1$ supersymmetric sigma model (if one goes to the order $O(1/N_f^2)$ there might be diagrams with propagators present from both multiplets $V$ and $H$).  
  Therefore the sum of this contributions should be such that anomalous scaling dimensions for $|\Phi|^2_{adj}$ and $\sigma$ are exactly zero. As one can see from (\ref{h21}, \ref{h22}) and (\ref{h23}, \ref{h24}) this is true. 
 
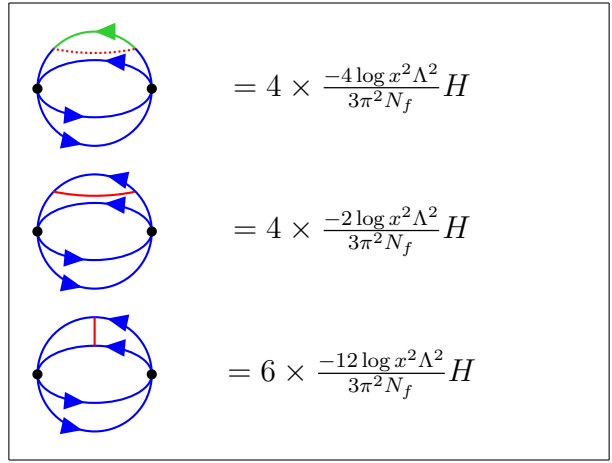
\begin{figure}[H]
  \centering
\begin{tikzpicture}[scale=0.38]

\draw[black] (-3,3) to (18,3);
\draw[black] (-3,3) to (-3,-13);
\draw[black] (18,3) to (18,-13);
\draw[black] (-3,-13) to (18,-13);

\begin{scope}[shift={(0,0)}]
\draw[scalar] (0,0) (0:2) arc (0:180:2 and 1);
\draw[scalar] (0,0) (-180:2) arc (-180:0:2 and 1);
\draw[thick,blue] (0,0) (0:2) arc (0:45:2);
\draw[fermion] (0,0) (45:2) arc (45:135:2);
\draw[thick,blue] (0,0) (135:2) arc (135:180:2);
\draw[scalar] (0,0) (-180:2) arc (-180:0:2);
\draw[thick, red, densely dotted] (0,3.35) [partial ellipse=-69:-112:3.9cm and 2.1cm];
\draw[thick,fill] (-2,0) circle (4pt);
\draw[thick,fill] (2,0) circle (4pt);
\node at (9,0) {$=4\times  \frac{ -4\log {x^2\Lambda^2} }{  3 \pi^2 N_f } H$};
\end{scope}

\begin{scope}[shift={(0,-5)}]
\draw[scalar] (0,0) (0:2) arc (0:180:2 and 1);
\draw[scalar] (0,0) (-180:2) arc (-180:0:2 and 1);
\draw[scalar] (0,0) (0:2) arc (0:180:2);
\draw[scalar] (0,0) (-180:2) arc (-180:0:2);
\draw[thick, red] (0,3.35) [partial ellipse=-69:-112:3.9cm and 2.1cm];
\draw[thick,fill] (-2,0) circle (4pt);
\draw[thick,fill] (2,0) circle (4pt);
\node at (9,0) {$=4\times  \frac{ -2 \log {x^2\Lambda^2} }{  3 \pi^2 N_f } H$};
\end{scope}

\begin{scope}[shift={(0,-10)}]
\draw[scalar] (0,0) (0:2) arc (0:180:2 and 1);
\draw[scalar] (0,0) (-180:2) arc (-180:0:2 and 1);
\draw[scalar] (0,0) (0:2) arc (0:180:2);
\draw[scalar] (0,0) (-180:2) arc (-180:0:2);
\draw[thick,red ] ([shift=(90:2)]0,0) to ([shift=(90:2 and 1)]0,0);
\draw[thick,fill] (-2,0) circle (4pt);
\draw[thick,fill] (2,0) circle (4pt);
\node at (9,0) {$=6\times  \frac{ -12  \log {x^2\Lambda^2} }{ 3 \pi^2 N_f } H$};
\end{scope}

\end{tikzpicture}
\caption{($\mathcal{N}=2$ SQED) quartic adjoint-2 operator renormalization. Dotted red line stands for the effective gaugino $\lambda_2$ propagator. Thick red line stands for effective D-field propagator.} \label{F19}
\end{figure}

 Finally, for the completeness, we compute the scaling dimensions of the scalar mesonic operators in $\mathcal{N}=2$ SQED, which at leading order have dimension 2, but are not protected.  One such operator is the quartic adjoint-2: $|\Phi|^4_{adj-2}$, to calculate its scaling dimension one uses graphs in the tables (\ref{F18}, \ref{F19}). 
 The operator $\sigma^2$ has its scaling dimension twice the scaling dimension of $\sigma$ field (which is exactly equal to 1) plus the contributions of the last two graphs in the table (\ref{F12}). The dimension of $\sigma |\Phi|^2_{adj}$ equals to the sum of: $\Delta[|\Phi|^2_{adj}]=1$, $\Delta[\sigma]=1$, plus the contribution of the graph (\ref{F20}). The final results are
  \begin{align}
 &\Delta[|\Phi|^4_{adj-2}] = 2 + \frac{48}{3\pi^2 N_f} + O(1/N_f^2) \\
 & \Delta[\sigma^2] =  2 + \frac{48}{3\pi^2 N_f} + O(1/N_f^2) \\
 &\Delta[ \sigma  |\Phi|^2_{adj}] =  2 + \frac{48}{3\pi^2 N_f} + O(1/N_f^2) \ .
 \end{align}

 \begin{figure}[H]
 \centering
\begin{tikzpicture}[scale=0.58]

\draw[blue,thick,  postaction={decorate}, decoration={markings, mark=at position .53 with {\arrow[blue]{triangle 45}}}] (0,0) (-180:2 and 1.7) arc (-180:0:2 and 1.7 );
\draw[blue,thick,  postaction={decorate}, decoration={markings, mark=at position .53 with {\arrow[blue]{triangle 45}}}] ([shift=(0:1 and 0.7)]-1,0)   arc (0:-180:1 and 0.7);
\draw[blue,thick,  postaction={decorate}, decoration={markings, mark=at position .53 with {\arrow[blue]{triangle 45}}}] ([shift=(0:1 and 0.7)],0)   arc (0:-180:1 and 0.7);
\draw[red,dashed, thick] ([shift=(0:1 and 0.7)]-1,0)   arc (0:180:1 and 0.7);
\draw[red, dashed,thick] ([shift=(0:1 and 0.7)],0)   arc (0:180:1 and 0.7);
\draw[thick,fill] (-2,0) circle (4pt);
\draw[thick,fill] (2,0) circle (4pt);
\node at (8,0) {$= \frac{-48 \log{x^2\Lambda^2}}{3 \pi^2 N_f} \times \Big( \frac{-i}{N_f (2\pi^2|x|^2)^2} \Big)$};

 \end{tikzpicture}
\caption{($\mathcal{N}=2$ SQED) $\sigma|\Phi|^2_{adj}$ operator renormalization. Dashed line stands for effective $\sigma$ field propagator.} \label{F20}
 \end{figure}

\acknowledgments{We are grateful to Francesco Benini, Pasquale Calabrese, Sara Pasquetti and Silviu Pufu for useful discussions, and to John Gracey for correspondence. This work is supported in part by the MIUR-SIR grant RBSI1471GJ ``Quantum Field Theories at Strong Coupling: Exact Computations and Applications". S.B. is partly supported by the INFN Research Projects GAST and ST$\&$FI. }

\appendix

\section{Bosonic QED's in the $4-\epsilon$ expansion} \label{sec:eps}

We consider the bosonic QED's with global symmetry $SU(N_f/2)\times SU(N_f/2) \times U(1)$ in  dimension $d=4-\epsilon$ 
\begin{align} \nonumber 
\CL=&\frac{1}{4} F^{\mu\nu}F_{\mu\nu}+ \sum_{i=1}^{N_f/2} \overline{D^\mu \Phi_i }D _\mu \Phi^i+ \sum_{i=1}^{N_f/2} \overline{D^\mu \tilde\Phi_i }D _\mu \tilde\Phi^i+\lambda_{ep} \Big(\big(\sum_{i=1}^{N_f/2} |\Phi^i|^2 \big)^2+\big(\sum_{i=1}^{N_f/2} |\tilde\Phi^i|^2 \big)^2\Big)\\
&+\lambda \sum_{i=1}^{N_f/2} |\Phi^i|^2 \sum_{j=1}^{N_f/2} |\tilde\Phi^j|^2 + \text{(gauge fixing term)} \ .
\end{align}
Where the $D_\mu=\partial_\mu+i e A_\mu$, and the kinetic term for the photon is canonically normalized. The beta functions of the gauge and quartic couplings at one loop order are
\begin{align} \label{betaf1}
&\beta_e= \frac{de}{dl}=\frac{\epsilon e}{2}-\frac{1}{(4\pi)^2} \frac{2N_f e^3}{6}\\
&\beta_{\lambda_{ep}}\! =\! \frac{d \lambda_{ep}}{dl}\! =\! \epsilon \lambda_{ep}-\frac{1}{(4\pi)^2} \big[ 8(N_f+8) \lambda_{ep}^2 +2N_f \lambda^2 -12\lambda_{ep} e^2 +\frac{3}{2} e^4 \big] \label{betaf2} \\
&\beta_{\lambda}=\frac{d\lambda}{dl}=\epsilon \lambda-\frac{1}{(4\pi)^2} \big( 16 \lambda^2+16(N_f+2) \lambda \lambda_{ep}-12 \lambda e^2+3 e^4 \big) \label{betaf3} \ .
\end{align}
The beta function of the gauge coupling has two zeroes. One trivial zero is when the gauge coupling vanishes, then we have the ungauged $O(N_f)\times O(N_f)$ vector model. See \cite{Benvenuti:2018cwd, Calabrese:2002bm} for discussions about the ungauged fixed points and RG flow. The other zero is when
\begin{align}
e^2=\frac{24 \pi^2}{N_f} \epsilon \ .
\end{align}
Plugging this value into the beta functions (\ref{betaf2}, \ref{betaf3}) we generically expect to find four fixed points. There are two fixed points (which we identify with bQED and bQED$_+$ discussed in the main text) for which $\lambda=2\lambda_{ep}$ with a global symmetry group $SU(N_f)$. The values of the quartic couplings at those fixed points are 
\begin{align} \label{fixedpoint}
\lambda = 2\lambda_{ep}=\frac{N_f+18\pm \sqrt{N_f^2-180N_f-540}}{N_f(N_f+4)} \pi^2 \epsilon \ .
\end{align}
It follows from (\ref{fixedpoint}) that for $N_f<182.95$ the quartic couplings become complex.  Similarly one writes solutions for the remaining two fixed points (which we identify with ep-bQED and bQED$_-$ discussed in the main text), for which global symmetry is $SU(N_f/2)\times SU(N_f/2)$:
\begin{align} \label{fixedpointep}
&\lambda= \frac{-(N+18)(N-4) \mp \sqrt{\mathcal{D}}}{N(N^2+8)}\pi^2 \epsilon \\
&\lambda_{ep} =\frac{288+160N+62N^2+3N^3 \pm (4-N)\sqrt{\mathcal{D}}}{2N(N+8)(N^2+8)} \pi^2 \epsilon \ . \label{fixedpointep1}
\end{align}
Where we defined the discriminant 
\begin{align} \label{discrim}
\mathcal{D}=N^4-188N^3-1676N^2-3744N-8640
\end{align}
It follows from (\ref{discrim}) that at these two fixed points the quartic couplings become complex when $N_f<196.22$.  

We provide the expansions of solutions (\ref{fixedpoint}, \ref{fixedpointep}, \ref{fixedpointep1}) in the large N$_f$ limit ($\epsilon=1$)
\begin{align}
& \text{bQED (tricritical): \  }\lambda_{ep}= \frac{54\pi^2}{N_f^2}+\frac{1944 \pi^2}{N_f^3}+O\Big (\frac{1}{N_f^4}\Big),   \ \lambda= 2\lambda_{ep}\\
& \text{bQED$_+$: \ } \lambda_{ep}= \frac{\pi^2}{N_f}-\frac{40\pi^2}{N_f^2}-\frac{2000 \pi^2}{N_f^3}+O\Big(\frac{1}{N_f^4}\Big),  \ \lambda=2 \lambda_{ep}\\
& \text{bQED$_-$: \ } \! \! \lambda_{ep}\!=\! \frac{\pi^2}{N_f}\!+\!\frac{72\pi^2}{N_f^2}\!+\!\frac{1936 \pi^2}{N_f^3}\!+\!O\Big(\! \frac{1}{N_f^4}\! \Big), \!  \!  \ \lambda\!=\! -\frac{2\pi^2}{N_f}\!+\!\frac{80\pi^2}{N_f^2}\!+\!\frac{5344\pi^2}{N_f^3}\!+\!O\! \Big(\!\frac{1}{N_f^4}\! \Big)\\
& \text{ep-bQED: \  } \! \lambda_{ep}\!=\! \frac{2\pi^2}{N_f}\!-\!\frac{34\pi^2}{N_f^2}\!-\!\frac{2104 \pi^2}{N_f^3}\!+\!O\Big(\! \frac{1}{N_f^4}\! \Big), \!  \ \lambda\!=\!-\frac{108\pi^2}{N_f^2}\!-\! \frac{5184\pi^2}{N_f^3}\!+\!O\! \Big( \!\frac{1}{N_f^4}\!\Big) \ .
\end{align}
Notice that at the tricritical point: $\lambda \sim 1/N_f^2$, while in bQED$_+$: $\lambda \sim 1/N_f$. Similarly, in bQED$_-$: $\lambda,  \lambda_{ep} \sim 1/N_f$, while in ep-bQED: $\lambda_{ep}\sim 1/N_f$, $\lambda \sim 1/N_f^2$. This justifies our identification with the four fixed points discussed at large $N_f$ in the main text.

\begin{figure}[H]
\caption{RG flow in scalar QED: $N_f=250, \ e^2=\frac{ 24\pi^2}{N_f}$ ($\epsilon=1$) . }
\centering
\includegraphics[width=0.7\textwidth]{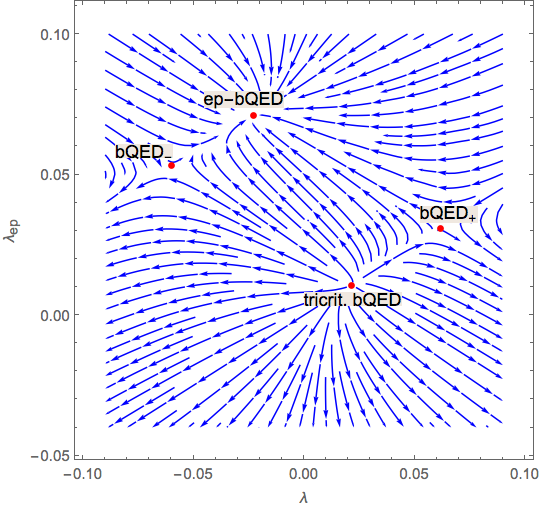} \label{flow}
\end{figure}

The figure (\ref{flow}) is an example of RG flow and fixed points in the  space of quartic couplings.

\section{Useful formulae} \label{usefulformulae}

\begin{align}
  G(x)= \int \frac{d^d p} {(2\pi)^d} e^{-ipx}  G(p) \label{a1} 
\end{align}
\begin{align}
 G(p) = \int \frac{d^d x}{1} e^{ipx} G(x) \label{a2}
\end{align}
\begin{align} \label{a3}
\frac{1}{|x|^{2\alpha}} = \frac{\Gamma(\frac{d}{2}-\alpha)}{ \pi^{\frac{d}{2}} 2^{2\alpha} \Gamma(\alpha)} \int d^d p \frac{e^{-i px}}{|p|^{d-2 \alpha}}
\end{align}
\begin{align} \label{a9}
\frac{x_\mu}{|x|^{2(\alpha+1)}} = \frac{\Gamma(\frac{d}{2}-\alpha)}{\pi^{\frac{d}{2}} 2^{2\alpha+1} \alpha \Gamma(\alpha) } \int d^d p \frac{ e^{-ipx} i p_\mu}{|p|^{d-2\alpha}}
\end{align}
\begin{align} \label{a4}
 \int \frac{d^3 q}{(2 \pi)^3} \frac{1}{q^2(q+p)^2} = \frac{1}{8|p|} 
\end{align}
\begin{align} \label{a5}
 \int \frac{d^3 q}{(2 \pi)^3} \frac{q_\mu}{q^4(q+p)^2} = -\frac{p _\mu}{16 |p|^3} 
\end{align}

\section{Feynman graphs} \label{sec:diag}
In this appendix we give some examples of computation of the Feynman diagrams, using an approach similar for instance to \cite{Chester:2016ref}. We read the graphs using position space Feynman rules, then we identify the region from where UV logarithmic divergences appear.  

  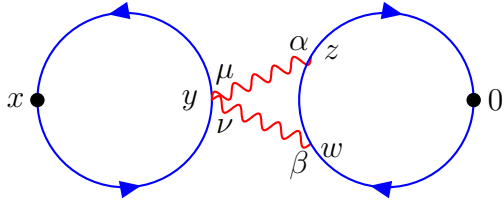
\begin{figure}[H]
  \centering
\begin{tikzpicture}[scale=0.58]

\draw[thick,blue,postaction={decorate}, decoration={markings, mark=at position .56 with {\arrow[blue]{triangle 45}}}]  ([shift=(0:2cm)]0,-2)  arc (0:180:2cm);
\node at (-2.5,-2) {$x$};
\node at (-2.5+4,-2) {$y$};
\node at (-2.5+4.8,-2+0.6) {$\mu$};
\node at (-2.5+4.8,-2-0.6) {$\nu$};

\node at (-2.5+7.25,-2+1.1) {$z$};
\node at (-2.5+7.25,-2-1.15) {$w$};
\node at (-2.5+11,-2) {$0$};
\node at (-2.5+6.5,-2+1.3) {$\alpha$};
\node at (-2.5+6.5,-2-1.5) {$\beta$};

\draw[thick,blue,postaction={decorate},decoration={markings, mark=at position .56 with {\arrow[blue]{triangle 45}}}]  ([shift=(-180:2cm)]0,-2) arc (-180:0:2cm);
\draw[photon, red] (0,-2) ([shift=(0:2cm)]0,-2)  to   ([shift=(-210:2cm)]6,-2);
\draw[photon, red] (0,-2) ([shift=(0:2cm)]0,-2)  to   ([shift=(-150:2cm)]6,-2);
\draw[thick,blue ,postaction={decorate}, decoration={markings, mark=at position .56 with {\arrow[blue]{triangle 45}}}] ([shift=(180:2cm)]6,-2) arc (180:0:2cm);
\draw[thick,blue,postaction={decorate}, decoration={markings, mark=at position .56 with {\arrow[blue]{triangle 45}}}] ([shift=(0:2cm)]6,-2) arc (0:-180:2cm);
\draw[thick,fill] (-2,-2) circle (4pt);
\draw[thick,fill] (8,-2) circle (4pt);

  \end{tikzpicture}
\caption{ Symmetry factor is 4 .}  \label{A1}
  \end{figure}
  
  The Feynman graph (\ref{A1}) using the Feynman rules  (\ref{F2}) can be read as follows 
  \begin{align} \nonumber
\text{Graph} \ (\ref{A1}) =   4 N_f\int d^3 y d^3 z d^3 w &\Big( \frac{1}{4\pi |x-y|}\Big)^2 (-\delta_{\mu\nu})\frac{8\delta_{\mu\alpha}}{N_f \pi^2 |y-z|^2} \frac{8\delta_{\nu\beta}}{N_f \pi^2 |y-w|^2} \\
  & \times  \Big [ \frac{1}{4\pi |w|} i \overset{\leftrightarrow}{\partial_\beta^w}  \frac{1}{4\pi|w-z|} i \overset{\leftrightarrow}{\partial_\alpha^z}\frac{1}{4\pi |z|} \Big ] \ . \label{a6}
  \end{align}
  where the 4 is the symmetry factor of the graph (\ref{A1}). Each blue loop in the graph gives a factor $N_f$, and since we normalized singlet bilinear operator as follows $\frac{1}{\sqrt{N_f}} \sum_{i=1}^{N_f} |\Phi_i|^2$ we also get a factor $1/N_f$. After cancelation we obtain the factor $N_f$ in (\ref{a6}). Also we choose to work in the $\xi=0$ gauge since the graph (\ref{A1}) turns to be independent from the choice of the gauge parameter.  
  
The logarithmic divergences come from the region where $y,z,w$ are close to 0. 
  \begin{align} \nonumber
\text{Graph} \ (\ref{A1})\! = &  \frac{4N_f}{(4\pi |x|)^2} \! \! \int \! d^3 y d^3 z d^3 w\! \frac{8}{N_f \pi^2 |y-z|^2} \frac{8}{N_f \pi^2 |y-w|^2} \! \Big [ \frac{1}{4\pi |w|} i \overset{\leftrightarrow}{\partial_\mu^w}  \frac{1}{4\pi|w-z|} i \overset{\leftrightarrow}{\partial_\mu^z}\frac{1}{4\pi |z|} \! \Big ] \\
 =&  - \frac{4N_f}{(4\pi |x|)^2} \Big(\frac{16}{N_f}\Big)^2 \int \frac{d^3p}{(2\pi)^3} \int \frac{d^3q}{(2\pi)^3} \frac{(p+q)^2}{p^4 q^2 (q-p)^2} \ .
  \end{align}
Where in the last line we performed Fourier transformation (\ref{a1}, \ref{a2}, \ref{a3}) to pass to the momentum space. First we perform integration over momentum $p$ using formulas (\ref{a4}, \ref{a5}) and we obtain 
\begin{align}
\text{Graph} \ (\ref{A1})=- \frac{4N_f}{(4\pi |x|)^2} \Big(\frac{16}{N_f}\Big)^2  \int \frac{d^3 q}{(2\pi)^3} \frac{1}{4 |q|^3} \ .
\end{align}
The integral over $q$ is logarithmically divergent. We regularize it by putting a UV cut-off $\Lambda$. The final result is
\begin{align}
\text{Graph} \ (\ref{A1})=4\times \frac{-16 \log (x^2 \Lambda^2)}{\pi^2 N_f}\Big( \frac{1}{4\pi |x|}\Big)^2 \ .
\end{align}

  \begin{figure}[H]
  \centering
\begin{tikzpicture}[scale=0.58]

\draw[thick,mygreen] ([shift=(180:2cm)]0,0)   arc (180:0:2cm);
\draw[thick,mygreen] ([shift=(0:2cm)]0,0)   arc (0:-180:2cm);
\draw[dashed,red,thick] (0,0) ([shift=(90:2cm)]0,0)  to   ([shift=(-90:2cm)]0,0);
\draw[fill]  (-2,0) circle (5pt);
\draw[fill]  (2,0) circle (5pt);
\node at (-2.5,0) {$x$};
\node at (2.5,0) {$0$};
\node at (0,2.5) {$y$};
\node at (0,-2.5) {$z$};
  \end{tikzpicture}
\caption{ Symmetry factor is 1 .}  \label{A3}
  \end{figure}
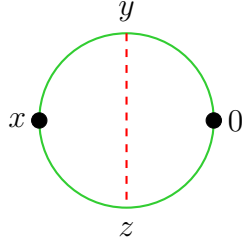

The Feynman graph (\ref{A3}) corresponds to the following expression 
\begin{align} \label{a8}
\text{Graph} \ (\ref{A3})=-\int d^3 yd^3 z \Tr \Big[ \frac{\slashed x - \slashed y}{4\pi |x-y|^3}   \frac{ \slashed y}{4\pi |y|^3}  \frac{ -\slashed z}{4\pi |z|^3}  \frac{\slashed z - \slashed x}{4\pi |z-x|^3}  \Big] \frac{4}{\pi^2 N_f |y-z|^2} \ .
\end{align}
Where the minus sign stands for the fermion loop. The logarithmic divergence of the integral (\ref{a8}) comes from the regions where $y,z$ are close either to 0 or to x.  We will consider the region $y,z$ close to 0 and multiply the answer by 2, since the other region gives the same contribution. 
\begin{align}
\text{Graph} \ (\ref{A3}) =-2 \Big( \frac{\slashed x}{4\pi |x|^3} \Big)^2 \int d^3y d^3z  \Tr \Big[  \frac{ \slashed y}{4\pi |y|^3}  \frac{ \slashed z}{4\pi |z|^3} \Big] \frac{4}{\pi^2 N_f |y-z|^2} \ .
\end{align}
Now we pass to the momentum space using (\ref{a1}, \ref{a2}, \ref{a3}, \ref{a9}). 
\begin{align}
\text{Graph} \ (\ref{A3})=-2 \Big( \frac{\slashed x}{4\pi |x|^3} \Big)^2 \int \frac{d^3 p}{(2\pi)^3} \Tr \Big[\frac{i \slashed p}{p^2} \frac{-i \slashed p}{p^2} \Big] \frac{8}{N_f |p|} \ .
\end{align}
This integral is logarithmically divergent. We regularize with a UV cutoff $\Lambda$ .
\begin{align}
\text{Graph} \ (\ref{A3})=\frac{4 \log{x^2 \Lambda^2}}{\pi^2 N_f}\times 2 \Big( \frac{\slashed x}{4\pi |x|^3} \Big)^2  \ .
\end{align}

\begin{figure}[H]
\centering
\begin{tikzpicture}[scale=0.58]
\node at (-1+0.3,-0) {$x$};
\draw[dashed ,thick, red] (0,0) to (2,2);
\node at (2,2.5) {$y$};
\draw[dashed ,thick, red] (0,0) to (2,-2);
\node at (2,-2-0.5) {$v$};
\draw[fill, red] (0,0) circle (5pt);
\draw[mygreen, thick] (2,2) to (2,-2);
\draw[mygreen, thick] (2,-2) to (5,-2);
\draw[mygreen, thick] (2,2) to (5,2);
\node at (5,2.5) {$z$};
\node at (5,-2.5) {$w$};
\draw[mygreen, thick] (5,2) to (5,-2);
\draw[dashed ,thick, red] (5,2) to (7,0);
\draw[dashed ,thick, red] (5,-2) to (7,0);
\draw[fill, red] (7,0) circle (5pt);
\node at (7+0.7,0) {$0$};

  \end{tikzpicture}
\caption{ Symmetry factor is 4. }  \label{A2}
  \end{figure}

Using the Feynman rules we can read the graph (\ref{A2})  as follows 
\begin{align} \nonumber 
& \text{Graph}  \ (\ref{A2})= -4N_f \int d^3 y d^3z d^3 w d^3 v \frac{4}{\pi^2 N_f |x-y|^2} \frac{4}{\pi^2 N_f |x-v|^2} \\
& \ \ \ \ \ \  \times \Tr \Big[ \frac{(\slashed y-\slashed z)}{4\pi |y-z|^3} \frac{(\slashed z-\slashed w)}{4\pi |z-w|^3}  \frac{(\slashed w-\slashed v)}{4\pi |w-v|^3} \frac{(\slashed v-\slashed y)}{4\pi |v-y|^3} \Big] \frac{4}{\pi^2 N_f |z^2|} \frac{4}{\pi^2 N_f |w^2|} \ . \label{a7}
\end{align}
The minus stands for the fermion (green) loop in (\ref{A2}), factor $N_f$  comes from summing over the number of fermion flavors in the same loop, $4$ is the symmetry factor of the graph. The logarithmic divergences of the integral (\ref{a7}) come from the regions where $y,z,v,w$ are close either to 0 or to x. Let us inspect the region $y,z,v,w$ close to 0 and multiply the answer by 2, since it is obvious that the other region gives the same logarithmic divergence. 
\begin{align} \nonumber
 \text{Graph}  \ (\ref{A2}) = &2\times(-4N_f) \Big(\frac{4}{\pi^2 N_f |x^2|} \Big)^2 \int d^3 y d^3z d^3 w d^3 v \frac{4}{\pi^2 N_f |z|^2} \frac{4}{\pi^2 N_f |w|^2}\\
&\ \ \ \ \ \ \times \Tr \Big[ \frac{(\slashed y-\slashed z)}{4\pi |y-z|^3} \frac{(\slashed z-\slashed w)}{4\pi |z-w|^3}  \frac{(\slashed w-\slashed v)}{4\pi |w-v|^3} \frac{(\slashed v-\slashed y)}{4\pi |v-y|^3} \Big] \ .
\end{align}
Now using (\ref{a1}, \ref{a2}, \ref{a3}, \ref{a9}) we perform a Fourier transformation to the momentum space. 
\begin{align}
 \text{Graph}  \ (\ref{A2}) = 2\Big(\frac{4}{\pi^2 N_f |x^2|} \Big)^2 \times \frac{-8^2 \cdot 4}{N_f} \int \frac{d^3 p}{(2\pi)^3}  \int \frac{d^3 q}{(2\pi)^3}  \frac{2 q(p+q)}{p^2 q^2 (p+q)^4} \ .
\end{align}
First one performs integral over the momentum $q$
\begin{align}
 \text{Graph}  \ (\ref{A2}) = 2\Big(\frac{4}{\pi^2 N_f |x^2|} \Big)^2 \times \frac{-8^2 \cdot 4}{N_f} \int \frac{d^3 p}{(2\pi)^3}   \frac{1}{8 |p|^3} \ .
\end{align}
The integral over $p$ is logarithmically divergent. We regularize it by putting a UV cutoff $\Lambda$ and perform integration over the $p$. The final answer is
\begin{align}
 \text{Graph}  \ (\ref{A2}) =  \frac{- 8 \log{x^2\Lambda^2}}{\pi^2 N_f}  \times 2\Big(\frac{4}{\pi^2 N_f |x^2|} \Big)^2  \ .
\end{align}

\section{Scaling dimensions of monopole operators in $\mathcal{N}=1$ SQED} \label{N=1mon}
The scaling dimensions of the monopole operators $\M^{2q}$ with topological charge $2q$ ($2q$ is an integer) in $\mathcal{N}=1$ SQED, at the leading order in the large $N_f$ expansion, have been computed in (\cite{Chester:2017vdh}). We use formula $2.59$ of (\cite{Chester:2017vdh})
 \begin{align} \label{h28}
 \frac{\Delta[\mathfrak{M}^{2q}]}{N_f}= \sum_{j \geq q-1/2} (2j+1) \sqrt{(j+1/2)^2-q^2}-   \widehat{\sum_{j \geq q-1/2}} (2j+1) \sqrt{(j+1/2)^2-q^2} \ .
 \end{align}
 where in the first sum $j \geq q-1/2$ runs over the values for which $(j-q)$ is a non-negative integer, while in the second sum $j \geq q-1/2$ runs over the values for which $(j-q-1/2)$ is a non-negative integer. Both sums are divergent, since for large values of $j$ the expressions under the sum scale like $j^2$. We follow the approach of (\cite{Chester:2017vdh}) to regularize the sums and extract the scaling dimensions of monopole operators. First we shift the power of the energy mode as follows
 \begin{align} 
 \Big( (j+1/2)^2-q^2 \Big)^{\frac{1}{2}} \rightarrow \Big( (j+1/2)^2-q^2 \Big)^{\frac{1}{2}-s} \ .
 \end{align}
 It is clear that by choosing large values for $s$ one makes the sum (\ref{h28}) convergent. Next we add and subtract quantities that are divergent when $s=0$
 \begin{align} \nonumber 
 \frac{\Delta[\mathfrak{M}^{2q}]}{N_f}\!=&\displaystyle{\lim_{s \to 0}}\!\! \! \sum_{j \geq q-1/2} \!\big[ (2j+1) \! \big( (j+1/2)^2\!-\!q^2 \big)^{\frac{1}{2}-s} \!-\!2(j+1/2)^{2-2s}\!+\!q^2 (1-2s) (j\!+\!1/2)^{-2s}\big]\\
 -&\displaystyle{\lim_{s \to 0}} \! \!  \sum_{j \geq q-1/2} \big[  -2(j+1/2)^{2-2s}+q^2 (1-2s) (j+1/2)^{-2s}\big] \nonumber \\
 -&\displaystyle{\lim_{s \to 0}} \! \! \! \widehat{\sum_{j \geq q-1/2}} \! \big[ (2j+1) \! \big( (j+1/2)^2\!-\!q^2 \big)^{\frac{1}{2}-s}\! -\!2(j+1/2)^{2-2s}+q^2 (1-2s) (j\!+\!1/2)^{-2s}\big] \nonumber \\
 +&\displaystyle{\lim_{s \to 0}} \! \! \widehat{\sum_{j \geq q-1/2}} \big[  -2(j+1/2)^{2-2s}+q^2 (1-2s) (j+1/2)^{-2s}\big] \ . \label{h29}
 \end{align}
 Notice that the first and the third lines of (\ref{h29}) are convergent, this is true since for large values of $j$ the expressions under sum scale like $1/j^2$.  One can evaluate them in the limit $s \rightarrow 0$ numerically. The second and forth lines are divergent and one needs to regularise them using zeta functions. Finally we obtain
 \begin{align} \nonumber
  \frac{\Delta[\mathfrak{M}^{2q}]}{N_f}=\sum_{j \geq q-1/2}  & \big[ (2j+1) \sqrt{ (j+1/2)^2-q^2 } -2(j+1/2)^2+q^2\big] + \frac{q(1+2q^2)}{6}\\
-  \widehat{\sum}_{j \geq q-1/2} & \big[ (2j+1) \sqrt{ (j+1/2)^2-q^2 } -2(j+1/2)^2+q^2 \big]-\frac{q(q+2)(2q-1)}{6} \ . \label{h30}
 \end{align}
Using (\ref{h30}) one evaluates scaling dimensions of monopole operators with charges $(\pm 1,\pm 2, \pm 3, \pm 4)$ as follows 
\begin{align} \label{h31}
&\frac{\Delta[\mathfrak{M}^{\pm 1}]}{N_f}= 0.3619 +O(1/N_f) \\ 
&\frac{\Delta[\mathfrak{M}^{\pm 2}]}{N_f}= 0.8996+O(1/N_f)  \\
&\frac{\Delta[\mathfrak{M}^{\pm 3}]}{N_f}= 1.5708+O(1/N_f) \\
&\frac{\Delta[\mathfrak{M}^{\pm 4}]}{N_f}=2.3534+O(1/N_f) \ .
\end{align}

\bibliographystyle{ytphys}

\begin{thebibliography}{100}

\bibitem{Benvenuti:2018cwd}
  S.~Benvenuti and H.~Khachatryan,
  ``QED's in $2{+}1$ dimensions: complex fixed points and dualities,''
  arXiv:1812.01544 [hep-th].
  
\bibitem{QCP1}
  T.~Senthil, A.~Vishwanath, L.~Balents, S.~Sachdev, M.P.A.~Fisher,
  ``"Deconfined" quantum critical points,''
  Science 303, 1490 (2004)
doi:10.1126/science.1091806
[arXiv:cond-mat/0311326 [cond-mat.str-el]].
  
\bibitem{QCP2}
T.~Senthil, L.~Balents, S.~Sachdev, A.~Vishwanath, and M.P. A.~Fisher,
``Quantum criticality beyond the Landau-Ginzburg-Wilson paradigm,''
Phys. Rev. B 70, 144407
doi:10.1103/PhysRevB.70.144407
[arXiv:cond-mat/0312617 [cond-mat.str-el]].
  
\bibitem{Motrunich:2003fz}
  O.~I.~Motrunich and A.~Vishwanath,
  Phys.\ Rev.\ B {\bf 70} (2004) 075104
  doi:10.1103/PhysRevB.70.075104
  [cond-mat/0311222].

\bibitem{Gorbenko:2018ncu} 
  V.~Gorbenko, S.~Rychkov and B.~Zan,
  ``Walking, Weak first-order transitions, and Complex CFTs,''
  JHEP {\bf 1810}, 108 (2018)
  doi:10.1007/JHEP10(2018)108
  [arXiv:1807.11512 [hep-th]].
  
\bibitem{Kubota:2001kk}
  K.~I.~Kubota and H.~Terao,
  ``Dynamical symmetry breaking in QED(3) from the Wilson RG point of view,''
  Prog.\ Theor.\ Phys.\  {\bf 105} (2001) 809
  doi:10.1143/PTP.105.809
  [hep-ph/0101073].

\bibitem{Kaveh:2004qa} 
  K.~Kaveh and I.~F.~Herbut,
  ``Chiral symmetry breaking in QED(3) in presence of irrelevant interactions: A Renormalization group study,''
  Phys.\ Rev.\ B {\bf 71}, 184519 (2005)
  doi:10.1103/PhysRevB.71.184519
  [cond-mat/0411594].
  
\bibitem{Herbut:2016ide} 
  I.~F.~Herbut,
  ``Chiral symmetry breaking in three-dimensional quantum electrodynamics as fixed point annihilation,''
  Phys.\ Rev.\ D {\bf 94}, no. 2, 025036 (2016)
  doi:10.1103/PhysRevD.94.025036
  [arXiv:1605.09482 [hep-th]].
  
\bibitem{Gusynin:2016som} 
  V.~P.~Gusynin and P.~K.~Pyatkovskiy,
  ``Critical number of fermions in three-dimensional QED,''
  Phys.\ Rev.\ D {\bf 94}, no. 12, 125009 (2016)
  doi:10.1103/PhysRevD.94.125009
  [arXiv:1607.08582 [hep-ph]].

\bibitem{Kotikov:2019rww} 
  A.~V.~Kotikov and S.~Teber,
  ``Addendum to "Critical behaviour of ($2+1$)-dimensional QED: $1/N_f$-corrections in an arbitrary non-local gauge",''
  arXiv:1902.03790 [hep-th].
 
\bibitem{Giombi:2015haa} 
  S.~Giombi, I.~R.~Klebanov and G.~Tarnopolsky,
  ``Conformal QED$_d$, $F$-Theorem and the $\epsilon$ Expansion,''
  J.\ Phys.\ A {\bf 49}, no. 13, 135403 (2016)
  doi:10.1088/1751-8113/49/13/135403
  [arXiv:1508.06354 [hep-th]].
  
    
\bibitem{Pisarski:1984dj} 
  R.~D.~Pisarski,
  ``Chiral Symmetry Breaking in Three-Dimensional Electrodynamics,''
  Phys.\ Rev.\ D {\bf 29}, 2423 (1984).
  doi:10.1103/PhysRevD.29.2423

   
\bibitem{DiPietro:2015taa} 
  L.~Di Pietro, Z.~Komargodski, I.~Shamir and E.~Stamou,
  ``Quantum Electrodynamics in d=3 from the $\epsilon$-Expansion,''
  Phys.\ Rev.\ Lett.\  {\bf 116}, no. 13, 131601 (2016)
  doi:10.1103/PhysRevLett.116.131601
  [arXiv:1508.06278 [hep-th]].
    
\bibitem{DiPietro:2017kcd} 
  L.~Di Pietro and E.~Stamou,
  ``Scaling dimensions in QED$_3$ from the $\epsilon$-expansion,''
  JHEP {\bf 1712}, 054 (2017)
  doi:10.1007/JHEP12(2017)054
  [arXiv:1708.03740 [hep-th]].
  
  
    
\bibitem{Li:2018lyb}
  Z.~Li,
  ``Solving QED$_3$ with Conformal Bootstrap,''
  arXiv:1812.09281 [hep-th].
 


  
\bibitem{MarchRussell:1992ei} 
  J.~March-Russell,
  ``On the possibility of second order phase transitions in spontaneously broken gauge theories,''
  Phys.\ Lett.\ B {\bf 296}, 364 (1992)
  doi:10.1016/0370-2693(92)91333-5
  [hep-ph/9208215].
  
\bibitem{Nahum:2015jya} 
  A.~Nahum, J.~T.~Chalker, P.~Serna, M.~Ortuno and A.~M.~Somoza,
  ``Deconfined Quantum Criticality, Scaling Violations, and Classical Loop Models,''
  Phys.\ Rev.\ X {\bf 5}, no. 4, 041048 (2015)
  doi:10.1103/PhysRevX.5.041048
  [arXiv:1506.06798 [cond-mat.str-el]].
  
\bibitem{Nahum:2015vka} 
  A.~Nahum, P.~Serna, J.~T.~Chalker, M.~Ortuno and A.~M.~Somoza,
  ``Emergent SO(5) Symmetry at the N\`eel to Valence-Bond-Solid Transition,''
  Phys.\ Rev.\ Lett.\  {\bf 115}, no. 26, 267203 (2015)
  doi:10.1103/PhysRevLett.115.267203
  [arXiv:1508.06668 [cond-mat.str-el]].
  
\bibitem{Serna:2018tct}
  P.~Serna and A.~Nahum,
  ``Emergence and spontaneous breaking of approximate $O(4)$ symmetry at a weakly first-order deconfined phase transition,''
  arXiv:1805.03759 [cond-mat.str-el].
  
\bibitem{Nogueira:2013oza} 
  F.~S.~Nogueira and A.~Sudbo,
  ``Deconfined Quantum Criticality and Conformal Phase Transition in Two-Dimensional Antiferromagnets,''
  EPL {\bf 104}, no. 5, 56004 (2013)
  doi:10.1209/0295-5075/104/56004
  [arXiv:1304.4938 [cond-mat.str-el]].

\bibitem{Karch:2016sxi} 
  A.~Karch and D.~Tong,
  ``Particle-Vortex Duality from 3d Bosonization,''
  Phys.\ Rev.\ X {\bf 6}, no. 3, 031043 (2016)
  doi:10.1103/PhysRevX.6.031043
  [arXiv:1606.01893 [hep-th]].
  
\bibitem{Wang:2017txt} 
  C.~Wang, A.~Nahum, M.~A.~Metlitski, C.~Xu and T.~Senthil,
  ``Deconfined quantum critical points: symmetries and dualities,''
  Phys.\ Rev.\ X {\bf 7}, no. 3, 031051 (2017)
  doi:10.1103/PhysRevX.7.031051
  [arXiv:1703.02426 [cond-mat.str-el]].
  
\bibitem{Aharony:2011jz} 
  O.~Aharony, G.~Gur-Ari and R.~Yacoby,
  ``d=3 Bosonic Vector Models Coupled to Chern-Simons Gauge Theories,''
  JHEP {\bf 1203}, 037 (2012)
  doi:10.1007/JHEP03(2012)037
  [arXiv:1110.4382 [hep-th]].
  
\bibitem{Giombi:2011kc} 
  S.~Giombi, S.~Minwalla, S.~Prakash, S.~P.~Trivedi, S.~R.~Wadia and X.~Yin,
  ``Chern-Simons Theory with Vector Fermion Matter,''
  Eur.\ Phys.\ J.\ C {\bf 72}, 2112 (2012)
  doi:10.1140/epjc/s10052-012-2112-0
  [arXiv:1110.4386 [hep-th]].

\bibitem{Aharony:2012nh} 
  O.~Aharony, G.~Gur-Ari and R.~Yacoby,
  ``Correlation Functions of Large N Chern-Simons-Matter Theories and Bosonization in Three Dimensions,''
  JHEP {\bf 1212}, 028 (2012)
  doi:10.1007/JHEP12(2012)028
  [arXiv:1207.4593 [hep-th]].

\bibitem{Son:2015xqa} 
  D.~T.~Son,
  ``Is the Composite Fermion a Dirac Particle?,''
  Phys.\ Rev.\ X {\bf 5}, no. 3, 031027 (2015)
  doi:10.1103/PhysRevX.5.031027
  [arXiv:1502.03446 [cond-mat.mes-hall]].
      
\bibitem{Aharony:2015mjs} 
  O.~Aharony,
  ``Baryons, monopoles and dualities in Chern-Simons-matter theories,''
  JHEP {\bf 1602}, 093 (2016)
  doi:10.1007/JHEP02(2016)093
  [arXiv:1512.00161 [hep-th]].

\bibitem{Seiberg:2016gmd} 
  N.~Seiberg, T.~Senthil, C.~Wang and E.~Witten,
  ``A Duality Web in 2+1 Dimensions and Condensed Matter Physics,''
  Annals Phys.\  {\bf 374}, 395 (2016)
  doi:10.1016/j.aop.2016.08.007
  [arXiv:1606.01989 [hep-th]].
  
\bibitem{Karch:2016aux} 
  A.~Karch, B.~Robinson and D.~Tong,
  ``More Abelian Dualities in 2+1 Dimensions,''
  JHEP {\bf 1701}, 017 (2017)
  doi:10.1007/JHEP01(2017)017
  [arXiv:1609.04012 [hep-th]].
  
\bibitem{Metlitski:2016dht} 
  M.~A.~Metlitski, A.~Vishwanath and C.~Xu,
  ``Duality and bosonization of (2+1) -dimensional Majorana fermions,''
  Phys.\ Rev.\ B {\bf 95}, no. 20, 205137 (2017)
  doi:10.1103/PhysRevB.95.205137
  [arXiv:1611.05049 [cond-mat.str-el]].
 
\bibitem{Hsin:2016blu} 
  P.~S.~Hsin and N.~Seiberg,
  ``Level/rank Duality and Chern-Simons-Matter Theories,''
  JHEP {\bf 1609}, 095 (2016)
  doi:10.1007/JHEP09(2016)095
  [arXiv:1607.07457 [hep-th]].
  
\bibitem{Aharony:2016jvv} 
  O.~Aharony, F.~Benini, P.~S.~Hsin and N.~Seiberg,
  ``Chern-Simons-matter dualities with $SO$ and $USp$ gauge groups,''
  JHEP {\bf 1702}, 072 (2017)
  doi:10.1007/JHEP02(2017)072
  [arXiv:1611.07874 [cond-mat.str-el]].
  
\bibitem{Benini:2017dus} 
  F.~Benini, P.~S.~Hsin and N.~Seiberg,
  ``Comments on global symmetries, anomalies, and duality in (2 + 1)d,''
  JHEP {\bf 1704}, 135 (2017)
  doi:10.1007/JHEP04(2017)135
  [arXiv:1702.07035 [cond-mat.str-el]].

\bibitem{Benini:2017aed} 
  F.~Benini,
  ``Three-dimensional dualities with bosons and fermions,''
  JHEP {\bf 1802}, 068 (2018)
  doi:10.1007/JHEP02(2018)068
  [arXiv:1712.00020 [hep-th]].
  
\bibitem{Jensen:2017bjo} 
  K.~Jensen,
  ``A master bosonization duality,''
  JHEP {\bf 1801}, 031 (2018)
  doi:10.1007/JHEP01(2018)031
  [arXiv:1712.04933 [hep-th]].
  
\bibitem{Komargodski:2017keh} 
  Z.~Komargodski and N.~Seiberg,
  ``A symmetry breaking scenario for QCD$_{3}$,''
  JHEP {\bf 1801}, 109 (2018)
  doi:10.1007/JHEP01(2018)109
  [arXiv:1706.08755 [hep-th]].
  
\bibitem{Gomis:2017ixy} 
  J.~Gomis, Z.~Komargodski and N.~Seiberg,
  ``Phases Of Adjoint QCD$_3$ And Dualities,''
  SciPost Phys.\  {\bf 5}, 007 (2018)
  doi:10.21468/SciPostPhys.5.1.007
  [arXiv:1710.03258 [hep-th]].

\bibitem{Bashmakov:2018wts} 
  V.~Bashmakov, J.~Gomis, Z.~Komargodski and A.~Sharon,
  ``Phases of $ \mathcal{N}=1 $ theories in 2 + 1 dimensions,''
  JHEP {\bf 1807}, 123 (2018)
  doi:10.1007/JHEP07(2018)123
  [arXiv:1802.10130 [hep-th]].
  
\bibitem{Benini:2018umh} 
  F.~Benini and S.~Benvenuti,
  ``$\mathcal{N}{=}1$ dualities in 2+1 dimensions,''
  arXiv:1803.01784 [hep-th].
 
\bibitem{Choi:2018ohn} 
  C.~Choi, M.~Rocek and A.~Sharon,
  ``Dualities and Phases of $3D N=1$ SQCD,''
  JHEP {\bf 1810}, 105 (2018)
  doi:10.1007/JHEP10(2018)105
  [arXiv:1808.02184 [hep-th]].
           
\bibitem{Choi:2018tuh} 
  C.~Choi, D.~Delmastro, J.~Gomis and Z.~Komargodski,
  ``Dynamics of QCD$_{3}$ with Rank-Two Quarks And Duality,''
  arXiv:1810.07720 [hep-th].
 
\bibitem{Senthil:2018cru} 
  T.~Senthil, D.~T.~Son, C.~Wang and C.~Xu,
  ``Duality between $(2+1)d$ Quantum Critical Points,''
  arXiv:1810.05174 [cond-mat.str-el].


\bibitem{LSK} 
J.~Lou, A.W.~Sandvik and N.~Kawashima,
 ``Antiferromagnetic to valence-bond-soild transitions in two-dimensional SU(N) Heisenberg models with multi-spin interactions,''
Phys. Rev. B 80, 180414(R)
doi:10.1103/PhysRevB.80.180414 
  [arXiv:0908.0740 [cond-mat.str-el]].
  
\bibitem{Kaul:2011dqx} 
  R.~K.~Kaul and A.~W.~Sandvik,
  ``Lattice Model for the SU$(N)$ N\`eel to Valence-Bond Solid Quantum Phase Transition at Large $N$,''
  Phys.\ Rev.\ Lett.\  {\bf 108}, no. 13, 137201 (2012)
  doi:10.1103/PhysRevLett.108.137201
  [arXiv:1110.4130 [cond-mat.str-el]].
  
\bibitem{DEmidio:2016wwg} 
  J.~D'Emidio and R.~K.~Kaul,
  ``New easy-plane $\mathbb{CP}^{N-1}$ fixed points,''
  Phys.\ Rev.\ Lett.\  {\bf 118}, no. 18, 187202 (2017)
  doi:10.1103/PhysRevLett.118.187202
  [arXiv:1610.07702 [cond-mat.str-el]].
  
\bibitem{Zhang:2018bfc} 
  X.~F.~Zhang, Y.~C.~He, S.~Eggert, R.~Moessner and F.~Pollmann,
  ``Continuous Easy-Plane Deconfined Phase Transition on the Kagome Lattice,''
  Phys.\ Rev.\ Lett.\  {\bf 120}, no. 11, 115702 (2018)
  doi:10.1103/PhysRevLett.120.115702
  [arXiv:1706.05414 [cond-mat.str-el]].
   
\bibitem{Karthik:2016ppr} 
  N.~Karthik and R.~Narayanan,
  ``Scale-invariance of parity-invariant three-dimensional QED,''
  Phys.\ Rev.\ D {\bf 94}, no. 6, 065026 (2016)
  doi:10.1103/PhysRevD.94.065026
  [arXiv:1606.04109 [hep-th]].

\bibitem{Nakayama:2016jhq} 
  Y.~Nakayama and T.~Ohtsuki,
  ``Conformal Bootstrap Dashing Hopes of Emergent Symmetry,''
  Phys.\ Rev.\ Lett.\  {\bf 117}, no. 13, 131601 (2016)
  doi:10.1103/PhysRevLett.117.131601
  [arXiv:1602.07295 [cond-mat.str-el]].
  
\bibitem{DSD} 
  D.~Simmons-Duffin, unpublished.
  
\bibitem{IliesiuTALK} 
L.~Iliesiu, talk at Simons Center for Geometry and Physics: "The N\`eel-VBA quantum phase transition and the conformal bootstrap", 2018-11-05.
    
\bibitem{Poland:2018epd} 
  D.~Poland, S.~Rychkov and A.~Vichi,
  ``The Conformal Bootstrap: Theory, Numerical Techniques, and Applications,''
  arXiv:1805.04405 [hep-th].
  

    
    
\bibitem{Klebanov:2011td} 
  I.~R.~Klebanov, S.~S.~Pufu, S.~Sachdev and B.~R.~Safdi,
  ``Entanglement Entropy of 3-d Conformal Gauge Theories with Many Flavors,''
  JHEP {\bf 1205}, 036 (2012)
  doi:10.1007/JHEP05(2012)036
  [arXiv:1112.5342 [hep-th]].
  
     
 \bibitem{Pufu:2013vpa} 
  S.~S.~Pufu,
  ``Anomalous dimensions of monopole operators in three-dimensional quantum electrodynamics,''
  Phys.\ Rev.\ D {\bf 89}, no. 6, 065016 (2014)
  doi:10.1103/PhysRevD.89.065016
  [arXiv:1303.6125 [hep-th]].
  
\bibitem{Dyer:2015zha} 
  E.~Dyer, M.~Mezei, S.~S.~Pufu and S.~Sachdev,
  ``Scaling dimensions of monopole operators in the $ \mathbb{C}{\mathrm{\mathbb{P}}}^{N_b-1} $ theory in 2 $+$ 1 dimensions,''
  JHEP {\bf 1506}, 037 (2015)
  Erratum: [JHEP {\bf 1603}, 111 (2016)]
  doi:10.1007/JHEP03(2016)111, 10.1007/JHEP06(2015)037
  [arXiv:1504.00368 [hep-th]].
  
\bibitem{Diab:2016spb} 
  K.~Diab, L.~Fei, S.~Giombi, I.~R.~Klebanov and G.~Tarnopolsky,
  ``On ${C}_{J}$ and ${C}_{T}$ in the Gross-Neveu and O(N) models,''
  J.\ Phys.\ A {\bf 49}, no. 40, 405402 (2016)
  doi:10.1088/1751-8113/49/40/405402
  [arXiv:1601.07198 [hep-th]].
  
\bibitem{Giombi:2016fct} 
  S.~Giombi, G.~Tarnopolsky and I.~R.~Klebanov,
  ``On $C_{J}$ and $C_{T}$ in Conformal QED,''
  JHEP {\bf 1608}, 156 (2016)
  doi:10.1007/JHEP08(2016)156
  [arXiv:1602.01076 [hep-th]].
 
\bibitem{Chester:2017vdh} 
  S.~M.~Chester, L.~V.~Iliesiu, M.~Mezei and S.~S.~Pufu,
  ``Monopole Operators in $U(1)$ Chern-Simons-Matter Theories,''
  JHEP {\bf 1805}, 157 (2018)
  doi:10.1007/JHEP05(2018)157
  [arXiv:1710.00654 [hep-th]].
  
 
\bibitem{Murthy:1989ps} 
  G.~Murthy and S.~Sachdev,
  ``Action of Hedgehog Instantons in the Disordered Phase of the (2+1)-dimensional {CP}**($^{1}$N) Model,''
  Nucl.\ Phys.\ B {\bf 344}, 557 (1990).
  doi:10.1016/0550-3213(90)90670-9

\bibitem{Borokhov:2002ib} 
  V.~Borokhov, A.~Kapustin and X.~k.~Wu,
  ``Topological disorder operators in three-dimensional conformal field theory,''
  JHEP {\bf 0211}, 049 (2002)
  doi:10.1088/1126-6708/2002/11/049
  [hep-th/0206054].
  
\bibitem{Gaiotto:2018yjh} 
  D.~Gaiotto, Z.~Komargodski and J.~Wu,
  ``Curious Aspects of Three-Dimensional ${\cal N}=1$ SCFTs,''
  JHEP {\bf 1808}, 004 (2018)
  doi:10.1007/JHEP08(2018)004
  [arXiv:1804.02018 [hep-th]].
  
\bibitem{Benini:2018bhk} 
  F.~Benini and S.~Benvenuti,
  ``$N=1$ QED in 2+1 dimensions: Dualities and enhanced symmetries,''
  arXiv:1804.05707 [hep-th].
  
 
 
\bibitem{Halperin:1973jh}
  B.~I.~Halperin, T.~C.~Lubensky and S.~K.~Ma,
  ``First order phase transitions in superconductors and smectic A liquid crystals,''
  Phys.\ Rev.\ Lett.\  {\bf 32} (1974) 292.
  doi:10.1103/PhysRevLett.32.292
  
  
\bibitem{Hikami:1979ih}
  S.~Hikami,
  ``Renormalization Group Functions of $CP^{N-1}$ Nonlinear Sigma Model and $N$ Component Scalar {QED} Model,''
  Prog.\ Theor.\ Phys.\  {\bf 62} (1979) 226.
  doi:10.1143/PTP.62.226
  
\bibitem{Vas:1983}
A.N.~Vasil'ev, M.Yu.~Nalimov, 
  ``The $CP^{N-1}$ model: Calculation of anomalous dimensions and the mixing matrices in the order 1/N,''
Theor. Math. Phys. (1983) 56: 643.  
https://doi.org/10.1007/BF01027537

  
      
\bibitem{Kaul:2008xw} 
  R.~K.~Kaul and S.~Sachdev,
  ``Quantum criticality of U(1) gauge theories with fermionic and bosonic matter in two spatial dimensions,''
  Phys.\ Rev.\ B {\bf 77}, 155105 (2008)
  doi:10.1103/PhysRevB.77.155105
  [arXiv:0801.0723 [cond-mat.str-el]].

  

 \bibitem{Braun:2014wja} 
  J.~Braun, H.~Gies, L.~Janssen and D.~Roscher,
  ``Phase structure of many-flavor QED$_3$,''
  Phys.\ Rev.\ D {\bf 90}, no. 3, 036002 (2014)
  doi:10.1103/PhysRevD.90.036002
  [arXiv:1404.1362 [hep-ph]].

\bibitem{Janssen:2017eeu} 
  L.~Janssen and Y.~C.~He,
  ``Critical behavior of the QED$_3$-Gross-Neveu model: Duality and deconfined criticality,''
  Phys.\ Rev.\ B {\bf 96}, no. 20, 205113 (2017)
  doi:10.1103/PhysRevB.96.205113
  [arXiv:1708.02256 [cond-mat.str-el]].
    

\bibitem{Ihrig:2018ojl} 
  B.~Ihrig, L.~Janssen, L.~N.~Mihaila and M.~M.~Scherer,
  ``Deconfined criticality from the QED$_3$-Gross-Neveu model at three loops,''
  Phys.\ Rev.\ B {\bf 98}, no. 11, 115163 (2018)
  doi:10.1103/PhysRevB.98.115163
  [arXiv:1807.04958 [cond-mat.str-el]].
  

\bibitem{Zerf:2018csr} 
  N.~Zerf, P.~Marquard, R.~Boyack and J.~Maciejko,
  ``Critical behavior of the QED$_3$-Gross-Neveu-Yukawa model at four loops,''
  Phys.\ Rev.\ B {\bf 98}, no. 16, 165125 (2018)
  doi:10.1103/PhysRevB.98.165125
  [arXiv:1808.00549 [cond-mat.str-el]].
  
   
\bibitem{Xu:2008} 
C.~Xu,
  ``Renormalization group studies on four-fermion interaction instabilities on algebraic spin liquids,''
Phys. Rev. B 78, 054432  
doi:10.1103/PhysRevB.78.054432
[arXiv:0803.0794 [hep-th]].
  
\bibitem{Chester:2016ref} 
  S.~M.~Chester and S.~S.~Pufu,
  ``Anomalous dimensions of scalar operators in QED$_{3}$,''
  JHEP {\bf 1608}, 069 (2016)
  doi:10.1007/JHEP08(2016)069
  [arXiv:1603.05582 [hep-th]].
  

  
\bibitem{Rantner:2002zz}
  W.~Rantner and X.~G.~Wen,
  ``Spin correlations in the algebraic spin liquid: Implications for high-Tc superconductors,''
  Phys.\ Rev.\ B {\bf 66} (2002) 144501.
  doi:10.1103/PhysRevB.66.144501

\bibitem{Hermele:2005dkq}
  M.~Hermele, T.~Senthil and M.~P.~A.~Fisher,
  ``Algebraic spin liquid as the mother of many competing orders,''
  Phys.\ Rev.\ B {\bf 72} (2005) no.10,  104404
  doi:10.1103/PhysRevB.72.104404
  [cond-mat/0502215 [cond-mat.str-el]].


 \bibitem{Erratum}
 Michael Hermele, T. Senthil, and Matthew P. A. Fisher
Phys. Rev. B 76, 149906

  
\bibitem{Boyack:2018zfx}
  R.~Boyack, A.~Rayyan and J.~Maciejko,
  ``Deconfined criticality in the $\text{QED}_{3}$-Gross-Neveu-Yukawa model: the $1/N$ expansion revisited,''
  arXiv:1812.02720 [cond-mat.str-el].
  
    
\bibitem{Gracey:1993ka} 
  J.~A.~Gracey,
  ``Gauged Nambu-Jona-Lasinio model at O(1/N) with and without a Chern-Simons term,''
  Mod.\ Phys.\ Lett.\ A {\bf 8}, 2205 (1993)
  doi:10.1142/S0217732393001938
  [hep-th/9306105].
  
  
\bibitem{Gracey:1991ry}
  J.~A.~Gracey,
  ``Critical point analysis of various fermionic field theories in the large N expansion,''
  J.\ Phys.\ A {\bf 25} (1992) L109.

  
\bibitem{Gracey:1993zn}
  J.~A.~Gracey,
  ``Gauge independent critical exponents for QED coupled to a four fermi interaction with and without a Chern-Simons term,''
  Annals Phys.\  {\bf 224} (1993) 275
  doi:10.1006/aphy.1993.1047
  [hep-th/9301113].
  
\bibitem{Gracey:2018fwq} 
  J.~A.~Gracey,
  ``Fermion bilinear operator critical exponents at $O(1/N^2)$ in the QED-Gross-Neveu universality class,''
  Phys.\ Rev.\ D {\bf 98}, no. 8, 085012 (2018)
  doi:10.1103/PhysRevD.98.085012
  [arXiv:1808.07697 [hep-th]].
  
  
\bibitem{Gates:1983nr}
  S.~J.~Gates, M.~T.~Grisaru, M.~Rocek and W.~Siegel,
  Front.\ Phys.\  {\bf 58} (1983) 1
  [hep-th/0108200].
  
  

  
\bibitem{Intriligator:1996ex}
  K.~A.~Intriligator and N.~Seiberg,
  ``Mirror symmetry in three-dimensional gauge theories,''
  Phys.\ Lett.\ B {\bf 387} (1996) 513
  doi:10.1016/0370-2693(96)01088-X
  [hep-th/9607207].
  
\bibitem{Kapustin:1999ha}
  A.~Kapustin and M.~J.~Strassler,
  ``On mirror symmetry in three-dimensional Abelian gauge theories,''
  JHEP {\bf 9904} (1999) 021
  doi:10.1088/1126-6708/1999/04/021
  [hep-th/9902033].

\bibitem{Gremm:1999su}
  M.~Gremm and E.~Katz,
  ``Mirror symmetry for N=1 QED in three-dimensions,''
  JHEP {\bf 0002} (2000) 008
  doi:10.1088/1126-6708/2000/02/008
  [hep-th/9906020].
 
  
\bibitem{Gracey:1990ac}
  J.~A.~Gracey,
  ``Critical Exponents for the Supersymmetric $\sigma$ Model,''
  J.\ Phys.\ A {\bf 23} (1990) 2183.
  
\bibitem{Calabrese:2002bm} 
  P.~Calabrese, A.~Pelissetto and E.~Vicari,
  ``Multicritical phenomena in O(n(1)) + O(n(2)) symmetric theories,''
  Phys.\ Rev.\ B {\bf 67}, 054505 (2003)
  doi:10.1103/PhysRevB.67.054505
  [cond-mat/0209580].

  
  

     
\end{thebibliography}

\end{document}